\documentclass[a4paper,onecolumn,accepted=2022-10-26]{quantumarticle}

\pdfoutput=1

\usepackage[utf8]{inputenc}
\usepackage{caption}
\usepackage{relsize}
\usepackage{amsmath,amssymb,amsfonts}
\usepackage{csquotes}
\usepackage{wrapfig}
\usepackage{physics}
\usepackage{dsfont}
\usepackage{graphicx}
\usepackage{cite} 
\usepackage{tikz}
\usepackage{subcaption}
\usepackage{bbm}

\usepackage{mathtools}

\usepackage{empheq}

\definecolor{cof}{RGB}{219,144,71}
\definecolor{pur}{RGB}{186,146,162}
\definecolor{greeo}{RGB}{91,173,69}
\definecolor{greet}{RGB}{52,111,72}
\definecolor{bluey}{RGB}{72,73,120}
\usetikzlibrary{arrows,shapes,trees}
\usetikzlibrary{positioning,calc}
\usetikzlibrary{decorations.markings}

\newcommand{\ra}{\rightarrow}

\newcommand{\End}{{\rm End}}

\newcommand{\im}{{\rm im}}

\newcommand{\ZZ}{{\mathbb Z}}

\newcommand{\cH}{{\mathcal H}}

\newcommand{\cE}{\mathcal E}
\newcommand{\cL}{\mathcal L}

\newcommand{\cO}{\mathcal O}

\newcommand{\cU}{\mathcal U}
\newcommand{\cS}{\mathcal S}
\newcommand{\cT}{{\mathcal T}}

\newcommand{\Q}{{\mathcal Q}}

\newcommand{\EE}{{\mathbb E}}

\usepackage{hyperref}
\usepackage{cleveref}

\crefformat{section}{\S#2#1#3}
\crefformat{subsection}{\S#2#1#3}
\crefformat{subsubsection}{\S#2#1#3}

\usepackage{tikz-cd}

\hypersetup{colorlinks=true,urlcolor=[rgb]{0,0,0.5},citecolor=[rgb]{0,0.5,0},linkcolor=[rgb]{0,0.5,0}}

\numberwithin{equation}{section}

\usepackage[affil-it]{authblk}
\author[1,2]{Caroline de Groot}\email{caroline.de.groot@mpq.mpg.de}
\author[1,2]{Alex Turzillo}\email{alex.turzillo@mpq.mpg.de}
\author[3,4,1,2]{Norbert Schuch}\email{norbert.schuch@gmail.com}
\affil[1]{Max-Planck-Institut f\"ur Quantenoptik, Hans-Kopfermann-Stra{\ss}e 1, 85748 Garching, Germany}
\affil[2]{Munich Center for Quantum Science and Technology, Schellingstra{\ss}e 4, 80799 M\"unchen, Germany}
\affil[3]{University of Vienna, Faculty of Mathematics, Oskar-Morgenstern-Platz 1, 1090 Wien, Austria}
\affil[4]{University of Vienna, Faculty of Physics, Boltzmanngasse 5, 1090 Wien, Austria}
\title{Symmetry Protected Topological Order in Open Quantum Systems}

\begin{document}

\maketitle

\begin{abstract}

We systematically investigate the robustness of symmetry protected topological (SPT) order in open quantum systems by studying the evolution of string order parameters and other probes under noisy channels. We find that one-dimensional SPT order is robust against noisy couplings to the environment that satisfy a strong symmetry condition, while it is destabilized by noise that satisfies only a weak symmetry condition, which generalizes the notion of symmetry for closed systems. We also discuss ``transmutation'' of SPT phases into other SPT phases of equal or lesser complexity, under noisy channels that satisfy twisted versions of the strong symmetry condition.

\end{abstract}

\tableofcontents

%%%%%%%%%%%%%%%%%%%%%%%%
%%%%%%%%%%%%%%%%%%%%%%%%

\section{Introduction}

The realization that the interplay of symmetries and entanglement can give rise to novel physics beyond the Landau paradigm has led to an expanding zoo of topologically ordered phases of matter. A particularly prominent role among those phases, in particular in one dimension (1D), is played by symmetry protected topological (SPT) phases, which have their root in Haldane's original work elucidating the gapped nature of the spin-1 Heisenberg chain and its topological origin, nowadays known as the Haldane phase~\cite{HALDANE1983464,PhysRevLett.50.1153,PhysRevLett.59.799}. SPT phases do not possess intrinsic topological order as their ground states lack long-range entanglement, yet they nevertheless exhibit topological phenomena such as non-local order parameters.

SPT phases consist of systems with a unique ground state and a gap, yet which are distinct from the trivial (mean-field) gapped phase, as witnessed by a number of characteristic fingerprints: most prominently, string order \cite{PhysRevB.40.4709,PhysRevB.45.304,pollmann2012detection}, specific degeneracies in the entanglement spectrum~%
\cite{pollmann:1d-sym-protection-prb}, and fractionalized edge excitations \cite{PhysRevLett.59.799}. A key step toward the comprehensive understanding of SPT phases was made by using Matrix Product State (MPS) representations~\cite{schollwock,cirac2021matrix} of their ground states. This step was based on the fact that MPS faithfully approximate ground states of gapped systems~\cite{hastings2007area,verstraete:faithfully,schuch:mps-entropies} and that they allow one to realize global symmetries locally on tensors  that carry physical and  entanglement degrees of  freedom~\cite{cirac2021matrix,molnar:normal-peps-fundamentalthm}. Namely, it was  understood that in  nontrivial SPT phases, the physical symmetry that protects the phase acts on entanglement as a projective, rather than linear, representation.
This insight was key in several ways. First, it provided a unified explanation for the aforementioned fingerprints of SPT phases in terms of this projective action. Second, it allowed one to obtain a comprehensive classification of SPT phases, based on the classification of projective representations by group cohomology. And finally, it connected the characterization of SPT phases based on fingerprints like string order and edge modes to the characterization based on the equivalence relation by which two systems are in the same phase if they can be connected by a path of gapped, symmetric Hamiltonians~\cite{PhysRevB.85.075125,chen2011classification,schuch2011classifying}.
Altogether, the representation of SPT states by MPS clarified and unified the various definitions for SPT order (SPTO) for the ground states of gapped Hamiltonian systems and allowed for their complete classification~\cite{chen:spt-bosons,cirac2021matrix}.

The situation becomes much less clear when moving from pure ground states to mixed states, which are the states we expect to appear in realistic physical systems. Several questions arise.
\emph{First}, which states should we consider? Depending on the scenario, relevant states might be thermal states of Hamiltonians \cite{PhysRevA.71.062313,Hastings_2011,Roberts_2017}, equilibrium states of dissipative evolutions (steady states of Lindbladians) \cite{Diehl_2011, Bardyn_2013,Kraus_2008,zhou2017symmetryprotected,PhysRevLett.125.240405}, or states -- for instance, originally pure states with SPT order -- which have been subjected to noise, which could be either Markovian or discrete-time non-Markovian noise.
\emph{Second}, what is the correct generalization of the symmetry condition? For Lindbladian noise, at least two different symmetry conditions have been considered~\cite{albert2018lindbladians,Bu_a_2012,PhysRevA.89.022118,lieu2020albertQECsym};
 they differ in how the symmetry is imposed on the joint system-bath interaction, and for discrete noise, even further symmetry conditions are conceivable.
\emph{Third}, which notion should one use for SPTO? The various fingerprints of SPTO could give divergent results, or might even be ill-defined, on mixed states. String order parameters can be defined for any state, but it is a priori not clear whether the patterns they exhibit are meaningful. For other fingerprints, such as entanglement spectra or edge modes, it is even unclear how to define them for mixed states.
\emph{Fourth}, are any of these fingerprints, which are defined on individual systems, compatible with the notion of SPT phase based on equivalence relations, analogous to paths of symmetric gapped Hamiltonians? 
All in all, in the quest to understand SPTO in the presence of noise, any approach must address these questions.

In this paper, we systematically investigate the robustness of SPTO under various types of symmetric noise. To this end, we characterize SPTO through string order parameters. These are constructed by placing local order parameters (labeled by irreducible representations) at the endpoints of strings of symmetry operators (labeled by group elements). In gapped phases, any string order parameter either decays exponentially to zero as the separation of the endpoints is increased, or converges to a constant whose value depends on the specific order parameter chosen, and which is generically nonzero. For ground states, the resulting pattern of zeros and nonzeros, as a function of the irrep and group element labels of the string, is a fingerprint of the SPT phase. In many cases, including all abelian symmetry groups, the pattern is in one-to-one correspondence with the SPT phases protected by the symmetry~\cite{pollmann2012detection}. We say that a mixed state has some SPT order if it exhibits the same pattern of zeros and nonzeros as pure states with the same SPTO; if a symmetric mixed state (such as a mixture of different pure SPT phases) exhibits a pattern which cannot appear in pure symmetric states, it is said to have no SPTO at all (not even trivial SPTO). This definition has several advantages: it coincides with the pure state definition in the limit of pure states, and, being an expectation value of an operator with tensor product structure, it is both simple to compute and to measure.

We study the robustness of SPTO, as witnessed by string order, for systems subject to evolution by general symmetric and locality-preserving noise. We consider both discrete-time evolutions, that is, quantum channels, as well as continuous noise described by Lindbladians. The Lindbladian evolution forms a special case of quantum channels, where the semigroup structure constrains the possibilities for the action of the symmetry on the channels at finite times. Locality-preservation encompasses both noise obtained from local Lindbladians and locality-preserving evolutions which are not locally
generated but which appear, for example, at the boundaries of two-dimensional systems, in driven systems, and via coupling to non-Markovian baths. We introduce two different notions of symmetry of quantum channels -- strong symmetry and weak symmetry. Weakly symmetric  channels are invariant under the symmetry action, which seems the natural definition of a symmetric channel, and corresponds to symmetric system-bath interactions where the symmetry acts simultaneously on the system and the bath. Strongly symmetric channels, on the other hand, have the property that each Kraus operator
individually commutes with the symmetry up to a constant phase factor, and correspond to symmetric system-bath interactions where the symmetry only acts on the system. Specializing these concepts to Lindbladian channels, where the semigroup structure imposes additional restrictions, we recover the notions of strong and weak symmetry studied by Albert and
others~\cite{albert2018lindbladians,Bu_a_2012,PhysRevA.89.022118,lieu2020albertQECsym}.

Our main result is that SPTO is robust against locality-preserving noise that satisfies the stronger of these two symmetry conditions. To be precise, we prove that the (local) strong symmetry condition on locality-preserving noise is \emph{sufficient} to preserve SPTO (\hyperref[lemma2]{Lemma 2}) and, conversely, that strong symmetry is \emph{necessary} for channels generated in finite time by strictly local Lindbladian evolution (\hyperref[theorem1]{Theorem 1}), which we conjecture to hold for all local Lindbladian evolutions (\cref{loclindstring}). This result might appear surprising in light of the work of Coser and P\'erez-Garc\'ia, who show that symmetric local Lindbladian noise, applied for a short amount of time, destroys SPTO~\cite{coser2019classification}. As we demonstrate, this is due to the fact that their noise is only weakly symmetric, and not strongly symmetric. We thus find that SPTO is robust to noise that satisfies a sufficiently strong yet natural symmetry condition; namely that the system-bath coupling is invariant under the symmetries acting on the system alone, as opposed to jointly on the system and bath.

Finally, we extend our analysis to the scenario where the noise does not commute with the symmetry but rather acts by exchanging symmetries, either permuting them or identifying their group actions. We term channels with this property \emph{twisted strongly symmetric}. In the case of permutation of symmetries, we demonstrate that the noise acts by permuting SPT phases with the same complexity, whereas, in the other case, the noise reduces complexity (\hyperref[theorem2]{Theorem 2}). Symmetric channels with a nontrivial twist cannot be generated by continuous evolution by a symmetric Lindbladian in finite time; thus, they are only relevant to scenarios with discrete noise and to infinite time evolution.

The paper is structured as follows. In \cref{sec:states}, we review states
with SPT order, and how their SPT phase is characterized through string
order parameters. In \cref{sec:strongsymmetry}, we introduce the weak and
strong symmetry conditions on arbitrary quantum channels, discuss
interpretations of the conditions in terms of conservation laws,
purifications, and couplings between the system and the environment, and
finally investigate the form of the symmetry conditions when applied to
Lindbladians. In \cref{sec:preserveSPTO-singlesite}, we focus on
uncorrelated noise, for which there is a decomposition of the channel as a
tensor product over sites of the lattice. Several SPT order parameters --
string operators, twisted sector charges, edge modes, and irrep
probabilities -- are studied analytically and numerically, and it is
proven that strong symmetry is necessary and sufficient for a noisy
evolution to preserve SPTO. In \cref{sec:causalchannels}, we discuss the
extension of this result from uncorrelated noise to causal (that is,
locality-perserving) noise, which includes the case of fast, local
Lindbladians. In \cref{sec:twistedstrong}, we broaden our investigation to
causal channels which act by exchanging symmetry actions.  Channels
satisfying \emph{twisted symmetry conditions} are shown to transform
between SPT phases, and we state a necessary and sufficient condition for a channel to preserve a given SPTO.

%%%%%%%%%%%%%%%%%%%%%%%
%%%%%%%%%%%%%%%%%%%%%%%

\section{SPT states and their invariants}\label{sec:states}

Let us begin by reviewing the invariants of pure SPT states, including their manifestation in tensor networks and in patterns of zeros of string operators. After this review, we will discuss how these invariants appear in a special class of mixed states we dub \emph{coherent SPT mixtures}.

\subsection{SPTO of matrix product states}\label{sec:SPT_MPS}

Symmetry protected topological phases of gapped, local Hamiltonians in one dimension, for a symmetry group $G$, are classified by an invariant $[\omega]$, a class in the second group cohomology group $H^2(G,U(1))$ \cite{PhysRevB.85.075125,chen2011classification,schuch2011classifying}. A useful method for determining the SPT invariant of a given Hamiltonian is to represent its ground state as a tensor network and study the symmetries of its tensor \cite{cirac2021matrix}. Let us review this procedure.

Our analysis of one-dimensional SPT states makes use of matrix product states (MPS), which efficiently approximate states obeying an entanglement entropy area law, and as such are applicable to ground states of gapped, local Hamiltonians \cite{hastings2007area,verstraete2006matrix}. A translation-invariant MPS is defined by a single rank-three tensor $A$ as
\begin{equation}\label{mpstens}
    \ket{\Psi[A]}= \sum_{i_1, \dots, i_N} \Tr(A^{i_1} \dots A^{i_N}) \ket{i_1 \dots i_N},
\end{equation}
where $A^i$ are matrices such that $A=\sum_i A^i \otimes \ket{i}$. As a tensor diagram, the MPS is written as
\begin{equation}\begin{split}
    \includegraphics[height = 0.045\textheight]{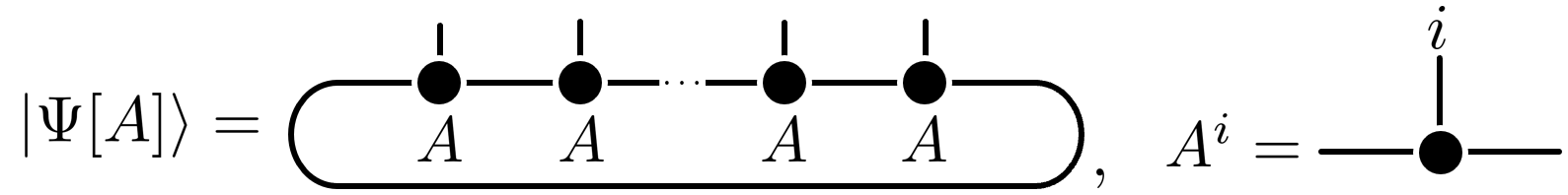}
    \quad\raisebox{0.01\textheight}{.}
\end{split}\end{equation}
An MPS tensor is said to be \emph{injective} if its transfer matrix $T=\sum_i A^i \otimes \overline{A^i}$ has a nondegenerate eigenspace of highest weight.  For a refresher on MPS technology, we refer readers to more comprehensive introductory literature \cite{biamonte2017tensor,ORUS2014117,handwaving} as well as to a recent review \cite{cirac2021matrix}.

Injective MPS satisfy a fundamental theorem \cite{cirac2021matrix}, which implies the following. If a state is invariant under a global symmetry $U_g^{\otimes N} \ket{\Psi[A]} = \ket{\Psi[A]}$, the action of the onsite symmetry $U_g$ on its local tensor $A$ results in an action on the virtual level
\begin{equation}\label{fundtheorem}
    \sum_j (U_g)_{ij} A^j= e^{i\phi_g} V_g A^i V_g^{\dagger}~,
\end{equation}
where $V_g$ is a projective representation of the symmetry \cite{perez2006matrix}, satisfying
\begin{equation}
     V_g V_h = \omega(g,h)\,V_{gh}~,
\end{equation}
for some values $\omega(g,h)$. As a tensor diagram, Eq. \eqref{fundtheorem} is written as
\begin{equation}\begin{split}
    \includegraphics[height = 0.055\textheight]{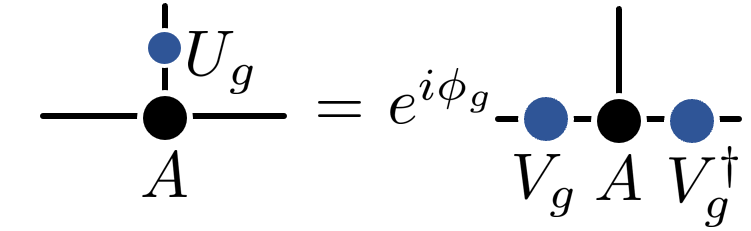}
    \quad\raisebox{0.024\textheight}{.}
\end{split}\end{equation}
For a given state, there is always a tensor in a canonical form, where $V_g$ is unitary and $\omega(g,h)$ is a phase \cite{cirac2021matrix}. The collection of phases $\omega(g,h)$ constitutes a group cocycle and is defined up to a group coboundary, meaning it determines a cohomology class $[\omega]$ \cite{schuch2011classifying,chen2011classification}. It turns out that $[\omega]$ is invariant along smooth paths of gapped, local, symmetric Hamiltonians, which is to say it is an SPT phase invariant; moreover, it is a complete invariant \cite{schuch2011classifying}. Physically, the virtual space of the MPS tensor may be interpreted as the space of edge modes; the fact that there is a minimal bond dimension on which $V_g$ can realize the invariant $[\omega]$ means that some of the edge modes are protected by the symmetry.

Essential to this definition of the SPT invariant $[\omega]$ is that the state is well-approximated by an MPS of bond dimension constant in the system size, a property which comes from the state being a ground state of a gapped, local Hamiltonian. Generic states in one dimension are not well-approximated by MPS, so for them a projective action $V_g$ -- and therefore an invariant $[\omega]$ -- cannot be defined this way.

%%%%%%%%%%%%%%%%%%%%%%%%%%%%%%%%

\subsection{String order}\label{stringsingle}

String order provides an alternate definition of the SPT invariant that does not rely on an MPS representation of the state. The string order parameter is a set of expectation values of string operators that can be defined on any state. On certain well-behaved states, such as the ground states of gapped Hamiltonians, it yields a well-defined pattern of zeros that is related to the SPT invariant $[\omega]$.

For the remainder of the paper, we take $G$ to be a finite abelian group. This assumption ensures that string order is a complete invariant of SPT phases, in that it uniquely determines $[\omega]$.

Assume $G$ is a finite abelian group, and let $U_g$ denote the action of $G$ on an individual site. The string operator is defined as
\begin{equation}\label{stringop}
    s(g,O^l_\alpha,O^r_\alpha)=\mathds{1}\otimes O^l_\alpha\otimes U_g^{\otimes j}\otimes O^r_\alpha\otimes\mathds{1}~,
\end{equation}
for some length $j$, where the end operators $O^{l,r}$ live in \emph{opposite} irreps of the adjoint action
\begin{equation}
    U_h^\dagger O^l_\alpha U_h=\chi_\alpha(h)O^l_\alpha~,\qquad U_h^\dagger O^r_\alpha U_h=\chi_\alpha^*(h)O^r_\alpha~.
\end{equation}
On some states, the string order parameter obeys a selection rule \cite{pollmann2012detection}. This rule says that, for each $g\in G$, there is a unique character $\alpha_g$ such that the expectation value $\langle s(g,O_\alpha^l,O_\alpha^r)\rangle$ vanishes, for all end operators, except for $\alpha=\alpha_g$. The values forced to vanish by the selection rule form what is called the \emph{pattern of zeros} of the state. An SPT state can be defined as a state for which this selection rule holds; that is, a state with a well-defined pattern of zeros. The SPT invariant of an SPT state is extracted by defining the ratios $\omega/\omega$ in terms of the unique character $\alpha_g$ for each $g$ as follows:
\begin{equation}\label{constraint}
    \frac{\omega(h,g)}{\omega(g,h)}=\chi_{\alpha_g}(h)~.
\end{equation}
The invariant $[\omega]$ may then be recovered from these ratios, as we argue in \cref{sec:patternsendos}. We remark that string order is also defined when $G$ is nonabelian; however, it may not determine $[\omega]$ uniquely \cite{pollmann2012detection}.

Now let us consider a state represented as an MPS and show that the definition of the SPT invariant $[\omega]$ by the projective action on edge modes agrees with its definition by string order parameters \cite{pollmann2012detection}. Injectivity may be achieved by blocking, which does not change the form of the string operator as long as its length is assumed to be large compared to the size of the blocks. On such a state, the string operator evaluates to
\begin{equation}\label{stringordeval}
    \langle s(g,O^l_\alpha,O^r_\alpha)\rangle=E_l(g,O_\alpha^l)E_r(g,O_\alpha^r)~,
\end{equation}
where
\begin{equation}
    E_{l,r}(g,O_\alpha^{l,r})=\Tr[N_{l,r}^gO_\alpha^{l,r}]
\end{equation}
are defined as
\begin{equation}\begin{split}
    \includegraphics[height = 0.09\textheight]{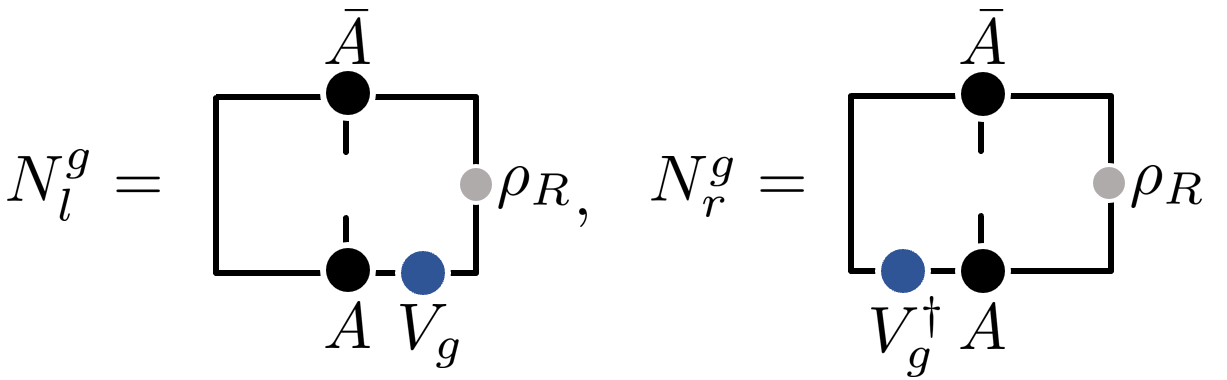}
    \quad\raisebox{0.04\textheight}{.}
\end{split}\end{equation}
Here, $\rho_R$ is the unique (by injectivity) right fixed point of the MPS transfer matrix, and we have used the canonical form where the left fixed point is the identity. The evaluation \eqref{stringordeval} can be seen with the following diagrammatic argument (due to Ref. \cite{pollmann2012detection}):
\begin{equation}\begin{split}\label{evaluatestring1}
    \includegraphics[height=0.09\textheight]{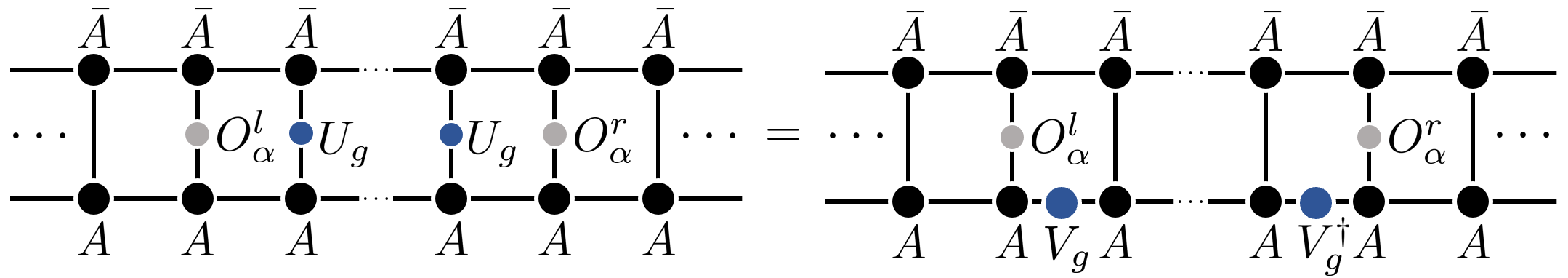}
    \quad\raisebox{0.04\textheight}{.}
\end{split}\end{equation}
Evaluating the above
\begin{equation}\begin{split}\label{evaluatestring2}
    \includegraphics[height=0.09\textheight]{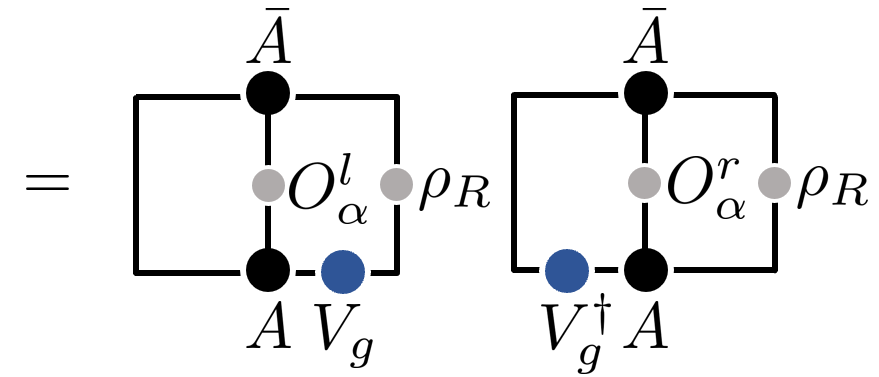}
    \quad\raisebox{0.04\textheight}{.}
\end{split}\end{equation}
The operators $N_{l,r}^g$ transform as
\begin{equation}
    U_h^\dagger N_l^gU_h=\frac{\omega(g,h)}{\omega(h,g)}N_l^g~,\qquad U_h^\dagger N_r^gU_h=\left(\frac{\omega(g,h)}{\omega(h,g)}\right)^*N_r^g~,
\end{equation}
The operators $N_{l,r}^{g\dagger}$ and $O_\alpha^{l,r}$ are orthogonal (and so $E_{l,r}$ is zero) unless they transform the same way; that is, unless the selection rule \eqref{constraint} is satisfied. This completes the argument.

If the operators $N_{l,r}^{g\dagger}$ and $O_\alpha^{l,r}$ transform the same way, then they are not orthogonal \emph{generically} (and so $E_{l,r}$ is nonzero generically). This is because $N_{l,r}^g$ picks out a single direction in the multiplicity space for $\alpha$, so the subspace orthogonal to this direction is codimension one in the full space of end operators $O_\alpha^{l,r}$. In other words, for a generic choice of end operators, the only expectation values $\langle s(g,O_\alpha^l,O_\alpha^r)\rangle$ that vanish are those that belong to the pattern of zeros determined by the SPT invariant. The nonzero values of the string order parameter depend on the choice of end operators, but the pattern of zeros does not.

%%%%%%%%%%%%%%%%

\subsection{Coherent SPT mixtures}

Now we turn to SPT invariants of open systems. In the formalism of Lindbladian evolution \cite{coser2019classification}, phases of open systems are defined in terms of states, as opposed to with some open systems analog of gapped paths of Hamiltonians. We do not attempt to answer the questions of which mixed states generalize the ground states of gapped, local Hamiltonians and what is their phase classification under an appropriate equivalence relation. Instead we seek to motivate the strong symmetry condition on Lindbladian evolution by focusing on a special class of mixed states for which we can define an invariant:
\begin{equation}\label{SPTmixed}
\fbox{
$\qquad$\parbox{0.55\linewidth}{
\centering
\textbf{Definition:} \emph{A coherent SPT mixture is a mixed state with a well-defined pattern of zeros.}
}$\qquad$
}
\end{equation}
Such states have a well-defined SPT invariant $[\omega]$ that can be extracted from the pattern of zeros, as discussed above. Mixed states that are ensembles
\begin{equation}\label{ensemble}
    \rho=\sum_ip_i|\psi_i^\omega\rangle\langle\psi_i^\omega|
\end{equation}
of SPT pure states $|\psi_i^\omega\rangle$ all in the same SPT phase $[\omega]$ are examples of coherent SPT mixtures. We leave open the possibility that there exist exotic coherent SPT mixtures that are not covered by this example.\footnote{Components of the ensemble that have a different SPTO or no SPTO at all could cancel exactly the expectation values of the string operators, yielding a well-defined pattern of zeros.}

The main claim of the paper is that translation-invariant pure SPT states (and more generally, states of the form \eqref{ensemble} where each component $|\psi_i^\omega\rangle$ is translation-invariant) are transformed into coherent SPT mixtures \eqref{SPTmixed} with the same SPT invariant by a Lindbladian evolution if and only if the evolution is strongly symmetric. We prove this claim for uncorrelated noise in \cref{sec:preserveSPTO-singlesite} and show its `if' direction (while conjecturing its `only if' direction) for fast, local Lindbladians in \cref{sec:causalchannels}. In particular, this result means that SPTO of pure states is robust in open systems described by strongly symmetric Lindbladians.

%%%%%%%%%%%%%%%%%%%%%%%%%%%%%%%%%%
%%%%%%%%%%%%%%%%%%%%%%%%%%%%%%%%%

\section{A strong symmetry condition on channels}\label{sec:strongsymmetry}

We begin by introducing the \emph{weak} \eqref{WSC} and \emph{strong} \eqref{SSC} \emph{symmetry conditions} and discussing their various formulations. The latter is motivated by showing that weak symmetry is insufficient to preserve SPTO. The argument that strong symmetry is necessary and sufficient to preserve SPTO is reserved for \cref{sec:preserveSPTO-singlesite}, \cref{sec:causalchannels}.

The symmetry conditions are first formulated and studied for arbitrary quantum channels. Then in \cref{sym-lind}, for the particularly important case of Lindbladian evolution $\cE_t=e^{t\cL}$, the conditions are reformulated in terms of $\cL$; the Lindbladian formulations have been discussed previously \cite{albert2018lindbladians,Bu_a_2012,PhysRevA.89.022118,lieu2020albertQECsym}. The definitions and results in this section apply to general systems on finite-dimensional Hilbert spaces, not just spin chains; $U_g$ denotes the action of the symmetry on the full system, not on a single site of a spin chain.

%%%%%%%%%%%%%%%%%%%%%%%%

\subsection{Weak and strong symmetry conditions}\label{sec:weakstrong}

A channel $\cE$ is said to satisfy the \emph{weak symmetry (WS) condition} if it commutes, as a superoperator, with the symmetry-implementing channels
\begin{equation}\label{symchannel}
    \cU_g(\rho)=U_g\rho\,U_g^\dagger~;
\end{equation}
that is, if
\begin{equation}\label{WSC}
    \boxed{\qquad\cU_g\circ\cE\circ\cU_g^\dagger=\cE~,\quad\forall\,g~.\qquad(\text{weak symmetry condition})\qquad}
\end{equation}
The channel $\mathcal E$ can be expressed in terms of a Kraus representation,  $\cE(\rho)=\sum K_i\rho K_i^\dagger$, where we can interpret the $K_i$ as representing different trajectories. 
In terms of a Kraus representation of $\cE$,  the weak symmetry condition reads
\begin{equation}\label{weakkraus}
    \sum_i (U_gK_iU_g^\dagger)\,\rho\,\,(U_gK_iU_g^\dagger)^\dagger=\sum_iK_i\rho\,K_i^\dagger~,\quad\forall\,g~.
\end{equation}
Since $K_i$ and $U_gK_iU_g^\dagger$ define Kraus representations of the same channel, they are related by a unitary $x^g$ \cite{nielsen2002quantum}:
\begin{equation}\label{xg}
    U_gK_iU_g^\dagger=\sum_jx_{ji}^gK_j~,\quad\forall\,i,g~.
\end{equation}
Since $U_g$ forms a representation of $G$, so does $x^g$:
\begin{equation}
    \sum_kx_{ki}^{gh}K_k = 
    U_{gh}K_iU_{gh}^\dagger=U_gU_hK_iU_h^\dagger U_g^\dagger=\sum_{jk}x_{ji}^hx_{kj}^gK_k
    \ ,\quad\forall\,i,g~.
\end{equation}
In a basis of Kraus operators $K_i^g$ that diagonalizes $x^g$ as a collection of phases $\theta_i(g)$, \eqref{xg} amounts to 
\begin{equation}\label{WSCbasis}
    U_gK_i^gU_g^\dagger=e^{i\theta_i(g)}K_i^g~,\quad\forall\,i,g~.
\end{equation}
The existence of a basis $K_i^g$, for each $g$, such that this relation holds is equivalent to the WS condition.

Observe that the phases $\theta_i$ in the WS condition \eqref{WSCbasis} may differ across the trajectories (labeled by $i$). It will be demonstrated in \cref{sec:preserveSPTO-singlesite} that this interference between the trajectories is the source of the destruction of SPTO, as eliminating it by setting the phases equal is sufficient to ensure that a channel preserves SPTO. Let us now take the phases to be equal: $\theta_i(g)=\theta(g)$ for all $i,g$. Under this restriction, the condition \eqref{WSCbasis} is independent of the basis of Kraus operators because $K_i^g{}'=\sum_jv_{ij}K_j^g$ (for any unitary $v_{ij}$) satisfies
\begin{equation}
    U_gK_i^g{}'U_{g}^\dagger=\sum_j v_{ij}U_gK_j^gU_{g}^\dagger=e^{i\theta(g)}\sum_j v_{ij}K_j^g=e^{i\theta(g)}K_i^g{}'~.
\end{equation}
Basis-independence means we can also drop the group label on the Kraus operators: $K_i^g=K_i$ for all $i,g$. We arrive at what we call the \emph{strong symmetry (SS) condition}:
\begin{equation}\label{SSC}
    \boxed{\qquad U_gK_iU_g^\dagger=e^{i\theta(g)}K_i~,\quad\forall\,i,g~.\qquad(\text{strong symmetry condition})\qquad}
\end{equation}
Note that there is no distinction between the WS and SS conditions for a reversible (unitary) channel, as such a channel is realized by a single Kraus operator, and so it has only a single phase $\theta(g)$.

By Schur's lemma, the SS condition \eqref{SSC} may be restated as the condition that each $K_i$ is block-diagonal in the irrep basis: $K_i=\oplus_\alpha K_i^\alpha$, where $K_i^\alpha$ acts on the multiplicity space of the (isomorphism class of the) irrep $\alpha$. The completeness relation $\sum_iK_i^\dagger K_i=\mathds{1}$ is equivalent to a completeness relation on each block, so the channel decomposes as $\cE=\oplus_\alpha\cE_\alpha$, where $\cE_\alpha$ is the channel with Kraus operators $K_i^\alpha$. This decomposition is a stronger constraint than the decomposition of WS channels, which, when viewed as matrices on the space of operators, have a block-diagonal form $\cE=\sum_\alpha\Phi_\alpha$ in the irrep basis of the action $U_g\otimes U_g^\dagger$; note that the operators $\Phi_\alpha$ are different from $\cE_\alpha$ and are not themselves channels.

%%%%%%%%%%%%%%%%%%%

\subsection{Charge conservation}\label{chargecons}

The strong symmetry condition may be alternatively characterized as
\begin{equation}\label{SSCalt}
    \boxed{\qquad \cE^\dagger(U_g)=e^{i\theta(g)}U_g~,\quad\forall\,g~.\qquad(\text{strong symmetry condition})\qquad}
\end{equation}
where the dual channel $\mathcal E^\dagger$ is the channel with Kraus operators $K_i^\dagger$, $\mathcal E^\dagger(X)=\sum K_i^\dagger X K_i$.
This alternative statement may be interpreted as conservation of symmetry charge. The charge of a state under a symmetry $g$ is the expectation value $\langle U_g\rangle_\rho=\Tr[\rho\,U_g]$ of the operator $U_g$ on the state, and the strong symmetry condition means this expectation value is the same (up to a phase) for $\rho$ and $\cE(\rho)$:
\begin{equation}
    \langle U_g\rangle_{\cE(\rho)}=\langle\cE^\dagger(U_g)\rangle_\rho\stackrel{SS}{=}e^{i\theta(g)}\langle U_g\rangle_\rho~.
\end{equation}
The connection between strong symmetry and conservation laws has been noted previously \cite{albert2018lindbladians}.

The equivalence of the two statements may be seen as follows. If a channel satisfies Eq.~\eqref{SSC}, then
\begin{equation}
    \cE^\dagger(U_g)=\sum_iK_i^\dagger U_g K_i=e^{i\theta(g)}\sum_iK_i^\dagger K_iU_g=e^{i\theta(g)}U_g~,
\end{equation}
which is Eq.~\eqref{SSCalt}. For the converse, we need a lemma: if $\cE^\dagger(X)=Y$ and $\cE^\dagger(X^\dagger X)=Y^\dagger Y$, then $XK_i = K_iY$. If this is true, the statement that Eq.~\eqref{SSCalt} implies Eq.~\eqref{SSC} follows from taking $X=U_g$ and $Y=e^{i\theta(g)}U_g$. The lemma is proved by borrowing the argument for Theorem 6.13 of Ref. \cite{wolf}: 
\begin{equation}
    \sum_i(XK_i-K_iY)^\dagger(XK_i-K_iY)=\cE^\dagger(X^\dagger X)-\cE^\dagger(X^\dagger)Y-Y^\dagger\cE^\dagger(X)+Y^\dagger\cE^\dagger(\mathds{1})Y=0~.
\end{equation}
Then, since the left hand side is a sum of positive terms, each of them must individually vanish: $XK_i=K_iY$.

%%%%%%%%%%%%%%%%%%%%%%%%

\subsection{Symmetric purifications}\label{sec:purif}

The symmetry conditions can be restated in terms of purifications:
\begin{equation}\label{purifclaim1}
\fbox{
$\qquad$\parbox{0.75\linewidth}{
\centering
\textbf{Claim:} \emph{A channel is weakly symmetric if it has a purification to a unitary that commutes, up to a phase, with some diagonal symmetry $U_g\otimes U_g^A$, for which the action $U_g^A$ on the ancillary space leaves the ancilla state invariant.}
}$\qquad$
}
\end{equation}
\begin{equation}\label{purifclaim2}
\fbox{
$\qquad$\parbox{0.75\linewidth}{
\centering
\textbf{Claim:} \emph{A channel is strongly symmetric if and only if it has a purification to a unitary that commutes, up to the phase $e^{i\theta(g)}$, with the symmetry $U_g\otimes\mathds{1}^A$.}
}$\qquad$
}
\end{equation}
We do not prove a converse to the first claim, though we expect it or a similar statement to hold. The second claim means that one may take $U_g^A=\mathds{1}^A$ precisely when the channel is strongly symmetric. These statements will come in handy in \cref{sec:causalchannels}, when we discuss causal channels in terms of their purifications to matrix product unitaries. The statements also have interpretations in terms of couplings between the system and environment, which we discuss here. First let us review the basics of purifications and justify the claims.

Let $\cE$ be a channel on a system with Hilbert space $\cH$. A purification of $\cE$ is a unitary $W$ on a space $\cH\otimes A$ -- the original space appended with an ancillary space -- such that, for some ancilla state $|a\rangle\in A$,
\begin{equation}\label{purif}
    \Tr_A(W(\rho\otimes|a\rangle\langle a|)W^\dagger)=\cE(\rho)~.
\end{equation}
A purification $W$ always exists. Given a set of Kraus operators $K_i$, indexed in a set $\mathcal I$, form the ancillary space $A$ spanned by an orthonormal basis $|e_i\rangle$, $i\in\mathcal I$ and the operator $V:\cH\ra\cH\otimes A$ that acts as
\begin{equation}
    V:|\psi\rangle\mapsto\sum_iK_i|\psi\rangle\otimes|e_i\rangle~.
\end{equation}
The operator $V$ is called a Stinespring dilation of $\cE$ and is an isometry since
\begin{equation}
    \langle\phi|V^\dagger V|\psi\rangle=\sum_{ij}\langle e_i|\langle\phi|K_j^\dagger K_i|\psi\rangle|e_j\rangle=\sum_i\langle\phi|K_i^\dagger K_i|\psi\rangle=\langle\phi|\psi\rangle~.
\end{equation}
Then use the fact that any isometry $V$ on $\mathcal H\cong \mathcal H\otimes \lvert a\rangle$, $\lvert a\rangle\in A$, can always be extended to a unitary $W$ on $\cH\otimes A$. Conversely, if we expand the expression \eqref{purif} in an orthonormal basis $|e_i\rangle$ of $A$ to obtain
\begin{equation}
    \sum_i\langle e_i|W|a\rangle\rho\langle a|W^\dagger|e_i\rangle = \cE(\rho)~,
\end{equation}
we see that $K_i:=\langle e_i|W|a\rangle$ are candidates for a set of Kraus operators for $\cE$. To see that they are actually Kraus operators, check completeness:
\begin{equation}
    \sum_i K_i^\dagger K_i=\sum_i\langle a|W^\dagger|e_i\rangle\langle e_i|W|a\rangle=\langle a|(\mathds{1}\otimes\mathds{1}^A)|a\rangle=\mathds{1}~.
\end{equation}
The purification can be expressed diagrammatically as a tensor
\begin{equation}\begin{split}
    \includegraphics[height=0.145\textheight]{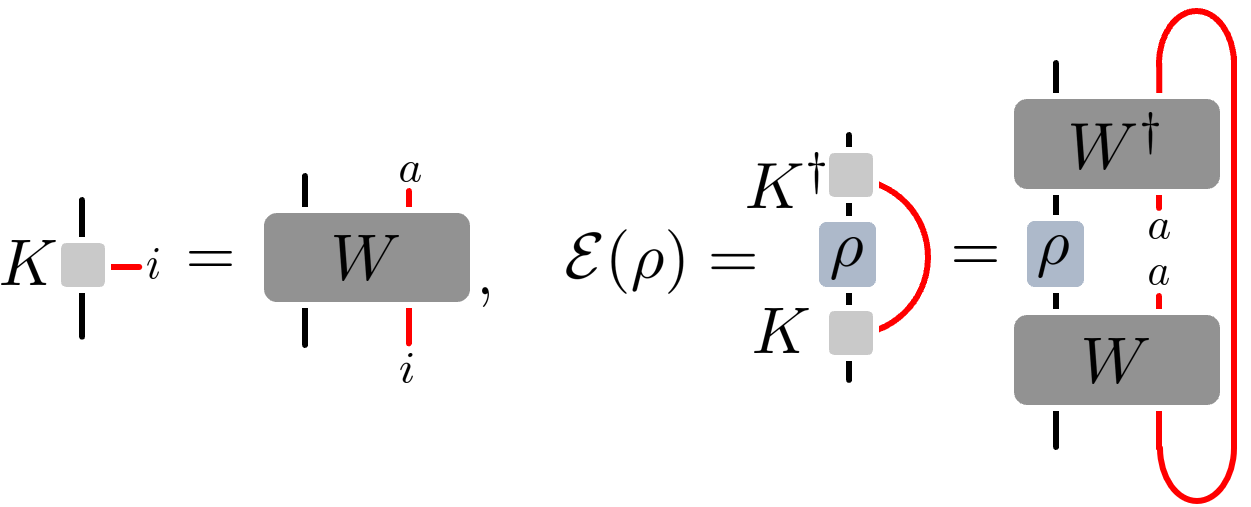}
    \quad\raisebox{0.065\textheight}{.}
\end{split}\end{equation} 
The claim \eqref{purifclaim1} about weak symmetry \eqref{WSC} may be expressed as
\begin{equation}\begin{split}
    \includegraphics[height=0.105\textheight]{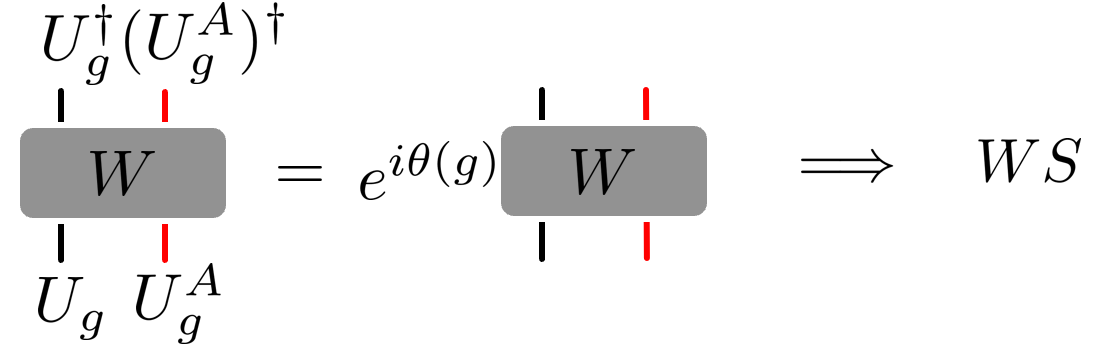}\raisebox{0.045\textheight}{,}
\end{split}\end{equation}
while the claim \eqref{purifclaim2} about strong symmetry \eqref{SSC} may be expressed as
\begin{equation}\begin{split}
    \includegraphics[height=0.105\textheight]{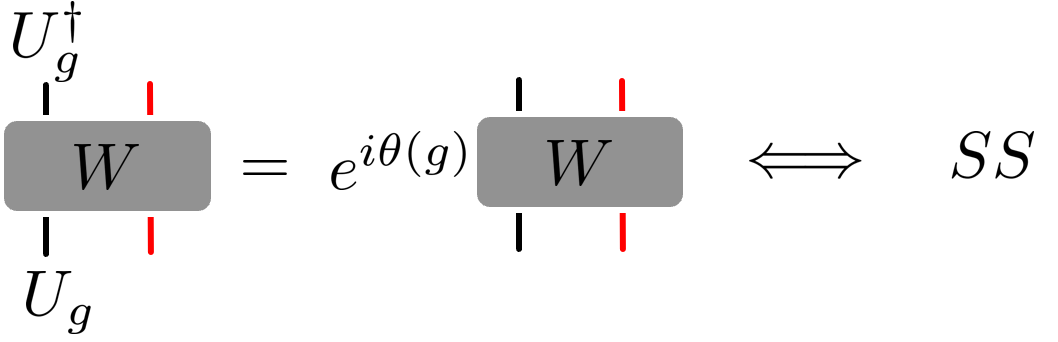}
    \raisebox{0.045\textheight}{.}
\end{split}\end{equation}
Let us now prove the claim \eqref{purifclaim1}. Suppose $W$ is symmetric with the symmetry $U_g\otimes U_g^A$ and the state $\lvert a\rangle$ is invariant:
$U_g^A|a\rangle=|a\rangle$. Then the $K_i=\langle e_i|W|a\rangle$ satisfy weak symmetry \eqref{xg}: 
\begin{equation}\begin{split}
    \includegraphics[height=0.09\textheight]{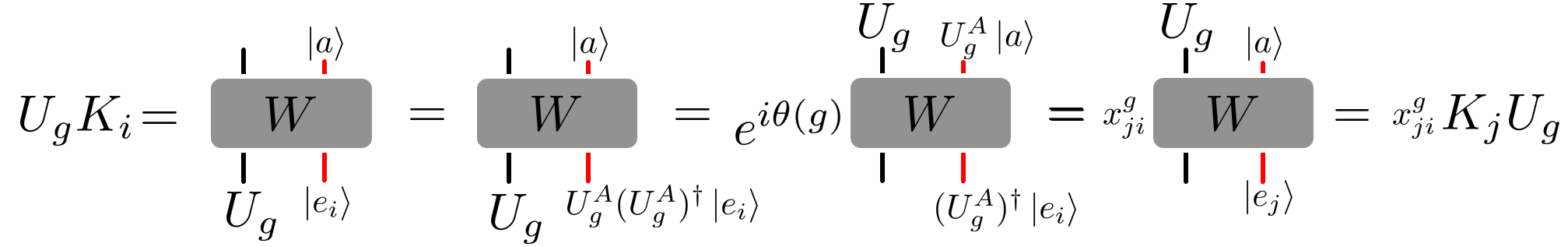}
    \raisebox{0.04\textheight}{,}
\end{split}\end{equation}
where $x^g_{ji}$ are the matrix elements $x^g_{ji}=e^{i\theta(g)}\langle e_i|(U_g^A)^\dagger|e_j\rangle=e^{i\theta(g)}(U_g^A)_{ji}^*$.

One direction of the claim \eqref{purifclaim2} follows from a similar argument. Suppose $W$ is symmetric with $U_g\otimes\mathds{1}^A$. Then the $K_i=\langle e_i|W|a\rangle$ satisfy the strong symmetry condition \eqref{SSC}:
\begin{equation}\begin{split}
    \includegraphics[height=0.09\textheight]{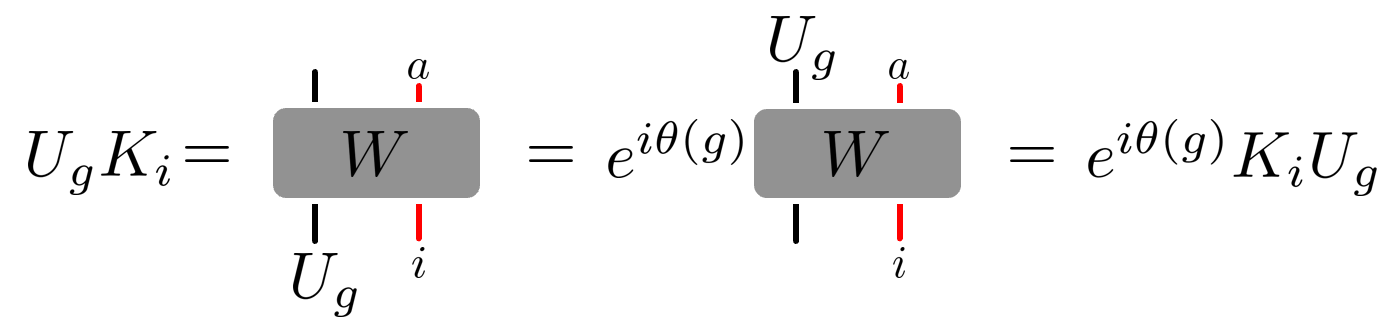}
    \raisebox{0.04\textheight}{.}
\end{split}\end{equation}

Conversely, suppose the Kraus operators satisfy the SS condition and construct a symmetric $W$ as follows. Without loss of generality, take $|a\rangle$ to be $|e_1\rangle$. A unitary extension of the Stinespring dilation $V$ is a square matrix $W$ consisting of blocks $K_i^j$, where the blocks of the first column are the Kraus operators $K_i^1:=K_i$, and we choose the remaining blocks so that $W$ is unitary:
\begin{equation}\label{unitary}
    \sum_iK_i^{j\dagger}K_i^k=\delta^{jk}\mathds{1}_\cH~.
\end{equation}
The remaining blocks may be chosen to be symmetric (so that $W$ is symmetric) as follows. Build linear independent columns by adding signs like $K_i^j:=(-1)^{\delta(i<j)}K_i$, then make them orthogonal by applying the Gram-Schimdt process, and finally normalize. The result is manifestly symmetric.

The claims \eqref{purifclaim1} and \eqref{purifclaim2} may be interpreted in terms of the coupling between the system and the environment (the ancillary space). Suppose $W$ represents unitary evolution by a Hamiltonian:
\begin{equation}
\label{eq:W-exp}
    W=e^{-itH/\hbar}~,\qquad H=\sum_i H_i^S\otimes H_i^E~,
\end{equation}
where the $H_i^S$ and $H_i^E$ are each assumed to be linearly independent. If the unitary evolution $W$ satisfies the formulation of the weak symmetry condition in the claim \eqref{purifclaim1} at all times $t$, then $U_g\otimes U_g^A$ is a symmetry of $H$. If the unitary evolution $W$ satisfies the strong symmetry condition \eqref{purifclaim2} at all times $t$, then
\begin{equation}
\label{eq:W-sym}
    0=(U_g\otimes\mathds{1}^A)H-e^{i\theta(g)}H(U_g\otimes\mathds{1}^A)=\sum_i(U_gH_i^S-e^{i\theta(g)}H_i^SU_g)\otimes H_i^E~,\,\forall\,g~,
\end{equation}
which, since the $H_i^E$ are linearly independent, implies
    $U_gH_i^S=e^{i\theta(g)}H_i^SU_g~,\,\forall\,i,g$. By conjugating both sides of the equation, we find that $e^{i\theta(g)}=e^{-i\theta(g)}$, so $\theta(g)=0,\pi$. Since $W$ is a continuous function of $t$, the phase $\theta(g)$ must vary continuously from zero at $t=0$ to its values at nonzero times (this can be formalized by including the constant order in the expansion of \eqref{eq:W-exp} in the symmetry condition \eqref{eq:W-sym}); this means it must be zero at all times. Therefore, for channels arising from a continuous coupling of system to environment,
\begin{equation}\label{systemcommute}
    U_gH_i^S=H_i^SU_g~,\,\forall\,i,g~,
\end{equation}
which is to say that the system alone, rather than merely its composite with the environment, is symmetric.

%%%%%%%%%%%%%%%%%%%%%%

\subsection{Symmetry conditions on Lindbladians}\label{sym-lind}

Let us now discuss semigroups of channels generated by continuous time evolution by a Lindbladian
\begin{equation}
\label{master}
    \mathcal L(\rho)=-\frac{i}{\hbar}[H^S,\rho]+\sum_{i=1}^\ell\left(L_i\rho L_i^\dagger-\frac{1}{2}L_i^\dagger L_i\rho-\frac{1}{2}\rho L_i^\dagger L_i\right)~.
\end{equation}
Here, the $L_i$ are jump operators and $H^S$ is the Hamiltonian of the system. If a semigroup consists of channels satisfying the weak or strong symmetry condition, the Lindbladian generating it satisfies, respectively,
\begin{equation}\label{WSClind}
    \boxed{\qquad \cU_g\circ\cL\circ\cU_g^\dagger=\cL~,\quad\forall\,g~.\qquad\text{(WS condition on $\cL$)}\qquad}
\end{equation}
\begin{equation}\label{SSClind}
    \boxed{\qquad U_gL_i=L_iU_g~,\qquad U_gH^S=H^SU_g~,\qquad\forall\, i,g~.\qquad\text{(SS condition on $\cL$)}\qquad}
\end{equation}
In particular, this implies that  
semigroups of strongly symmetric channels generated by Lindbladian evolution necessarily have $\theta(g)=0,\,\forall\,g$ at all times.

To see the weak symmetry condition \eqref{WSClind}, observe that the channels $\cE_t=e^{t\cL}$ commute with $\cU_g$ at all times $t$ if and only if $\cL$ commutes with $\cU_g$. To see the strong symmetry condition \eqref{SSClind}, observe that
\begin{align}\begin{split}
    \mathcal{E}_{\delta t}(\rho)&=\rho+\delta t\,\mathcal L(\rho) 
    \\
    & = \rho + \delta t\,\left[ 
        \left(-\frac{i}{\hbar}H^S-\frac12 \sum_{i=1}^\ell L^\dagger_iL_i\right)\rho + 
        \rho \left(\frac{i}{\hbar}H^S-\frac12 \sum_{i=1}^\ell L^\dagger_iL_i\right)
        \right]
        + \sum_{i=1}^\ell \delta t\,L_i\rho L_i^\dagger~,
\end{split}\end{align}
at small times $\delta t$, and thus
\begin{equation}
    \mathcal{E}_{\delta t}(\rho)=\sum_{i=0}^{\ell}K_i(\delta t)\,\rho\, K_i(\delta t)^\dagger
\end{equation}
with Kraus operators
\begin{equation}
    K_0(\delta t)=\mathds{1}+\delta t\,
        \left(-\frac{i}{\hbar}H^S-\frac12 \sum_{i=1}^\ell L^\dagger_iL_i\right)
    ~,\qquad K_{i>0}(\delta t)=\sqrt{\delta t}\,L_i\ .
\end{equation}
This relationship between Kraus operators and jump operators lets us translate our strong symmetry condition \eqref{SSC} on channels into the strong symmetry condition \eqref{SSClind} on the Lindbladians that generate them:
First, from the commutation relation $U_g K_{i>0} = e^{i\theta(g)}K_{i>0} U_g$, we infer that 
$U_gL_i = e^{i\theta(g)}L_iU_g$.
It follows that $X:=\mathds{1}-\delta t\,\tfrac12\sum L_i^\dagger L_i$ commutes with $U_g$.
Second, from the commutation relation $U_g K_{0} = e^{i\theta(g)} K_{0} U_g$, we get 
$U_g(-(i/\hbar)H^S\delta t +X)U_g^\dagger=
e^{i\theta(g)}
(-(i/\hbar)H^S\delta t +X)$.
Taking the Hermitian part and using that $X$ commutes with $U_g$, we obtain 
$X  = U_g X U_g^\dagger = \cos\theta(g)\, X + \sin\theta(g)\,(\delta t/\hbar)\,H^S$,
or
\begin{equation}
    (1-\cos\theta(g)) X = \sin\theta(g)\,(\delta t/\hbar)\,H^S\ .
\end{equation}
As this must hold for all $g\in G$ simultaneously, the proportionality factor $(1-\cos\theta(g))/\sin\theta(g)$
must not depend on $g$, 
which is only possible if $\theta(g)$ is constant. As $\theta(g)$ is a representation, this implies $\theta(g)\equiv 0$ for all $g\in G$ (and in particular, the above equation then imposes no constraints on $X$ and $H^S$).\footnote{There is an alternative topological argument for $\theta(g)=0$. Since the $K_i(t)$ are continuous functions in $t$, the phase $\theta(g)$ in their commutation relations must also be continuous in $t$. But $e^{i\theta}$ is a one-dimensional representation of $G$, and there are only discretely many of these (even if $G$ is continuous), so $\theta(g)$ must be a constant function in $t$. Then, since $\theta(g)=0$ for all $g$ at $t=0$ (because $W(t=0)=\mathds{1}$), the condition $\theta(g)=0$ must also be true at all times $t$.}
It thus follows that both $L_i$ and $H^S$ must commute with $U_g$.
We conclude that a Lindbladian generates a family of SS channels if and only if it satisfies the condition \eqref{SSClind}.

We also obtain a characterization of strong symmetry for Lindbladians by applying the charge conservation condition \eqref{SSCalt} to $\mathcal E_{\delta t} = e^{\delta t\mathcal L}$ (using that we now know that $\theta(g)\equiv 0$ for Lindbladian channels):
\begin{equation}
    U_g = \mathcal E_{\delta t}^\dagger(U_g) = U_g + \delta t\,
    \mathcal L^\dagger(U_g)\ ,
\end{equation}
and thus
\begin{equation}\label{SSC_lind_charge}
    \boxed{\qquad\mathcal L^\dagger(U_g) = 0~.\qquad\text{(SS condition on $\cL$)}\qquad}
\end{equation} 

The Lindblad master equation \eqref{master} can also be recovered from the Hamiltonian that couples the system to the environment, under the Born (weak coupling, large environment) and Markov (memoryless environment) approximations. We refer readers to Chapter 6.2.1 of Ref. \cite{benenti-casati-strini} for a detailed analysis of this procedure. The jump operators appear in the coupling Hamiltonian as
\begin{equation}
    H=H^S\otimes\mathds{1}^E+\mathds{1}^S\otimes H^E+\sum_{i>0}L_i\otimes B_i~.
\end{equation}
In this picture, the strong symmetry condition \eqref{SSClind} on Lindbladians is equivalent to our previous result about the strong symmetry condition on purifications \eqref{systemcommute}.

%%%%%%%%%%%%%%%%%%%%%%%%%%

\subsubsection{Destruction of SPTO by weakly symmetric coupling}\label{coserexample}

Let us now demonstrate that having merely weakly symmetric noise is insufficient to preserve SPTO. Specifically, the fast, local Lindbladian evolution defined by Coser and P\'erez-Garc\'ia \cite{coser2019classification}, which they showed to destroy SPTO, is weakly symmetric. However, as we will also see, it lacks the strong symmetry condition, leaving open the possibility that the latter preserves SPTO.

The local Lindbladian of Ref. \cite{coser2019classification} is given as a sum over sites on a spin chain
\begin{equation}\label{singlelind}
    \cL=\sum_s\cL_s~,
\end{equation}
where
\begin{equation}\label{coser}
    \cL_s=\cT_s-\mathds{1}_s~,\qquad\cT_s(\rho)=\Tr_s[\rho]|\phi\rangle_s\langle\phi|
\end{equation}
for some single-site state $|\phi\rangle$. It is shown to drive any one-dimensional SPT state toward the product state $|\phi\rangle^{\otimes L}$, approximating it well in short time. This result suggests that no SPTO is robust in open systems.

Any time $\cL$ is given as a sum of single site terms \eqref{singlelind}, the SS condition \eqref{SSC_lind_charge} reads
\begin{equation} \label{coser-notstrong}
    0=\cL^\dagger(U_g^{\otimes L})=\sum_s\cL_s^\dagger(U_g^{\otimes L})=\sum_s\cL_s^\dagger(U_g^{(s)}\otimes\mathds{1}^{(L
\backslash s)})\otimes U_g^{(L\backslash s)}~,
\end{equation}
where $U_g$ now denotes the action of the symmetry on a single site. This condition is equivalent to each of the single site terms satisfying the local condition $\cL_s^\dagger(U_g^{(s)}\otimes\mathds{1}^{(L
\backslash s)})=0$. The channel \eqref{coser} has $\cL_s^\dagger(X)=\langle\phi|X|\phi\rangle_s\otimes\mathds{1}^{(L\backslash s)}-X$, so it fails this condition for any $g\ne 1$ and therefore is not SS. On the other hand, $\cL$ commutes with $\cU_g$ and so is WS, as long as $|\phi\rangle$ is taken to be symmetric.

%%%%%%%%%%%%%%%%%%%%%%%%%%
%%%%%%%%%%%%%%%%%%%%%%%%%%

\section{Strongly symmetric uncorrelated noise}\label{sec:preserveSPTO-singlesite}

We begin by considering uncorrelated noise -- channels that decompose into onsite operations as $\cE=\otimes_s\cE_s$. If $\{K_{i_s}^s\}$ is a Kraus representation of $\cE_s$, a Kraus representation of $\cE$ is by operators
\begin{equation}\label{singlesitekraus}
    K_i=\otimes_sK_{i_s}^s~.
\end{equation}
The full channel satisfies the WS or SS condition if and only if all of the single site channels do. Uncorrelated noise is the simplest class of channels, and include the example of \cref{coserexample}, so they are a natural place to start.

The following subsections consider several typical probes of SPTO, in particular, string order, twisted sector charges, edge modes, and irrep probabilities. We demonstrate that these probes are preserved by strongly symmetric uncorrelated noise, indicating that SPTO is preserved. We show in \hyperref[theorem1]{Theorem 1} that, for semigroups of noise generated by Lindbladians, the strong symmetry condition on the Lindbladians is both necessary and sufficient for string order to be preserved at all finite times. We present analytical arguments as well as numerical investigations of example states and channels.

%%%%%%%%%%%%%%%%%%%%%%

\subsection{Preservation of string order by strongly symmetric channels}\label{stringsinglepres}

The first indicator of SPTO we consider is the string order parameter, which was introduced in \cref{stringsingle}. A channel $\cE$ preserves string order if the evolved state $\cE(\rho)$ has the same pattern of zeros as the initial state $\rho$. In the Heisenberg picture, this means that the collection of values $\langle\cE^\dagger(s(g,O_\alpha^l,O_\alpha^r))\rangle$ has the same pattern of zeros as $\langle s(g,O_\alpha^l,O_\alpha^r)\rangle$, where $\cE^\dagger$ is the dual channel to $\cE$. We save for \cref{necsuff} the question of precisely which channels preserve the string order of a state in a given SPT phase. For now, we show
\begin{center}\label{lemma1}
\fbox{
$\qquad$\parbox{0.55\linewidth}{
\centering
\textbf{Lemma 1:} \emph{A channel of uncorrelated noise maps string operators to other string operators of the same type $(g,\alpha)$ if and only if the channel is strongly symmetric.} 
}$\qquad$
}
\end{center}
Note that the evolved string operators are not guaranteed to be nonvanishing.\footnote{It is possible that a strongly symmetric channel annihilates some string operators by mapping their end operators to zero. This phenomenon is unrelated to the symmetry selection rules for string order, discussed in \cref{stringsingle}, which happens on the level of expectation values. For channels that are generic in the sense of Eq. \eqref{channel-generic}, it does not occur for generic end operators.}

To see the lemma, consider evolving the string operator \eqref{stringop} by the uncorrelated noise. It becomes
\begin{equation}\label{evolvedstring}
    \cE^\dagger(s(g,O_\alpha^l,O_\alpha^r))=\mathds{1}\otimes\cE_l^\dagger(O_\alpha^l)\otimes\left(\bigotimes\cE_s^\dagger(U_g)\right)\otimes\cE_r^\dagger(O_\alpha^r)\otimes\mathds{1}~.
\end{equation}
If $\cE_s$ is SS, each of the terms in the bulk of the string becomes $\cE_s^\dagger(U_g)=e^{i\theta_s(g)}U_g$. Since an SS channel is in particular WS \eqref{WSC}, it maps the end operators to other end operators with the same charge: $\tilde O_\alpha^{l,r}:=\cE_s^\dagger(O_\alpha^{l,r})$ has $U_g^\dagger\tilde O_\alpha^{l}U_g=\chi_\alpha(g)\tilde O_\alpha^l$ and similarly for $\tilde O_\alpha^r$. Then in total, we have
\begin{equation} \label{SSchannelstring}
    \cE^\dagger(s(g,O_\alpha^l,O_\alpha^r))=e^{i\sum_s\theta_s(g)}s(g,\tilde O_\alpha^l,\tilde O_\alpha^r)~.
\end{equation}
Conversely, asking that $\cE^\dagger(s(g,O_\alpha^l,O_\alpha^r))$ is a string operator of type $(g,\alpha)$ for all $(g,\alpha)$ requires,\footnote{The choice $O_\alpha^{l,r}=\mathds{1}$ always results in nonvanishing $\tilde O_\alpha^{l,r}$, so the bulk $\cE_s^\dagger(U_g)$ can be compared to $U_g$.} in particular, that $\cE_s^\dagger(U_g)$ is proportional to $U_g$, which is the strong symmetry condition.

%%%%%%%%%%%%%%%%%%%%%

\subsubsection{A necessary and sufficient condition on Lindbladians}\label{necsuff1}

Our discussion has so far focused on the string operators. We now turn toward analyzing their expectation values, which encode the invariant of SPT states. In the following, a coherent SPT mixture is a mixed state with a well-defined pattern of zeros (and thus a well-defined invariant $[\omega]$), in the sense of Definition \eqref{SPTmixed}. A ``coherent SPT phase'' is a class consisting of all coherent SPT mixtures with a given invariant. Preserving a phase means that every mixture of translation-invariant pure SPT states in the phase (states of the form Eq. \eqref{ensemble}) is mapped to a coherent SPT mixture in the same phase.\footnote{We expect that strongly symmetric Lindbladian evolution takes every coherent SPT mixture to a coherent SPT mixture in the same phase; however, we restrict ourselves to initial states that are mixtures of translation-invariant pure SPT states in order to make use of tensor network methods in the proof of the theorem.}

Our main result is stated as a theorem:
\begin{center}\label{theorem1}
\fbox{
$\qquad$\parbox{0.55\linewidth}{
\centering 
\textbf{Theorem 1:} \emph{Fix any coherent SPT phase. A semigroup of channels of uncorrelated noise generically preserves the phase at all finite times if and only if the semigroup is generated by a strongly symmetric Lindbladian. 
}
}$\qquad$
}
\end{center}
In other words, the notion of coherent SPT phase defined by patterns of zeros coincides with the notion of phase defined by strongly symmetric Lindbladian evolution.

To make a claim as strong as \hyperref[theorem1]{Theorem 1}, it is necessary to work on the level of phases rather than the states (equivalently, systems) that compose them. A phase protected by a symmetry $G$ consists of systems together with the data of embeddings of $G$ into the systems' full groups of symmetries. For example, the AKLT system may be regarded as lying in a $G=\ZZ_2\times\ZZ_2$ SPT phase if one specifies this group's embedding into the system's larger $SO(3)$ intrinsic symmetry group. Also, there is no restriction on the physical degrees of freedom, or the symmetry action on these degrees of freedom, that a system in a phase may have. For example, the $\ZZ_2\times\ZZ_2$ SPT phase to which the AKLT system belongs also contains systems that are not built of spin-$1$ degrees of freedom. Whether two systems lie in the same phase has not to do with their intrinsic symmetry groups or degrees of freedom but on the values taken by their order parameters. This means that MPS representations of states within the phase will have tensors of various physical dimensions and symmetries. The theorem concerns which symmetry condition channels must satisfy so that they preserve the string order of \emph{every system in the SPT phase}. To do this, we consider generic states, whose MPS tensors have exactly the symmetry $G$, are injective, and have physical Hilbert space no larger than the tensor's image. We emphasize that this result does not preclude the possibility of nongeneric systems within the phase for which a condition weaker than strong symmetry is sufficient.

With these remarks behind us, we are ready to prove the theorem, starting with the `if' direction. If the semigroup is generated by a strongly symmetric Lindbladian, each channel $\cE_t$ is itself strongly symmetric. Then by \hyperref[lemma1]{Lemma 1}, $\cE_t$ preserves the type of string operators. Moreover, finite time Lindbladian evolution defines channels that are invertible as linear maps\footnote{Note that invertibility of the channel as a linear map is a weaker condition than reversibility of the channel, as the inverse linear map is not required to be a channel.} since $\det(e^{t\cL}) = e^{\Tr(t\cL)} \ne 0$ for finite $t$, so these channels do not annihilate any string operators. Thus the pattern of zeros of the expectation values of string operators is preserved, as long as one uses end operators that are generic in the sense of \cref{stringsingle} -- namely, that $O_{\alpha}^{l,r}$ is orthogonal neither to $N_{l,r}^{g\dagger}$ nor to $\cE(N_{l,r}^{g\dagger})$. The assumption that the end operators are generic can be dropped if one is interested in generic evolution times. This is because the expectation value of a string operator is analytic as a function of time, and so it either vanishes at all times -- as occurs where the initial pattern of zeros has a zero -- or is zero only at isolated points in time. The assumption of finite time cannot be dropped, as an infinite time evolution may annihilate some string operators and therefore alter the pattern of zeros. An evolution for which this phenomenon occurs is the fully dephasing channel introduced below.

Next we turn to the `only if' direction. We show it for initial states that are translation-invariant pure SPT states. Then it also holds for their mixtures. If a channel $\cE$ preserves the string order of an SPT state, there is, for every symmetry $g$, a string operator with $g$ in the bulk whose expectation value is nonvanishing. This expectation value may be written as $\langle L|T_{\cE_s^\dagger(U_g)}^j|R\rangle$, where $T_{\cE_s^\dagger(U_g)}$ is the transfer matrix
\begin{equation}\begin{split}\label{transfer-single-eq}
    \includegraphics[height = 0.095\textheight]{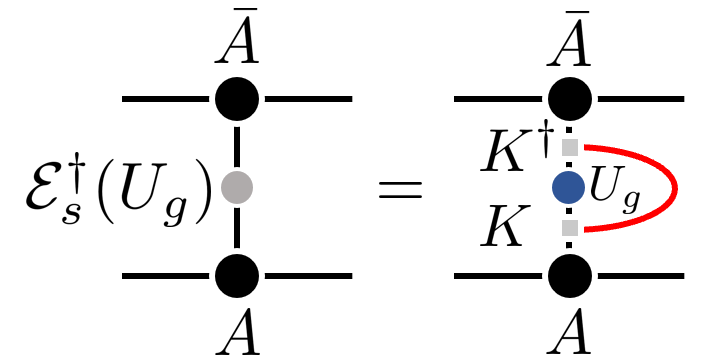}
    \raisebox{0.04\textheight}{.}
\end{split}\end{equation}
This operator must have $\lambda=1$ as its maximum eigenvalue in order for the expectation value to be nonvanishing in the thermodynamic limit.\footnote{At finite string lengths, the difference between strongly symmetric and non-strongly symmetric Lindbladians is reflected in the length-dependence of the order parameters: only for the latter is there decay as a function of length.} By Lemma 2 of Ref. \cite{handwaving}, for injective MPS, in order for $T_X$ to have $\lambda=1$, $X$ must be a symmetry of the MPS, which by assumption is some element $h_g\in G$. Thus $\cE_s^\dagger(U_g)=e^{i\theta(g)}U_{h_g}$. Taking $X=U_g$ and $Y=e^{i\theta(g)}U_{h_g}$ in the argument at the end of \cref{chargecons}, we see that $U_gK_i=e^{i\theta(g)}K_iU_{h_g}$ for all $i$. The map $\sigma:g\mapsto h_g$ is an endomorphism\footnote{An endomorphism is a map from the group to itself that is compatible with the group structure.} of $G$ since $U$ is faithful and, up to phases $\theta$,
\begin{equation}
    U_{\sigma(g)}U_{\sigma(h)}=\sum_iK_i^\dagger K_iU_{\sigma(g)}U_{\sigma(h)}\sim\sum_iK_i^\dagger U_gU_hK_i=\sum_iK_i^\dagger U_{gh}K_i\sim\sum_iK_i^\dagger K_iU_{\sigma(gh)}=U_{\sigma(gh)}~.
\end{equation}
If $\cE$ is connected to the trivial channel by a semigroup of channels $\cE_t$ satisfying the above at all times, the endomorphism $\sigma$ must be connected to the identity endomorphism $\sigma=1$ by a continuous path $\sigma_t$. Since we have assumed $G$ is abelian, the only identity-connected endormophism is $\sigma=1$ itself. Thus, we have $\cE_s^\dagger(U_g)=U_g$, which is the strong symmetry condition \eqref{SSCalt} on the site $s$. This holds for all $s$, so the full channel $\cE$ is strongly symmetric. It holds at all finite times, so by the discussion in \cref{sym-lind}, the Lindbladian that generates it is strongly symmetric. This completes the proof of \hyperref[theorem1]{Theorem 1}.

It is interesting to note that, by \hyperref[theorem1]{Theorem 1} and \hyperref[lemma1]{Lemma 1}, any semigroup that preserves an SPT phase, even the trivial SPT phase, also preserves the type of all string operators. This reflects how special SPT states, even trivial SPT states, are among mixed states, most of which lack valid patterns of zeros.

The proof of \hyperref[theorem1]{Theorem 1} involved only the identity-connected endomorphisms of the symmetry group. In \cref{sec:twistedstrong} we lift the restriction that the channel belongs to a semigroup of SPT-preserving channels; in this broader setting, more general group endomorphisms play an important role.

Let us comment on the generalization to a nonabelian symmetry group $G$. Recall that in this case, string order is not guaranteed to capture the full SPT invariant $\omega$; nevertheless, we can ask about which channels preserve string order. If $G$ is a finite group, the only identity-connected endomorphism is again $\sigma=1$ itself. On the other hand, if $G$ is a semisimple Lie group, the only identity-connected endomorphisms are \emph{inner automorphisms} \cite{fultonharris}. An automorphism $\sigma$ is said to be inner if there exists an element $h\in G$ such that
\begin{equation}\label{inneraut}
    \sigma(g)=\text{conj}_h(g):=h^{-1}gh~,\quad\forall g\in G~.
\end{equation}
The element $h$ is defined up to elements of the center. An example of a channel with automorphism $\sigma=\text{conj}_h$ is the symmetry-implementing channel $\cU_h$ \eqref{symchannel}, which has a single Kraus operator $K=U_h$ satisfying
\begin{equation}
    U_{\sigma(g)}:=K^\dagger U_gK=U_h^{-1}U_gU_h=U_{h^{-1}gh}~.
\end{equation}
Any channel $\cE$ with inner automorphism $\sigma=\text{conj}_h$ may be expressed as the composition of the symmetry-implementing channel $\cU_h$ and a strongly symmetric channel $\cE_{SS}$:
\begin{equation}\label{SSplussym}
    \mathcal{E}=\mathcal{E}_{SS}\circ\cU_h~.
\end{equation}
To see this, let $K_i$ denote the Kraus operators for $\cE$ and define Kraus operators for $\cE_{SS}$ as
\begin{equation}
    K_i'=U_h^{-1}K_i~.
\end{equation}
One can easily verify that the latter define a strongly symmetric channel $\cE_{SS}$. The semigroup of channels of the form \eqref{SSplussym} is generated by the sum of of a strongly symmetric Lindbladian $\cL_{SS}$ and a generator $\Q$ of the continuous symmetry $\cU_h=e^\Q$. We obtained this condition by asking that a string operator expectation value not vanish, which is weaker than asking that string order is preserved; in fact, string order may be modified by such channels. The $\cU_h$ factor changes the bulks of string operators from $U_g$ to $\cU_h^\dagger(U_g)=U_{h^{-1}gh}$ and the ends from $O_\alpha$ to $\cU_h^\dagger(O_\alpha)$, which transform as $(\sigma^{-1})^*\alpha$. It may happen, however, that string order is preserved despite the permutation of string operators; this phenomenon is discussed in \cref{sec:twistedstrong}.

%%%%%%%%%%%%%%%%%%%%%%%%%%%%%%

\subsection{Numerical study of string order under channels}\label{stringordersimulations}

In this subsection, we simulate the behavior of string order under the action of two example channels -- the depolarising channel and the dephasing channel. We find, in agreement with the analytical results, that evolutions with strong symmetry preserve string order, while weakly symmetric channels do not. We also use the example of the fully dephasing channel, which destroys SPTO despite being strongly symmetric, to demonstrate the importance of the finite time assumption in the theorem.

The depolarising channel is a severely noisy channel, as it contracts the Bloch sphere to the origin, driving states towards the maximally mixed state. This channel is written as
\begin{equation}\label{eq:depolarising}
        \cE(\rho) = \left( 1- \lambda \right)\,\rho + \lambda\,(\Trace{\rho}) \,\tfrac{1}{d}\mathbbm{1}~,
\end{equation}
with the decay rate parametrized by $\lambda \in [0,1]$ \cite{breuer2002theory}. The value $\lambda =1$ gives the fully depolarising channel. The channel satisfies the weak symmetry condition \eqref{WSC} for any symmetry.

The Kraus decomposition for this channel can be written as a twirling operation since the normalised $d$-identity can be decomposed as an average over the generators of the Lie algebra of $SO(d)$ \cite{nielsen2002quantum}. In $d=2$, these operators are Paulis, while in higher $d$ they are the Heisenberg-Weyl matrices: the shift operator $X\ket{j} = \ket{j+1 \mod d}$ and the phase operator $Z\ket{j} = e^{i 2 \pi  j/d}\ket{j}$, which have commutation relation $Z^m X^n = e^{i 2 \pi n m /d} X^n Z^m$. The Kraus decomposition for spin-$1$ ($d=3$) systems such as the AKLT state is given by
\begin{equation}\label{depolarising}
   \cE(\rho) =  (1-\lambda)\,\rho +  \frac{\lambda}{9}  \sum_i N_i \rho N_i^\dagger~,\qquad N_i = \{ \mathbbm{1}, Z, Z^2, X, ZX, Z^2X, X^2,ZX^2, Z^2X^2 \}~.
\end{equation}
A $G=\ZZ_2\times\ZZ_2$ symmetry acts on the spin-$1$ system as $U_g=e^{i\pi S_j}$, where $S_j$ are the spin-$1$ operators. Since the Kraus operators $K_i\sim N_i$ only commute with the symmetry action up to different phases, the channel does not satisfy the strong symmetry condition \eqref{SSC}; it is only weakly symmetric.

The second channel we discuss is given by
\begin{equation}\label{DEPHASING}
    \cE(\rho) = (1-\lambda) \rho + \frac{\lambda}{4} \sum_j N_j \rho N_j^\dagger~,\qquad N_j = \{\mathbbm{1}, e^{i \pi S_x}, e^{i \pi S_y}, e^{i \pi S_z} \}~,
\end{equation}
where $S_j$ are the spin-$1$ operators. Since the Kraus operators $K_j\sim N_j$ commute with the symmetry action, the channel is strongly symmetric \eqref{SSC}. The channel is the fully dephasing channel when $\lambda= 1$. 

%%%%%%%%%%%%%%

\subsubsection{String order after a single time-step}\label{expressionstringord}

Let us consider the AKLT state, which belongs to the Haldane SPT phase. This state has nontrivial $SO(3)$ SPT order, but can be protected by just the subgroup $\ZZ_2 \times \ZZ_2$ \cite{pollmann2012detection, Haegeman_2012}. The AKLT state has an exact tensor network representation given by the Pauli operators in the basis $\{ \ket{+}, \ket{0}, \ket{-} \}$.

The $|G|^2=16$ string operators $s_{ij}$ for the symmetry $G= \ZZ_2 \times \ZZ_2$ are built out of end operators $O_\alpha^{l,r} = S_i$ and bulk operators $U_g = e^{i \pi S_j}$, where $S_i$ are the spin-$1$ matrices with $j = \{ e,x,y,z\}$. On the AKLT state, the diagonal string operators take nonzero values $s_{zz},s_{xx}, s_{yy} = -4/9$ and $s_{\mathbbm{1}, \mathbbm{1}} = \mathbbm{1}$ independent of string length in the limit of infinite system size, while the off-diagonal string operators elements $s_{ij}, i \neq j$ have vanishing expectation values in the limit of long string length.

Recall that the string order parameter may be manipulated into the form \eqref{evaluatestring2}
\begin{equation}
     \langle s(g,O_\alpha^{l,r})  \rangle = \Tr(\rho_l T_{O_\alpha^l} (T_{U_g})^{N-2} T_{O_\alpha^r} \rho_r)~,
\end{equation}
where $T$ is the transfer matrix $T = \sum_i A^i \bigotimes \overline{A}^i$ for the MPS tensor $A$, and $\rho_{l,r}$ are its fixed points. Now consider evolving the string operator under a channel. The operator becomes
\begin{equation}
    \langle \cE^\dagger(s(g,O_\alpha^l,O_\alpha^r)) \rangle = \Tr(\rho_l T_{\cE^\dagger(O_\alpha^{l})} T_{\cE^\dagger(U_g)}^{N-2} T_{\cE^\dagger(O_\alpha^{r})}\rho_r)~.
\end{equation}

For the SS channel \eqref{DEPHASING}, the evolved string order pattern for the AKLT state as a function of the decay rate $\lambda$ is
\begin{equation}\label{patternSS}
     \langle \cE^\dagger(s(e^{i \pi S_j},S_i))\rangle = 
     {\small
     \begin{pmatrix}
  1 & -\frac{4}{9}(1- \lambda)^2( -\frac{1}{3})^N &  -\frac{4}{9}(1- \lambda)^2( -\frac{1}{3})^N  & -\frac{4}{9}(1- \lambda)^2( -\frac{1}{3})^N \\
(-\frac{1}{3})^N &  -\frac{4}{9}(1- \lambda)^2 &  -\frac{4}{9}(1- \lambda)^2( -\frac{1}{3})^N & -\frac{4}{9}(1- \lambda)^2( -\frac{1}{3})^N \\
(-\frac{1}{3})^N & -\frac{4}{9}(1- \lambda)^2( -\frac{1}{3})^N & -\frac{4}{9}(1- \lambda)^2 &-\frac{4}{9}(1- \lambda)^2( -\frac{1}{3})^N\\
(-\frac{1}{3})^N\ & -\frac{4}{9}(1- \lambda)^2( -\frac{1}{3})^N &  -\frac{4}{9}(1- \lambda)^2( -\frac{1}{3})^N  & -\frac{4}{9}(1- \lambda)^2  \end{pmatrix}
    }~.
\end{equation}
In the limit of large string length $N\ra\infty$, the diagonal entries are nonzero while the off-diagonal entries are zero, which indicates that the string order of the AKLT state was preserved, as is expected for SS channels. Note that if we were to repeatedly evolve the state by $\cE$ a large number of times, the entire string order set would decay exponentially to zero, so the preservation of the pattern is only visible at finite times. This exponential decay in time is visible in Fig. \ref{fig:strong,weakdepolarising,strongdephasing}.

In contrast, consider the WS depolarising channel \eqref{depolarising}. The string order for the evolved state is
\begin{equation}
     \langle \cE^\dagger(s(e^{i \pi S_j},S_i))\rangle =
     {\small
     \begin{pmatrix}
  1 & -\frac{4}{9}(1- \lambda)^2( -\frac{1}{3})^N &  -\frac{4}{9}(1- \lambda)^2( -\frac{1}{3})^N  & -\frac{4}{9}(1- \lambda)^2( -\frac{1}{3})^N \\
(-\frac{1}{3})^N &  - \frac{4 (1-\lambda)^2}{9}(1 - \frac{8\lambda}{9})^N &  - \frac{4(1-\lambda)}{9}( -\frac{1}{3}+\frac{4\lambda}{9})^N & - \frac{4(1-\lambda)}{9}( -\frac{1}{3}+\frac{4\lambda}{9})^N\\
(-\frac{1}{3})^N &  - \frac{4(1-\lambda)}{9}( -\frac{1}{3}+\frac{4\lambda}{9})^N & - \frac{4 (1-\lambda)^2}{9}(1 - \frac{8\lambda}{9})^N & - \frac{4(1-\lambda)}{9}( -\frac{1}{3}+\frac{4\lambda}{9})^N\\
(-\frac{1}{3})^N\ & - \frac{4(1-\lambda)}{9}( -\frac{1}{3}+\frac{4\lambda}{9})^N&  - \frac{4(1-\lambda)}{9}( -\frac{1}{3}+\frac{4\lambda}{9})^N &  - \frac{4 (1-\lambda)^2}{9}(1 - \frac{8\lambda}{9})^N \end{pmatrix}
    }~.
\end{equation}
For any $\lambda>0$, the string order goes to zero instantaneously (with a single time-step, application of $\cE$) in the limit $N\ra\infty$. This is because, in contrast to the SS channel, this WS channel receives an exponential contribution to its decay from the bulk of the string.

%%%%%%%%%%%%%%%%%%%%

\begin{figure}
    \centering
    \includegraphics[width= 0.4 \textwidth]{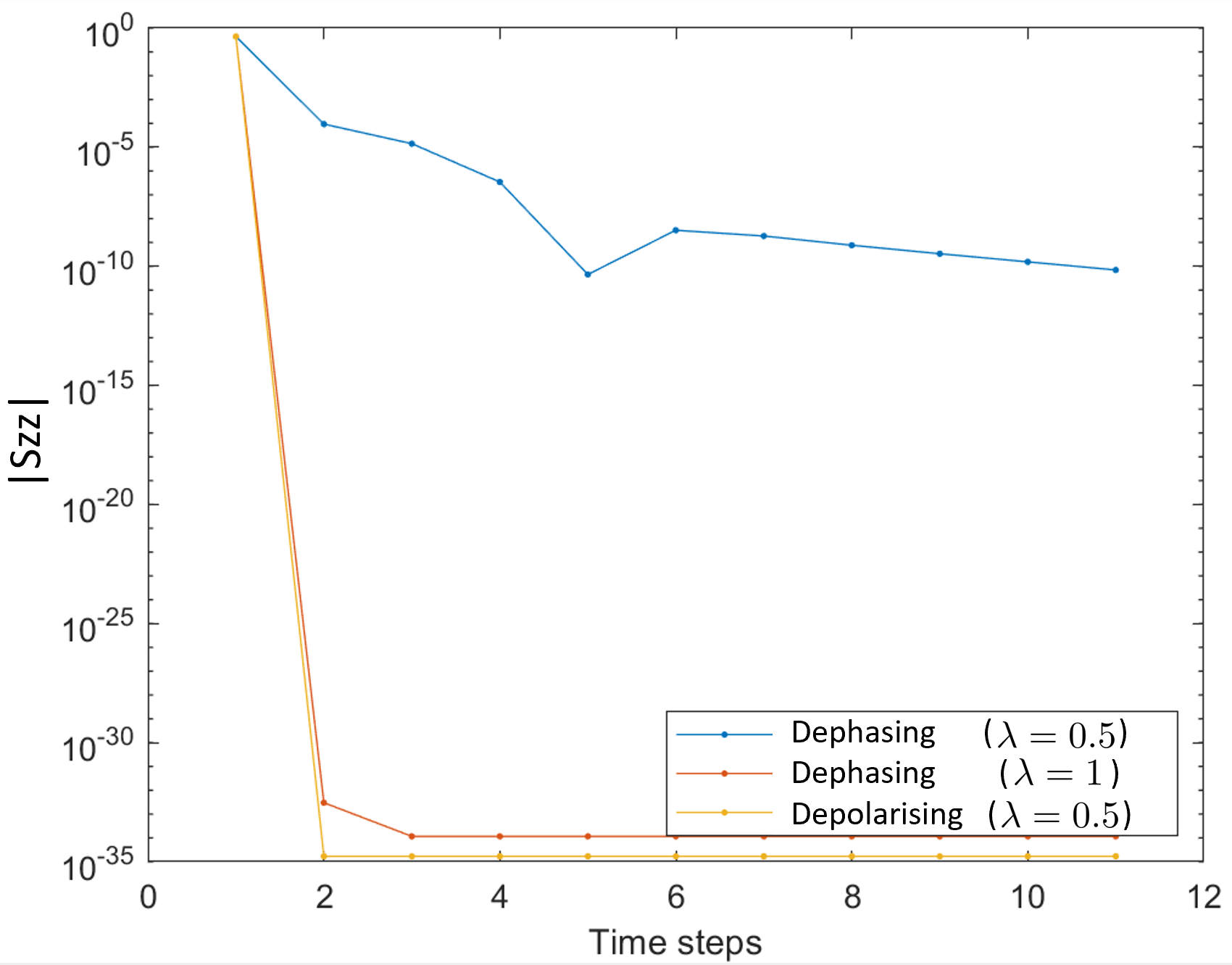}
    \caption{Evolution of the absolute value of the $s_{zz}$ component of the $\ZZ_2\times\ZZ_2$ string order parameter for the AKLT state (initial $|s_{zz}|>0$) under the dephasing channel \eqref{DEPHASING} and the depolarising channel \eqref{depolarising}.
    }\label{fig:strong,weakdepolarising,strongdephasing}
\end{figure}

\subsubsection{Master equation simulation of string order}\label{wsandsschannel}

We simulate the time-evolved master equation on pure SPT states by acting with the channel at each time-step
\begin{equation}
    \cE_t =  \cE_T \circ \cdots \circ \cE_{t1} \circ  \cE_{t_0} .
\end{equation}
Figure \ref{fig:strong,weakdepolarising,strongdephasing} depicts the results of simulating the evolution of the $s_{zz}$ component of the string order parameter under the following channels: the dephasing channel \eqref{DEPHASING} with $\lambda = 0.5$, the dephasing channel with $\lambda = 1$ (fully dephasing), and the depolarising channel \eqref{depolarising}. The first two of these channels are SS, while the third is only WS. The first exhibits slow decay, while the others exhibit instantaneous decay.

The reason that the fully dephasing channel fails to preserve SPT order despite being SS can be traced to how it annihilates the end operators $S_i$ with nontrivial labels and thus sets the corresponding rows of the pattern of zeros to zero. This phenomenon cannot occur for finite time evolutions, as these channels are invertible as linear maps, so \hyperref[theorem1]{Theorem 1} is safe. The fully dephasing channel, however, is only realized as an \emph{infinite} time evolution, so SS is insufficient to protect SPTO. In \cref{necsuff}, we discuss a genericness condition on SS channels, failed by the fully dephasing channel, that says whether or not they preserve SPTO.

%%%%%%%%%%%%

\subsection{Twisted sector charges}\label{twistedsector}

The SPTO of a state may alternatively be detected in the charges of its twisted sector states \cite{shiozakiryu,PhysRevB.96.075125}. Consider a state on a closed chain, represented as an MPS
\begin{equation}
    \langle i_1\cdots i_L|\psi\rangle = \Tr[\,A^{i_1}\cdots A^{i_L}\,]~,
\end{equation}
with the symmetry represented projectively on the virtual space by operators $V_g$. The states
\begin{equation}
    \langle i_1\cdots i_L|\psi_h\rangle = \Tr[\,V_h\,A^{i_1}\cdots A^{i_L}\,]
\end{equation}
are the twisted sector states. In closed systems, they appear as ground states of the Hamiltonians obtained by twisting the original Hamiltonian by the insertion of a symmetry flux through the closed chain.

The charge of a twisted sector state is obtained by acting on the state with the charge operator $U_g^{\otimes L}$:
\begin{equation}\begin{split}
    \includegraphics[height=0.085 \textheight]{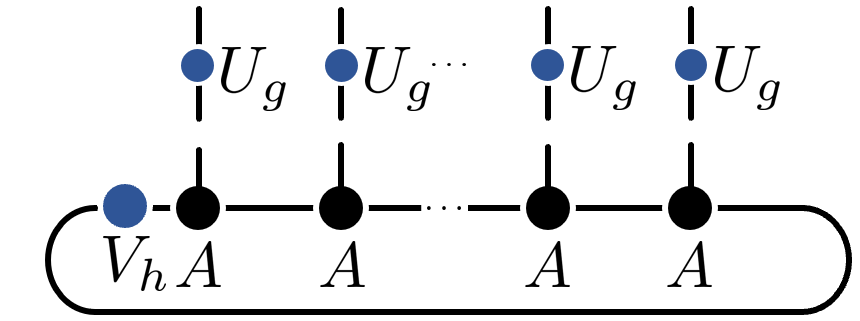}
    \quad\raisebox{0.02\textheight}{.}
\end{split}\end{equation}
For simplicity, assume $G$ is abelian. The charges of the twisted sector states are given by
\begin{equation}
    \langle i_1\cdots i_L|U_g^{\otimes L}|\psi_h\rangle=\Tr[\,V_g^{-1}V_hV_g\,A^{i_1}\cdots A^{i_L}\,]=\frac{\omega(h,g)}{\omega(g,h)}\Tr[\,V_h\,A^{i_1}\cdots A^{i_L}\,]~.
\end{equation}
As we saw in the previous subsection, this ratio $\omega/\omega$ determines the cohomology class $[\omega]$. This means that the collection of twisted sector charges completely characterizes the SPTO.

The twisted sector charges of an initial state and the state reached by evolving by a channel are the expectation values of $U_g^{\otimes L}$ and $\cE^\dagger(U_g^{\otimes L})$, respectively, on the initial twisted sector states $|\phi_h\rangle$. Since for a strongly symmetry channel \eqref{SSCalt} these are equal, such a channel preserves the SPTO.

%%%%%%%%%%%%%%%%%%%%%%%%%%%%

\subsection{Protected edge modes}\label{edgesingle}

SPT phases are also characterized by their topologically protected edge modes. Consider a pure state $|\psi^\omega\rangle$ in an SPT phase characterized by a cocycle $\omega$. The SPT invariant is encoded in the projective action of the symmetry on the edge introduced by cutting the system:
\begin{equation}
    (\mathds{1}^l\otimes U_g^r)|\psi^\omega\rangle=\sum_a|\psi^\omega_{l,a}\rangle\otimes U_g^r|\psi^\omega_{r,a}\rangle=\sum_{a,b}(V_g)_{ab}|\psi^\omega_{l,a}\rangle\otimes|\psi^\omega_{r,b}\rangle~,\qquad V_gV_h=\omega(g,h)V_{gh}~.
\end{equation}
After evolving through a channel, $|\psi^\omega\rangle$ becomes a mixture of the (unnormalized) states $K_i|\psi^\omega\rangle$. We claim that, if the channel is strongly symmetric and onsite, each of these states has SPT invariant $\omega$. To see this, write the Kraus operators as $K_i^l\otimes K_i^r$, so that $K_i|\psi^\omega_i\rangle=\sum_aK_i^l|\psi^\omega_{l,a}\rangle\otimes K_i^r|\psi^\omega_{r,a}\rangle$. Then, since $K_i^r$ commutes with $U_g^r$, the projective representation $V_g$ on this cut is the same as for $|\psi^\omega\rangle$; in particular, $\omega$ is the same.

%%%%%%%%%%%%%%%%%%%%%%%%%%%%%%

\subsection{Irrep probabilities and SPT complexity}\label{sec:irreps}

In this section, we investigate the behavior of irrep probabilities -- another probe of SPTO -- under channels, strongly symmetric and not. Irrep probabilities measure the weight of the state in each symmetry sector. They are given by Fourier transforms of the string operators with identity end operators
\begin{equation}\label{irrepprob}
    p_\alpha=\frac{1}{|G|}\sum_g\chi_\alpha(g)\langle s(U_g,\mathds{1},\mathds{1})\rangle~,
\end{equation}
Using orthogonality of characters, one can confirm that the irrep probabilities sum to $1$:
\begin{equation}
    \sum_\alpha p_\alpha=\frac{1}{|G|}\sum_g\langle s(U_g,\mathds{1},\mathds{1})\rangle\sum_\alpha\chi_\alpha(g)=\frac{1}{|G|}\sum_g\langle s(U_g,\mathds{1},\mathds{1})\rangle\,|G|\,\delta_{g,1}=\langle s(\mathds{1},\mathds{1},\mathds{1})\rangle=1~,
\end{equation}

Irrep probabilities partially distinguish SPT phases by capturing the \emph{SPT complexity}
\begin{equation}\label{complexity}
    D_\omega=\sqrt{|G|/|K_\omega|}~,
\end{equation}
where
\begin{equation}\label{projcenter}
    K_\omega=\{\,g\,:\,\omega(g,h)=\omega(h,g)~\,\forall\,h\,\}
\end{equation}
is subgroup of $G$ called the \emph{projective center}. A phase with $K_\omega=\{1\}$ has maximum complexity and is said to be \emph{maximally noncommutative} (MNC) \cite{else2012symmetry}. An example of an MNC phase is the Haldane phase of systems with $\ZZ_2\times\ZZ_2$ symmetry. MPS representations of states in this phase have virtual symmetries related to the Pauli operators, which do not commute, meaning that $K_\omega=\{1\}$. On the other end of the spectrum is the trivial phase $\omega=1$, with complexity $D_1=1$. The complexity of a generic MPS was shown numerically to appear in the degeneracy $D_\omega^2$ of the irrep probabilities \cite{de_Groot_2020}. In an MNC phase, the probabilities are all $p_\alpha=1/|G|$ with degeneracy $|G|$; in the trivial phase, they are generically distinct. The value $D_\omega$ also appears as the degree of the projective representation $V_g$ (c.f. \cite{berkovich1998}, theorem VI.6.39), which means $D_\omega^2$ is the number of topologically protected edge modes. For a given symmetry $G$, there may be multiple phases with the same complexity and these cannot be distinguished from each other by their irrep probabilities; nevertheless, because they can distinguish phases with different complexities, irrep probabilities are a useful tool.

The values of the irrep probabilities are preserved by strongly symmetric channels. To see this, observe that such channels preserve the string operators $s(U_g,\mathds{1},\mathds{1})$, and therefore their Fourier transforms, exactly. The exact preservation of irrep probabilities stands in contrast with general string order parameters, whose end operators $\cO_\alpha\ne\mathds{1}$ may cause decay toward zero in infinite time.

In contrast with strongly symmetric channels, non-SS Lindbladian channels map all irrep probabilities to the maximally degenerate values $p_\alpha=1/|G|$, for any phase. This is because, as was established in \cref{necsuff1}, such channels annihilate all of the string order parameters accept for those with $g=1$. The result is
\begin{equation}\label{maxdegen}
    p_\alpha=\frac{1}{|G|}\sum_g\chi_\alpha(g)\,\delta_{g,1}=\frac{1}{|G|}~.
\end{equation}
This means that, outside of states which already have maximally degenerate irrep probabilities (for example, states in MNC phases), the effect of a Lindbladian channel on irrep probabilities is a diagnostic for whether or not a channel is strongly symmetric. In \cref{twisted-probs}, we discuss non-Lindbladian channels, some of which preserve irrep probabilities and SPT complexity despite not being SS.

%%%%%%%%%%%%%%%%%%%%%%%%%%%%%
%%%%%%%%%%%%%%%%%%%%%%%%%%%%%

\section{Causal channels}\label{sec:causalchannels}

Let us extend our analysis from uncorrelated noise to causal channels. A channel is said to be \emph{causal} if there is a range $r$ such that it maps operators supported on a compact region $A$ to operators supported on the region of sites within distance $r$ of $A$ \cite{PhysRevLett.125.190402}.\footnote{The terms ``causal'' and ``locality-preserving'' have different meanings in Ref. \cite{PhysRevLett.125.190402}, and we are interested in the former.} Channels that are not causal can create long-range correlations, and so are expected to destroy topological order and SPTO, no matter the symmetry condition imposed on them. For this reason, we will not consider non-causal channels here.

A subset of causal channels, labeled ``dQC''\footnote{The ``d'' stands for ``dilation,'' as in the Stinespring dilation of the channel. Our ``sd'' stands for ``symmetric dilation.''} in Ref. \cite{PhysRevLett.125.190402}, have purifications that are causal.  In addition, one can consider their convex combinations, which are also causal. It has been suggested that these convex combinations might constitute all causal channels, but this remains an open question \cite{PhysRevLett.125.190402}. Another open question is whether every channel in dQC that has a symmetric purification (and so by Claim \eqref{purifclaim1} is weakly symmetric) has a purification that is \emph{both} causal and symmetric with respect to an on-site symmetry. Rather than attempting to answer this question, we let ``sdQC'' denote the channels with such a purification. We restrict our focus to channels in dQC (in \cref{TNcausal}) and sdQC (in \cref{sec:localSS} and \cref{stringlocal}) and leave open the possibility that channels outside of these classes exhibit different behaviors. As we remark in \cref{loclindstring}, our expectation is that only channels in these classes are relevant to local Lindbladian evolution.

Low depth circuits of local unitary gates are a special class of unitary causal channels. Causal unitaries have a topological index\footnote{This index may be computed locally from the tensor of the MPU discussed below \cite{Ignacio_Cirac_2017,PhysRevB.98.245122}.} that takes values $\log(p/q)$ for natural numbers $p, q$, capturing the flow of information to the left and to the right \cite{gnvw}, and low depth circuits of local unitary gates are precisely the causal unitaries for which this topological index vanishes. From the perspective of phase classification, circuits are the only causal unitaries one should care about, as they approximate fast local unitary evolution. Nevertheless, it is most convenient for us to work with causal unitaries in general -- forgetting whether or not they are circuits -- because causal unitaries have convenient tensor network representations (which we will describe shortly). Similarly, a special class of causal channels is given by low depth circuits of local channels.\footnote{By this, we mean that the channel consists of a small number of layers of disjoint channels supported on small intervals.} It may be the case that these are the causal channels which approximate fast local Lindbladian evolution, just as unitary circuits do for unitary evolution, and that they are in dQC and are characterized by a vanishing topological index of their purification. We do not attempt to prove this conjecture. Regardless, our analysis considers causal channels in general, even if most of them are unrelated to Lindbladian evolution.

%%%%%%%%%%%%

\subsection{Tensor network representations of causal channels}\label{TNcausal}

Causal unitary operators, which in this context are the purifications of channels in dQC, have finite bond dimension tensor network representations called matrix product unitaries (MPUs) \cite{Ignacio_Cirac_2017,PhysRevB.98.245122}:
\begin{equation}\begin{split}\label{mpu}
    \includegraphics[height=0.045\textheight]{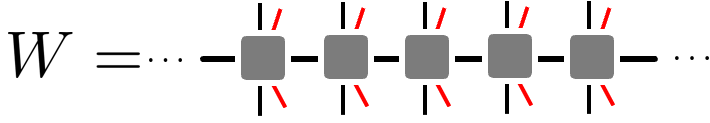}  \quad\raisebox{0.02\textheight}{.}
\end{split}\end{equation}
As in \cref{sec:purif}, the red legs of the tensor carry ancilla indices.

Here, unlike in the previous section, we simplify the analysis by restricting to channels that are translation invariant, which means that the tensor network representations of their purifications consist of the same tensor at every site. We expect, however, that our results hold without this assumption.

The theory of MPUs says that there exists a length $r\le \delta^4$ (where $\delta$ is the bond dimension of $W$) such that on blocks of $r$ sites, the tensor satisfies the following ``simpleness relations'' \cite{Ignacio_Cirac_2017}:
\begin{equation}\begin{split}
    \includegraphics[height=0.095\textheight]{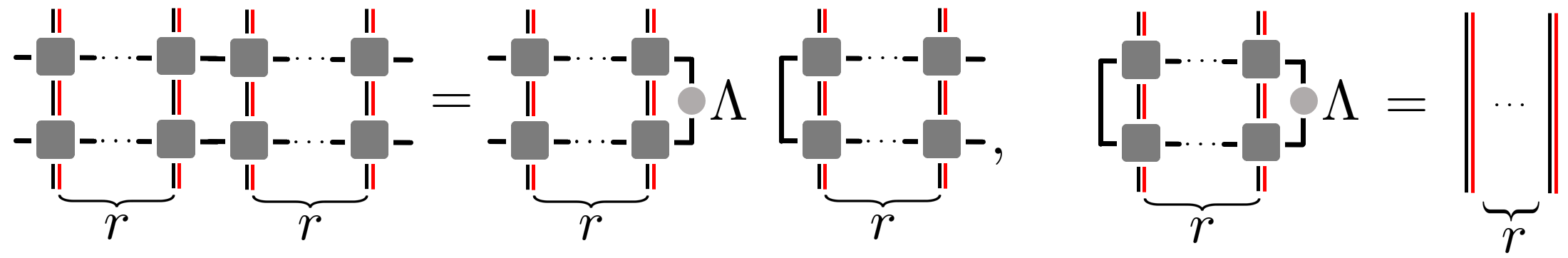}
     \raisebox{0.05\textheight}{,}
\end{split}\end{equation}
where $\Lambda$ denotes the right fixed point of the MPU transfer matrix.

The fact that the causal unitaries have tensor network representations means that channels in dQC do as well. The Kraus operator $K_i$ is realized as the matrix product operator obtained by plugging in the ancilla state $|a\rangle=\otimes_s|a_s\rangle$ (take all $a_s$ to be the same) and the ancillary space basis vector $|e_i\rangle=\otimes_s|e_{i_s}\rangle$:
\begin{equation}\begin{split}
    \includegraphics[height=0.06\textheight]{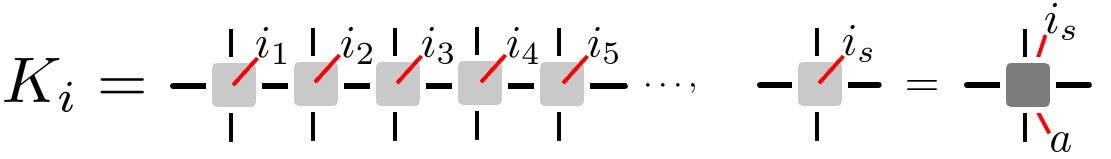}
    \raisebox{0.025\textheight}{.}
\end{split}\end{equation}
Then, in terms of the tensors for the Kraus operators, the simpleness relations become
\begin{equation}\begin{split}\label{simplekraus}
    \includegraphics[height=0.095\textheight]{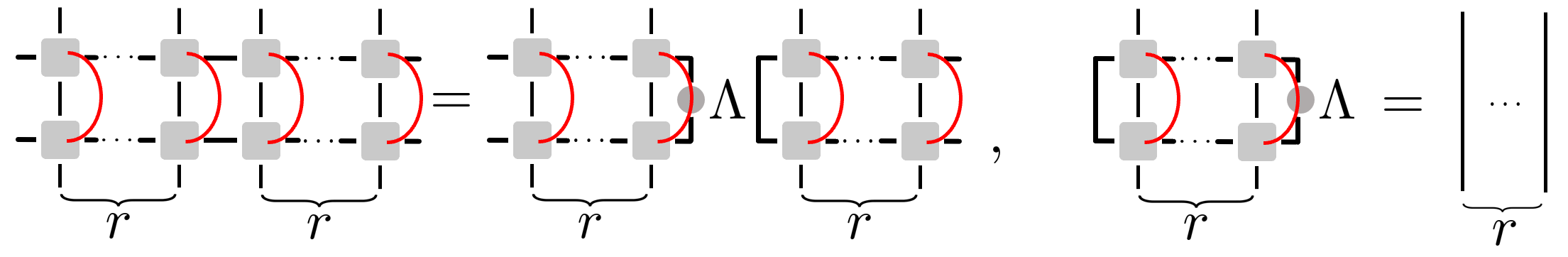}
     \raisebox{0.065\textheight}{.}
\end{split}\end{equation}

The uncorrelated noise considered in \cref{sec:preserveSPTO-singlesite} appears here as the channels whose MPUs have $\delta=1$:
\begin{equation}\begin{split}
    \includegraphics[height=0.11\textheight]{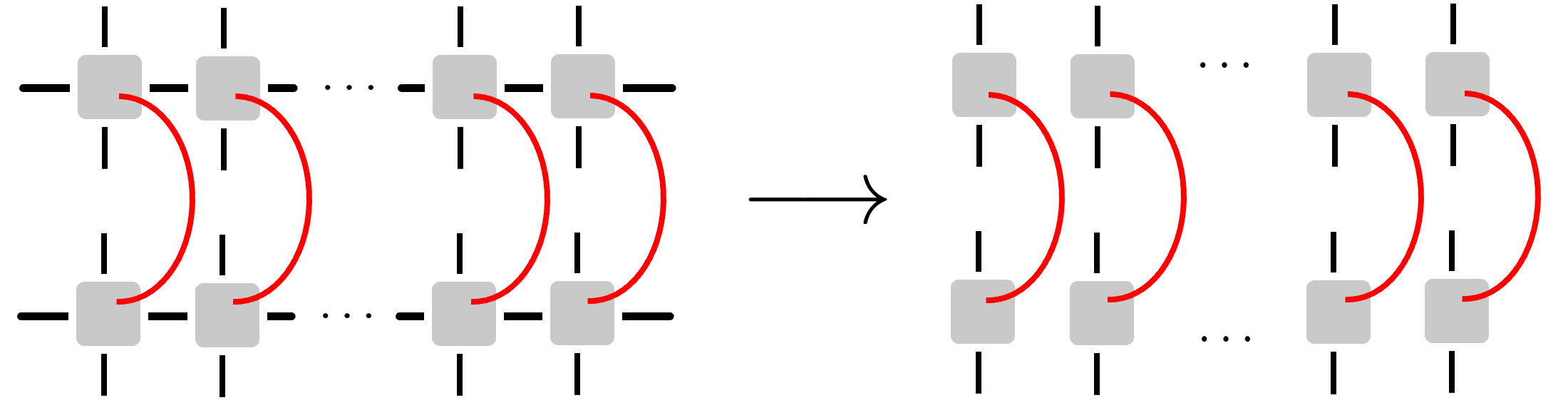}
    \raisebox{0.055\textheight}{\quad.}
\end{split}\end{equation}

%%%%%%%%%%%%%%%%%%%%%%%%%

\subsection{Local realization of the symmetry conditions}\label{sec:localSS}

It will be demonstrated in the following subsections that a causal channel preserves SPTO if it satisfies not just the strong symmetry condition, but the strong symmetry condition \emph{realized locally} \eqref{locSS}. The present subsection is dedicated to describing what is meant by local realization of the symmetry conditions and to motivating it. We consider channels in the class sdQC we defined earlier, meaning that the causal purification is symmetric under acting with the symmetry operator $U_g\otimes U_g^A$ on every site.

As always, the symmetry conditions refer to how the Kraus operators transform under conjugation by a symmetry. If the Kraus operators do not decompose as products of uncorrelated terms, we must work out what these global conditions mean locally, in terms of their tensor network representation. Local properties of a symmetry action may be studied by cutting the spin chain into two halves, so that the purified channel decomposes as $W=\sum_\mu W_l^\mu\otimes W_r^\mu$, and acting with the symmetry on the right half:
\begin{equation}\label{cutkraus}
    (\mathds{1}^l\otimes (U_g\otimes U_g^A)^r)W(\mathds{1}^l\otimes (U_g\otimes U_g^A)^r)^\dagger=\sum_{\mu,\nu}(Q_g)_{\mu\nu}W_l^\mu\otimes W_r^\nu~.
\end{equation}
The operators $Q_g$ are defined up to redefinition by phases and form a projective representation: $Q_gQ_h=\nu(g,h)Q_{gh}$. By folding the MPU representing the purification $W$ into a normal MPS \cite{Ignacio_Cirac_2017} and applying the usual arguments \cite{cirac2021matrix,handwaving}, it can be shown that the $Q_g$ satisfy\footnote{The version of the Fundamental Theorem of MPS in Theorem IV.4 of Ref. \cite{cirac2021matrix} can be used to show that this condition holds not just on blocks of size $r$ but on individual sites; however, we will not require this stronger condition in our arguments.}
\begin{equation}\begin{split}\label{SSClocal}
  \includegraphics[height=0.1\textheight]{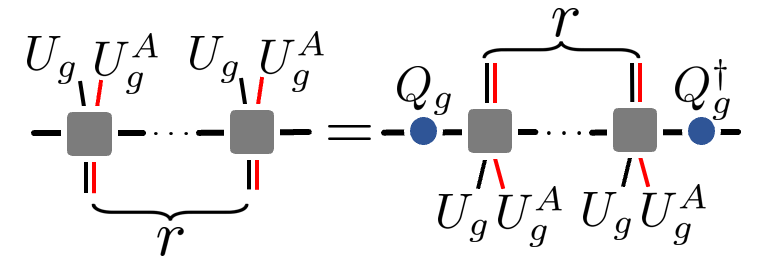}
  \raisebox{0.05\textheight}{.}
\end{split}\end{equation}
In particular, when the channel is SS, we have $U_g^A=\mathds{1}$, so the Kraus operators satisfy
\begin{equation}\begin{split}
 \includegraphics[height=0.10\textheight]{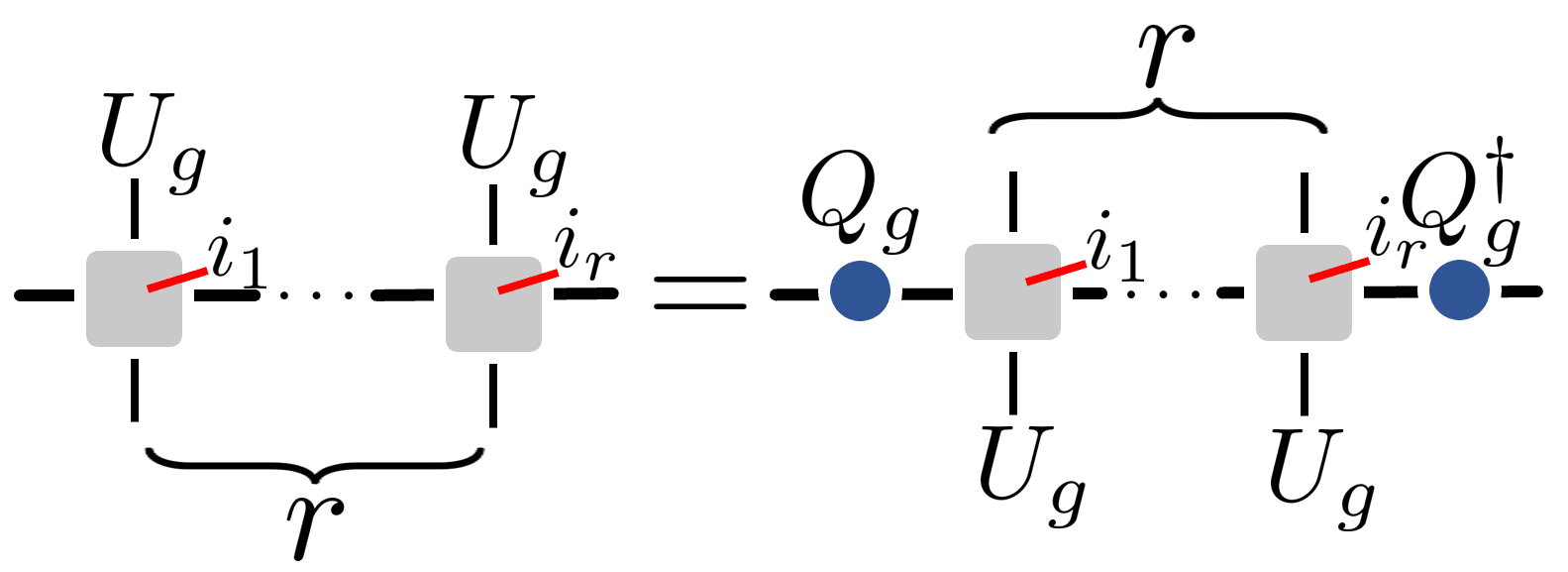}
    \raisebox{0.05\textheight}{.}
\end{split}\end{equation}
We note that symmetric MPUs have been studied previously \cite{PhysRevLett.124.100402}.

Now we can state the condition of local realization of the symmetry:
\begin{equation}\label{locSS}
    \boxed{\qquad Q_gQ_h=Q_{gh}~,\quad\forall\,g,h~,\quad\text{up to redefinition by phases.}\qquad(\text{locally realized symmetry})\qquad}
\end{equation}
Under redefinition of $Q$ by phases, the cocycle $\nu$ that captures the projectivity of $Q$ shifts by a coboundary, so local realization is the condition that $[\nu]$ is trivial in cohomology. In the case of uncorrelated noise, the symmetry conditions are automatically realized locally because $Q$ acts on a one-dimensional space. When a channel is WS or SS and its symmetry is realized locally, we say it is ``locally-WS'' or ``locally-SS''.

To build intuition for local realization, let us see that it is satisfied by the circuits of symmetric gates that define phase equivalence for states of closed systems. For a circuit, symmetry of the gates means that $Q$ can be extracted from a single gate by acting on half of its legs with the symmetry $g$:
\begin{equation}\begin{split}\label{circuit-locSS}
    \includegraphics[height=0.11\textheight]{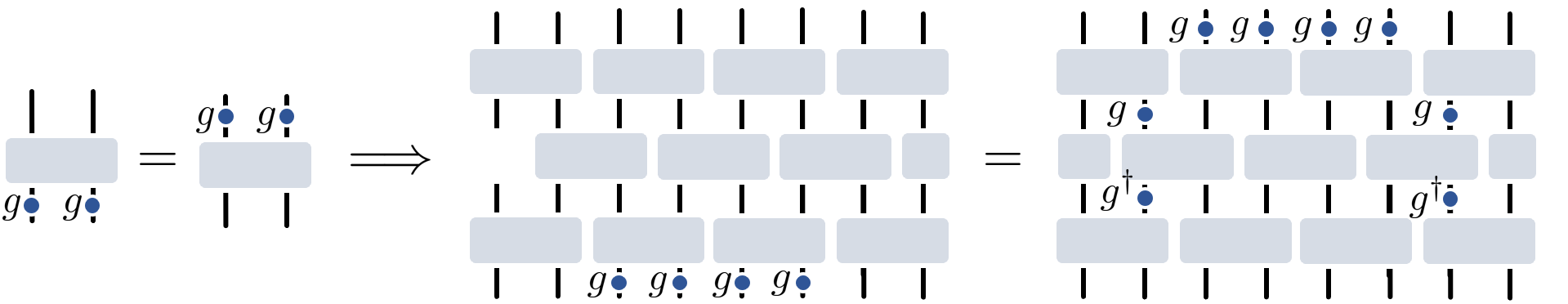}
    \raisebox{0.06\textheight}{.}
\end{split}\end{equation}
Since the adjoint action of $U_g$ on half of the gate is a linear (non-projective) representation, $Q$ is linear as well. More generally, consider the circuits of local channels that were mentioned briefly above. In analogy to the condition that the unitary gates are symmetric, these local channels can be made to satisfy the weak or strong symmetry condition. If they satisfy the WS or SS condition, the argument we just used for unitary circuits demonstrates that the causal channel as a whole is locally-WS or locally-SS, respectively.

%%%%%%%%%%%%%%%%%%%%%%%%%%

\subsection{String operators}\label{stringlocal}

Let us generalize \hyperref[lemma1]{Lemma 1} of \cref{stringsinglepres} to causal channels by showing the following lemma.
\begin{center}\label{lemma2}
\fbox{
$\qquad$\parbox{0.55\linewidth}{
\centering
\textbf{Lemma 2:} \emph{A translation-invariant channel in sdQC maps string operators to sums of string operators of the same type $(g,\alpha)$ if and only if the channel satisfies the local strong symmetry condition.}
}$\qquad$
}
\end{center}
Doing so requires working with a slight generalization of string operators where the end operators are supported intervals, rather than sites. It is assumed that the system size and the length of the string are large compared to the length of these intervals and the range $r$ of the channel.

After evolution by the channel, the string operator \eqref{stringop} becomes
\begin{equation}\begin{split}\label{local-evolvedstring}
    \includegraphics[height=0.09 \textheight]{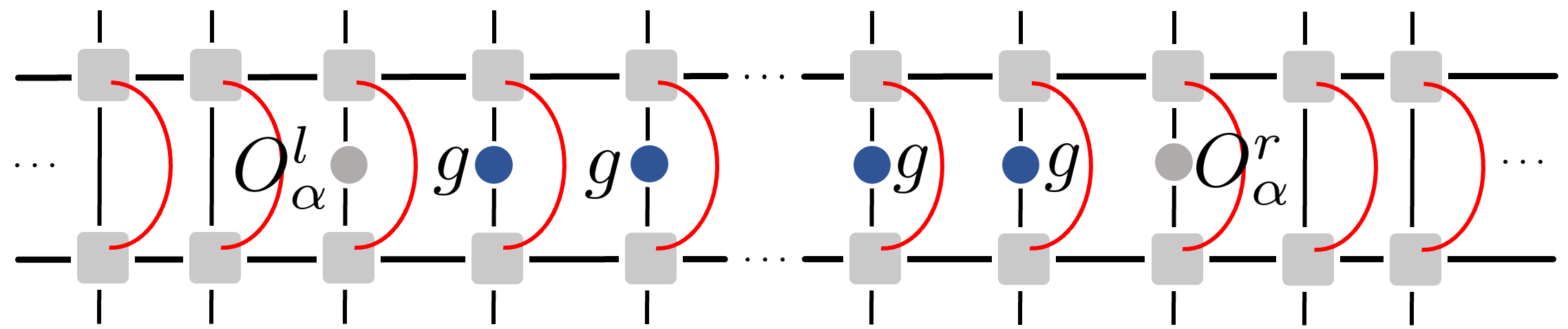}
      \quad\raisebox{0.02\textheight}{.}
\end{split}\end{equation}
If the SS condition is satisfied, the symmetry pulls through at the cost of operators $Q_g$:
\begin{equation}\begin{split}
    \includegraphics[height=0.11\textheight]{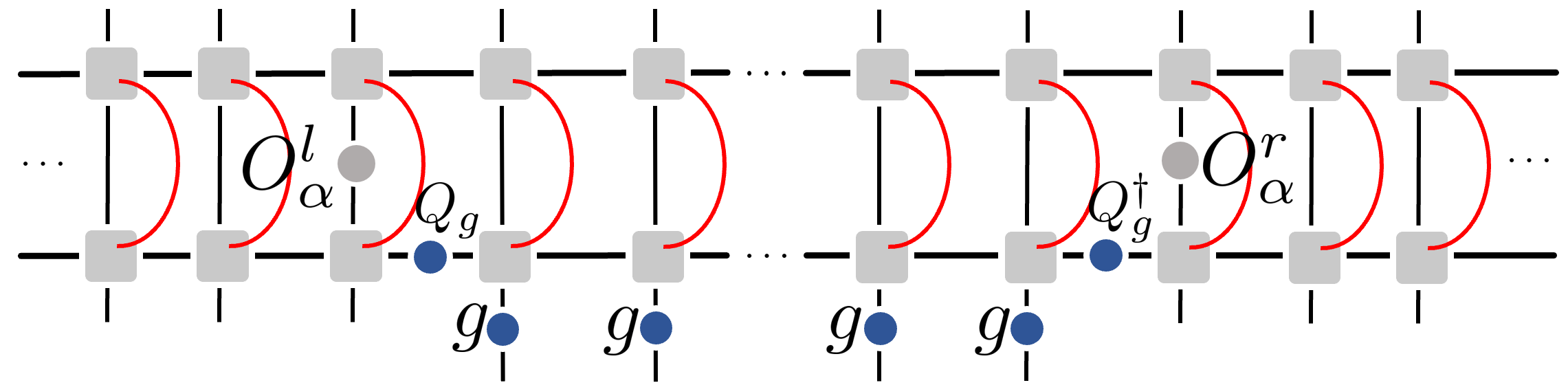}  \quad\raisebox{0.04\textheight}{,}
\end{split}\end{equation}
which cancel except near the ends of the string. Then the simpleness relations \eqref{simplekraus} can be applied to obtain
\begin{equation}\begin{split}\label{newstringop}
    \includegraphics[height=0.11\textheight]{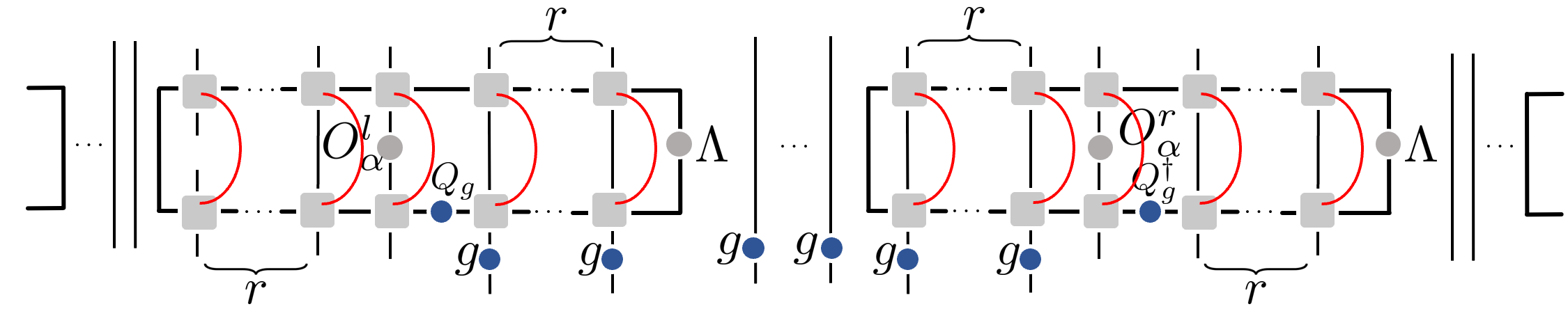}
      \quad\raisebox{0.06\textheight}{.}
\end{split}\end{equation}
This is a string operator with the symmetry $g$ in the bulk. The new end operators are the result of acting on the original end operators by the superoperators (generalizing the $\cE_{l,r}^\dagger$ of uncorrelated noise)
\begin{equation}\begin{split}\label{superop}
    \includegraphics[height=0.11\textheight]{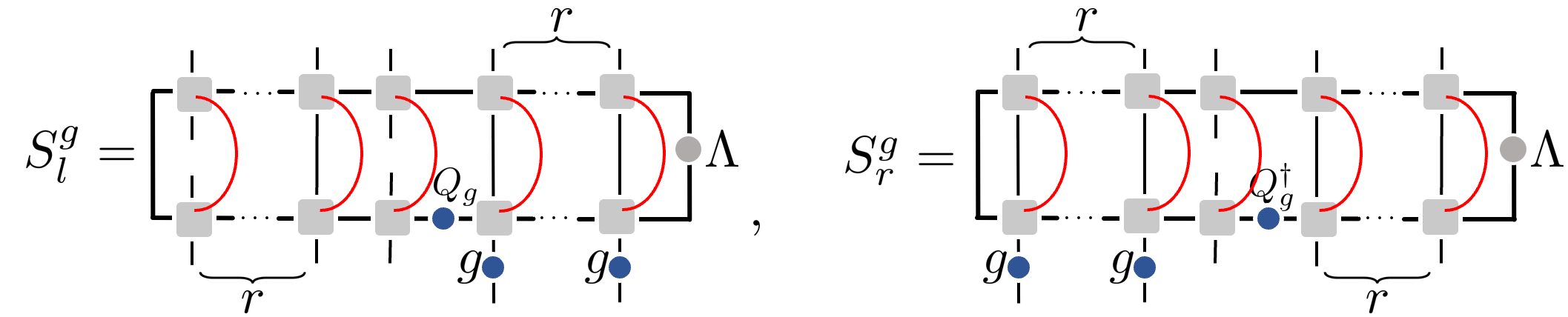}
      \quad\raisebox{0.04\textheight}{.}
\end{split}\end{equation}and are supported on $2r$ more sites than the original end operators. The superoperators $\cS_l^g$ and $\cS_r^g$ transform in the representations $h\mapsto \nu(g,h)/\nu(h,g)$ and $h\mapsto\left(\nu(g,h)/\nu(h,g)\right)^*$, respectively:
\begin{equation}\begin{split}\label{superoptrans}
    \includegraphics[height=0.26\textheight]{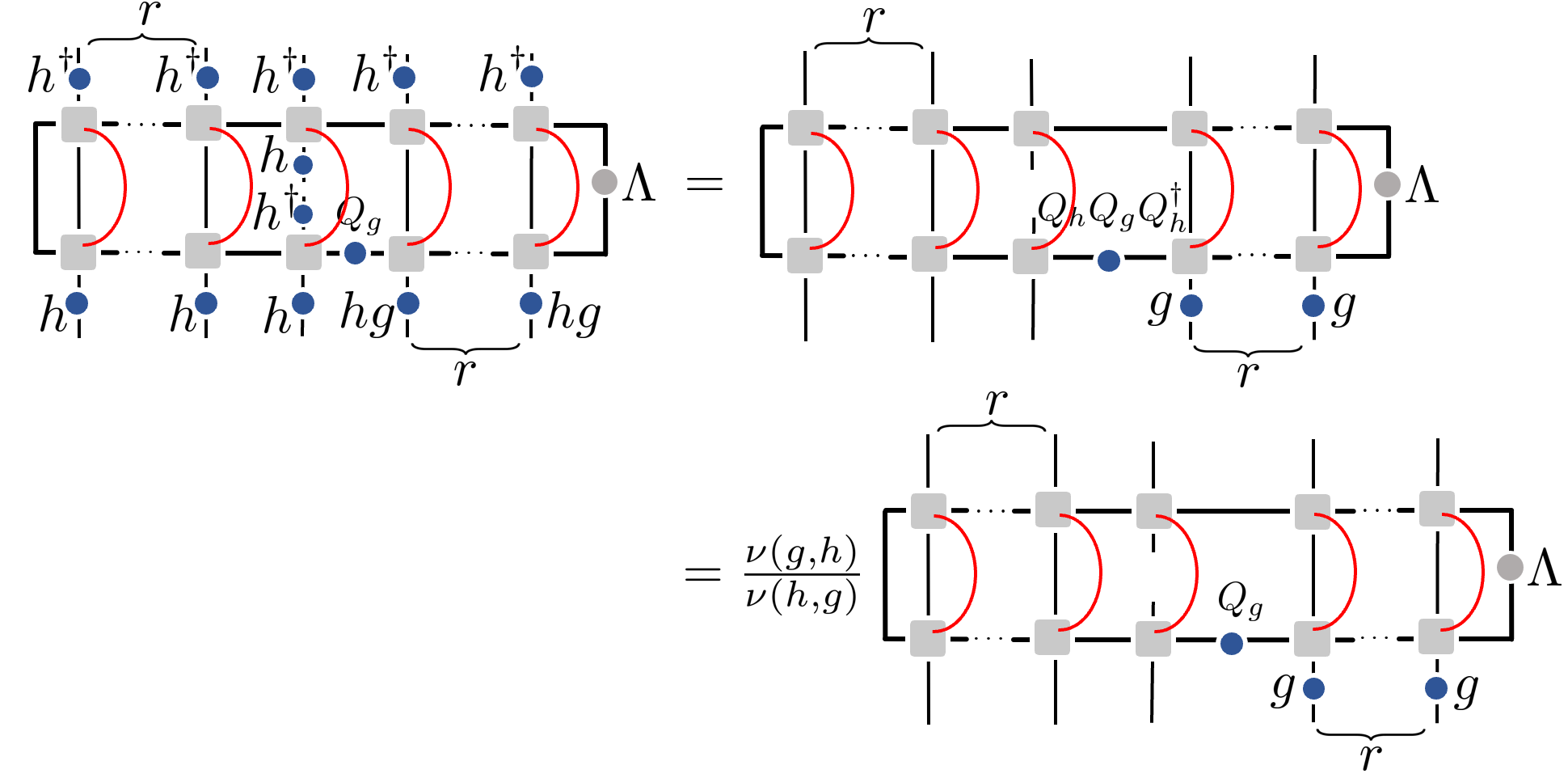}
      \quad\raisebox{0.05\textheight}{.}
\end{split}\end{equation}
This means that, if and only if $\nu$ is trivial, as is the case when the symmetry condition is locally realized, the new end operators $\cS_{l,r}^g(\cO_\alpha^{l,r})$ transform in the same representations as the original ones $\cO_\alpha^{l,r}$. We conclude that the evolved string operator is of type $(g,\alpha)$ if the channel satisfies the locally realized SS condition. Note that, in contrast with the case of uncorrelated noise, WS is not enough to ensure the correct transformation of the end operators. This is because the WS condition states only that the charge of operators is conserved \emph{globally}. When correlations between sites are present, charge can flow between regions of the system, such as between the two end operators, changing their individual charges.

It remains to show the converse: that, assuming the evolved string operator \eqref{local-evolvedstring} is a string operator of type $(g,\alpha)$, the channel must have been locally SS. The string operator with bulk $U_g^{\otimes j'}$ evolves into
\begin{equation}\begin{split}\label{stringopevolve1}
   \includegraphics[height=0.12\textheight]{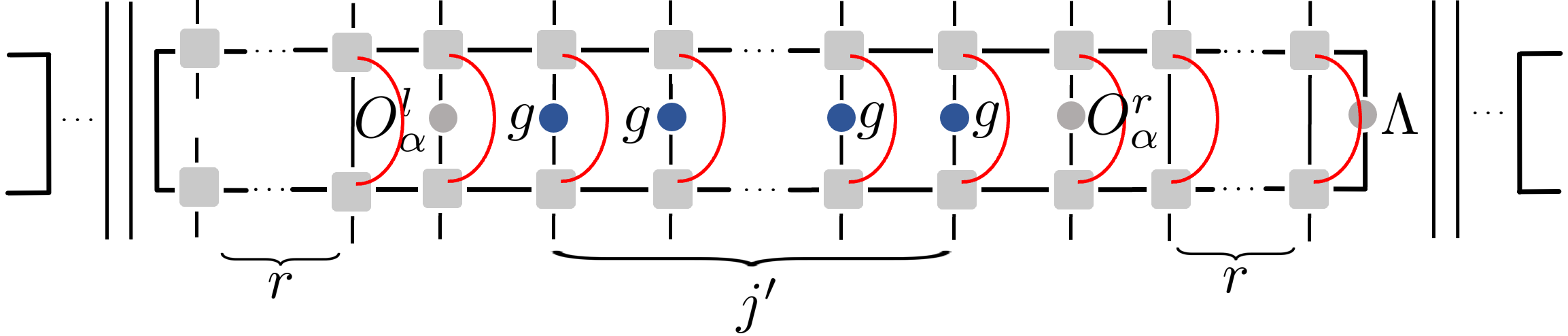}
     \quad\raisebox{0.035\textheight}{.}
\end{split}\end{equation}
Compose it with $(U_g^\dagger)^{\otimes j'}$ on the string bulk and take the trace to obtain 
\begin{equation}\begin{split}\label{eq:traced}
    \includegraphics[height=0.16\textheight]{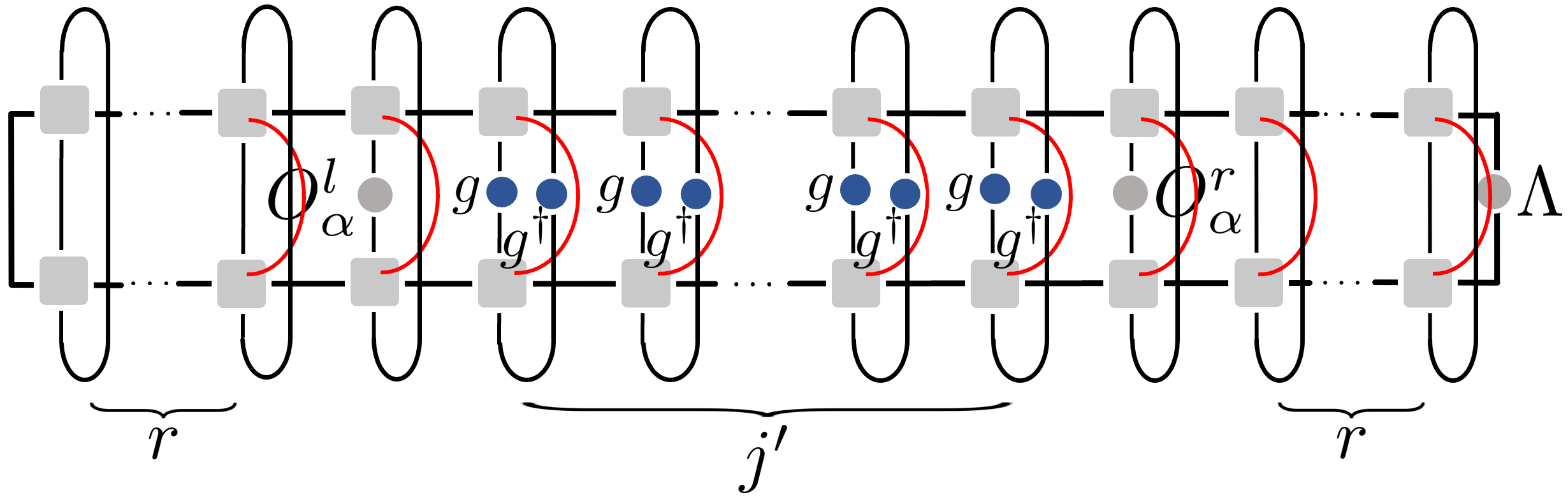}
      \quad\raisebox{0.04\textheight}{,}
\end{split}\end{equation}
neglecting the part of the tensor network outside the support of the evolved string. Meanwhile, doing the same to some string operator results in 
\begin{equation}\begin{split}\label{eq:traced2}
    \includegraphics[height=0.14\textheight]{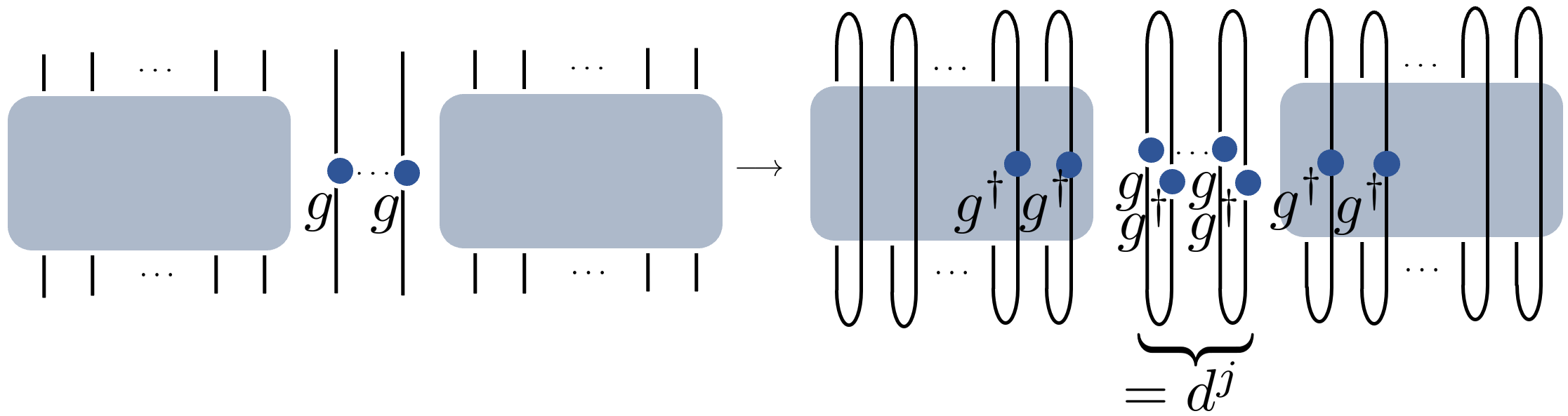}
      \quad\raisebox{0.04\textheight}{,}
\end{split}\end{equation}
where $j = j'-2r'$ for $r'$ the spread of the end operators under the channel (which turns out to be $r'=r$ due to SS). Setting \eqref{eq:traced} and \eqref{eq:traced2} equal (by our assumption), 
and defining 
\begin{equation}\begin{split}
    \includegraphics[height = 0.095\textheight]{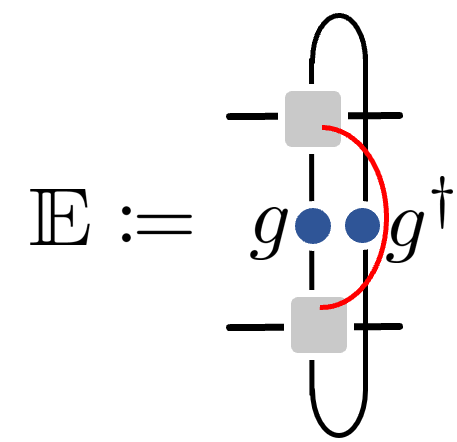}\quad\raisebox{0.04\textheight}{.}
\end{split}\end{equation}
we find that $\langle \rho_\ell|\EE^j|\rho_r\rangle = v d^j$ for some relative normalization $v\ne 0$, where $\langle \rho_\ell|$ and $|\rho_r\rangle$ are the boundary conditions imposed by the end operators in Eq.~\eqref{eq:traced}. Expressing $\EE$ in terms of its distinct nonzero eigenvalues $\lambda_k$, this amounts to the condition $\sum w_k \lambda_k^j=v d^j$ for all $j\ge1$, which implies that there must be an eigenvalue $\lambda_1=d$.%
\footnote{
First, note that by moving the $v d^j$ to the other side of the equation, one obtains $\sum w_k \lambda_k^j - v d^j=0$ for all $j\ge 1$, and thus, one is left with showing the following \\
\textbf{Lemma.}
\emph{
Given $K$ distinct $\mu_k\ne 0$, then
\begin{equation}\label{eq:vandermondesys}
    \sum_{k=1}^K c_k\mu_k^j = 0 \quad \forall\, j=J_0,\dots, K+J_0-1
    \quad \Rightarrow \quad
    c_k = 0 \quad \forall\,k=1,\dots,K\,.\tag{$\star$}
\end{equation}
}
\noindent\textbf{Proof.} The matrix with entries $M_{jk}\equiv\big(\mu_k^j\big)_{jk}$ is the product of the diagonal matrix $\mathrm{diag}(\mu_1^{J_0},\dots,\mu_K^{J_0})$ with the Vandermonde matrix $\big(\mu_k^{j-J_0}\big)_{jk}$, both of which are invertible. Thus, the linear system~\eqref{eq:vandermondesys}, $M\vec c = \vec 0$, has the unique solution $c_k\equiv 0$.\hspace*{\fill}$\square$\\
Note that this also provides a concise proof of the often-used Lemma in the MPS literature that $\sum a_k^j = \sum b_k^j$ implies that the $a_k$ and $b_k$ must be pairwise equal.
}
On the other hand, considering the MPU $W$ on a periodic ring of length $N$, we have -- using Cauchy-Schwarz -- that
\begin{equation}\label{cauchy}
    \big|\sum_k \lambda_k^N\big| = 
    \big|\mathrm{tr}\,\EE^N\big| =
    \big|\Tr[U_g^{\otimes N}W (U_g^{\otimes N})^\dagger W]\big|
    \le 
    \sqrt{\big|\Tr[(U_g^{\otimes N}W 
        (U_g^{\otimes N})^\dagger)(\cdots)^\dagger]\big|\,
    \big|\mathrm{tr}[WW^\dagger]\big|} = d^N\ .
\end{equation}
Thus, $\left| m_1 d^N + \sum_{k>1}\lambda_k^N\right|\le d^N$ for all $N$, where $m_1\ge1$ is the multiplicity of $\lambda_1=d$ and the $\lambda_{k>1}\ne d$ are the other eigenvalues. This implies that $\EE$ has one nondegenerate eigenvalue $\lambda_1=d$,
and all other eigenvalues are $0$.\footnote{To see this, write $\lambda_k = e^{2\pi i\xi_k}|\lambda_k|$, and fix $M=8^K$. Dirichlet's approximation theorem states that there are integers $p_k$ and $1\le q\le M$ such that $|\xi_k-p_k/q|\le 1/(qM^{1/K})$. Then, $|2\pi(q\xi_k-p_k)|\le \pi/4$, and thus $\mathrm{Re}\big[(e^{2\pi i \xi_k})^q\big]>0$. It follows that $\mathrm{Re}\big[\sum_{k>1}\lambda_k^q\big]>0$ and thus 
$\left| m_1 d^q + \sum_{k>1}\lambda_k^q\right|> d^q$, unless $m_1=1$ and there are no other nonzero eigenvalues $\lambda_k\ne0$.}
Thus, the Cauchy-Schwarz inequality \eqref{cauchy} is saturated, which implies that 
$U_g^{\otimes N}W (U_g^{\otimes N})^\dagger=e^{i\phi}W$, which is to say that the channel with purification $W$ is strongly symmetric. To see that the SS condition is realized locally, apply the `if' direction to obtain the evolved end operators. By assumption, they transform in the same irrep $\alpha$ as the inital end operators; therefore, the local-SS condition $[\nu]=0$ must hold.

%%%%%%%%%%%%%%%%%%%

\subsection{Preservation of string order by strongly symmetric local Lindbladians}\label{loclindstring}

Following Ref. \cite{coser2019classification}, we say a Lindbladian is \emph{local} if it can be written as a sum of local terms, each supported on an interval of length at most some constant.

The preservation of string operators by locally-SS channels in sdQC (\hyperref[lemma2]{Lemma 2}) means that we can state the following analog to \hyperref[theorem1]{Theorem 1}, where short times are times that are small compared to the system size.
\begin{center}
\fbox{
$\qquad$\parbox{0.75\linewidth}{
\centering 
\textbf{Conjecture:} \emph{Evolution generated by a local Lindbladian preserves SPTO at short times if and only if the Lindbladian is strongly symmetric.}
}$\qquad$
}
\end{center}

The conjecture is inspired by a plausible connection between local Lindbladians and causal channels. Just as local unitary evolution is approximated by locally-symmetric causal unitaries (in particular, circuits of symmetric local unitary gates) precisely when the generating Hamiltonian is symmetric, we expect that
\begin{equation}\label{statement}
\parbox{0.70\linewidth}{
\centering 
\emph{Evolution by a local Lindbladian is approximated by locally-WS/SS channels in sdQC precisely when the Lindbladian is WS/SS.}
}
\end{equation}

Let us motivate this statement nonrigorously. Local Lindbladian evolution is subject to Lieb-Robinson bounds \cite{coser2019classification}, so we expect it to be described by causal channels (with range $r$ linear in time), up to exponentially small errors outside of the lightcone. Moreover, we expect such causal channels to live in dQC since nontrivial convex combinations of channels in dQC (which plausibly are arbitrary causal channels \cite{PhysRevLett.125.190402}) seem to introduce unphysically long-range correlations. We established in \cref{sym-lind} that WS/SS of a Lindbladian implies WS/SS of the channels it generates, but the question remains whether locality of the symmetric Lindbladian implies that the channel is in sdQC and that the symmetry of the channel is locally realized. As mentioned previously, it may be the case that causal channels approximating local Lindbladian evolution are circuits of local channels, just as causal unitaries approximating local unitary evolution are circuits of local unitaries, and that these local `gates' are WS/SS precisely when the generating Lindbladian is WS/SS. Then an argument like Eq. \eqref{circuit-locSS} would translate the symmetry of the gates into locally realized symmetry of the channel.

Taking the statement \eqref{statement} for granted and neglecting the issue of approximation, the `if' direction of the conjecture follows from \hyperref[lemma2]{Lemma 2}. The strongly symmetric local Lindbladian generates locally-SS channels in sdQC, which by \hyperref[lemma2]{Lemma 2} preserve the types of string operators and therefore their patterns of zeros with generic end operators. The short time of the evolution is a crucial assumption, as it was necessary in \hyperref[lemma2]{Lemma 2} that the ranges of the causal channels were small compared to the string length; otherwise, the bulks of the strings were swallowed up by the end operators. The `only if' direction of the conjecture might require  analyzing the transfer matrix of the expectation value of the evolved string operator, as in \cref{necsuff1}.

%%%%%%%%%%%%%%%%%%%%%%%%%%%%%%%%

\subsection{Protected edge modes}

As in \cref{edgesingle}, consider a pure state $|\psi^\omega\rangle$ in an SPT phase characterized by the invariant $\omega$. Under the channel, it evolves into a mixture of states $K_i|\psi^\omega\rangle$. The SPT invariant will be obtained by cutting the system into left and right halves and acting on the right half by the symmetry. Across the cut, the state and Kraus operators decompose as $|\psi^\omega\rangle=\sum_a|\psi_{l,a}^\omega\rangle\otimes|\psi_{r,a}^\omega\rangle$ and $K_i=\sum_\mu K_{i,l}^\mu\otimes K_{i,r}^\mu$. The SS condition means that Eq.~\eqref{cutkraus} holds. Therefore, the states in the mixture transform as
\begin{equation}
    (\mathds{1}^l\otimes U_g^r)K_i|\psi^\omega\rangle=\sum_{a,b,\mu,\nu}(V_g)_{ab}(Q_g)_{\mu\nu}K_{i,l}^\mu|\psi_{l,a}^\omega\rangle\otimes K_{i,r}^\nu|\psi_{l,b}^\omega\rangle~,
\end{equation}
so their SPT invariants are captured by the projectivity class $[\omega\nu]$ of $V\otimes Q$. If the SS condition is realized locally, $[\nu]$ is trivial, so the SPT invariant $[\omega\nu]=[\omega]$ is unchanged.

This argument has a simple diagrammatic representation when the state $|\psi^\omega\rangle$ is an MPS. In this case, the MPO tensor for $K_i$ is contracted with the MPS tensor for $|\psi^\omega\rangle$ to obtain an MPS tensor for $K_i|\psi^\omega\rangle$. The virtual space of the new MPS tensor has symmetry action $V\otimes Q$ on blocks of size $r$.
\begin{equation}\begin{split}
    \includegraphics[width=0.65\textwidth]{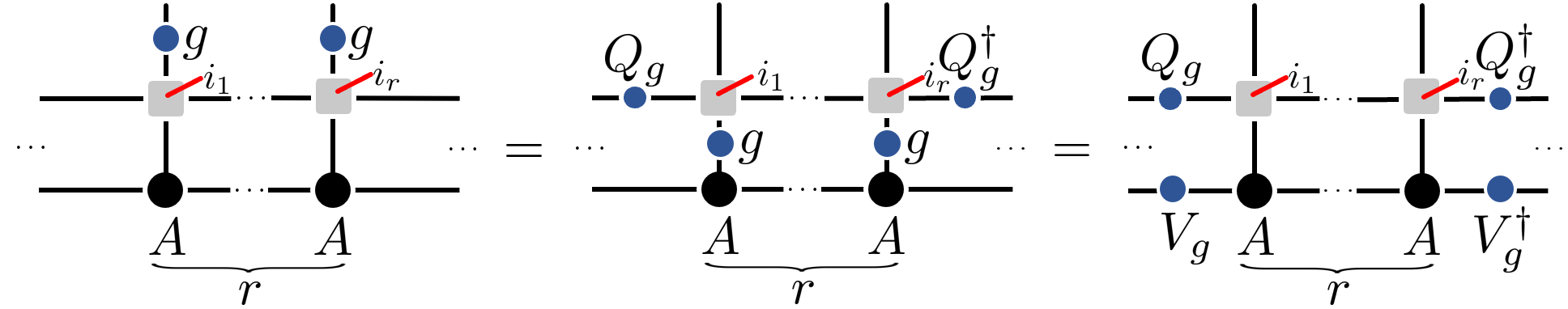}
      \quad\raisebox{0.04\textheight}{.}
\end{split}\end{equation}
Crucial to the preservation of protected edge modes is the fastness assumption, which means that the range $r$ of the MPO is small compared to the system size; without it, there is no invariant to be extracted locally.

%%%%%%%%%%%%%%%%%%%%%%%%
%%%%%%%%%%%%%%%%%%%%%%%%

\section{Transmutation of SPT phases by twisted symmetric channels}\label{sec:twistedstrong}

This section discusses versions of the symmetry conditions that are twisted by an endomorphism $\sigma:G\ra G$ of the symmetry group. The twisted conditions are stated in Eqs. \eqref{tWSC}, \eqref{tSSC}, and \eqref{tSSCalt}. For $\sigma$ that are not identity-connected (every nontrivial $\sigma$ when $G$ is finite and all but those of the form \eqref{inneraut} otherwise \cite{fultonharris}), channels twisted by $\sigma$ are not generated by continuous symmetric Lindbladian evolution in finite time. These channels therefore describe infinite time evolution (for example in \cref{coser-revisited}) and discrete noise.

First we discuss an action of group endomorphisms $\sigma$ on the SPT invariant $[\omega]$. We then argue in \hyperref[theorem2]{Theorem 2} that $\sigma$-twisted SS channels have the effect of changing the SPTO according to this action. In particular, when an endomorphism $\sigma$ does not change $[\omega]$, channels satisfying the $\sigma$-twisted SS condition preserve the phase with invariant $[\omega]$. This allows us to answer a question we had previously deferred -- of the \emph{necessary} condition for a channel to preserve a given SPTO. The answer is that the channel must be a mixture of $\sigma$-twisted SS channels for $\sigma$ that fix the SPTO. We also discuss the general situation where a channel does not preserve the phase but rather transmutes it into one of equal or lesser complexity, defined in Eq. \eqref{complexity}.

%%%%%%%%%%%%%%%%%%%%%%%

\subsection{The action of endomorphisms on SPT phases}\label{endaction}

\begin{figure}
    \centering
    \includegraphics[width = 0.8\textwidth]{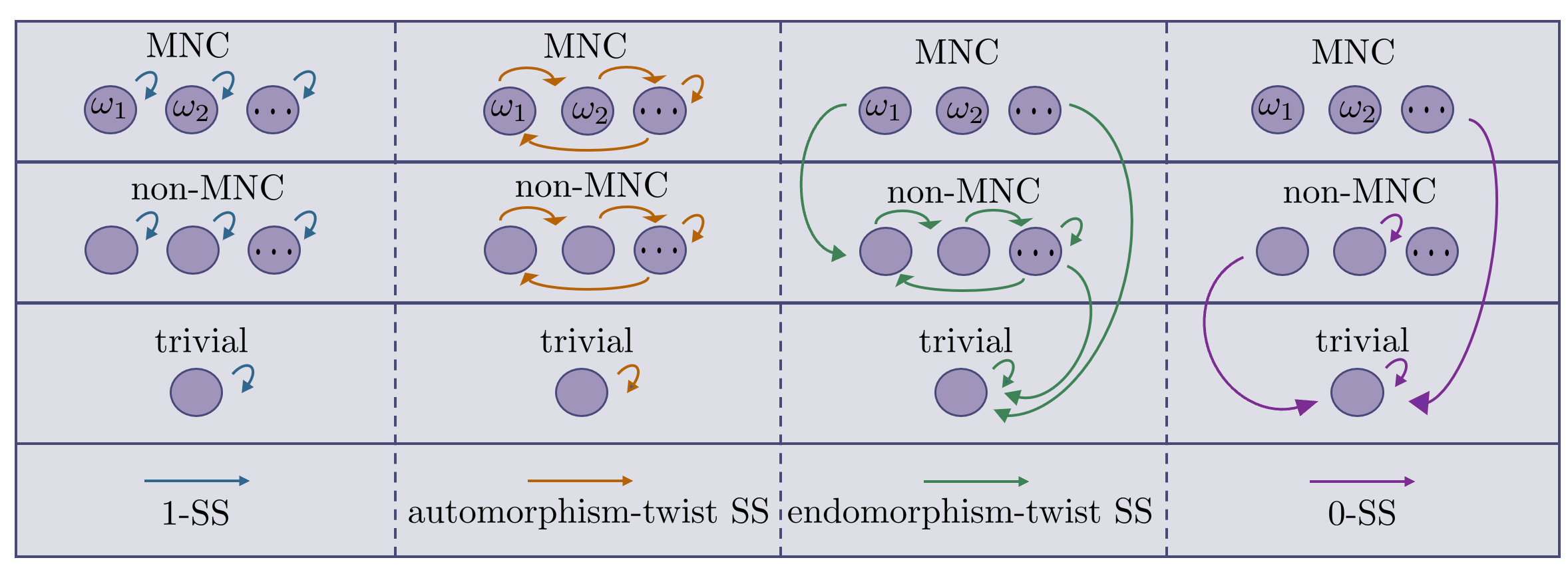}
    \caption{ \small Phases $\omega$ appear as nodes and the action of endomorphisms $\sigma$ as arrows between nodes. From left to right, the columns depict the identity endormorphism, automorphisms, noninvertible endomorphisms, and the constant endomorphism. In anticipation of the implementation of these endomorphism actions by twisted symmetric channels (c.f. \cref{necsuff}), the columns are labeled by the corresponding symmetry conditions.}
    \label{fig:phases_compare}
\end{figure}

\begin{figure}
    \centering
    \includegraphics[width=0.7\textwidth]{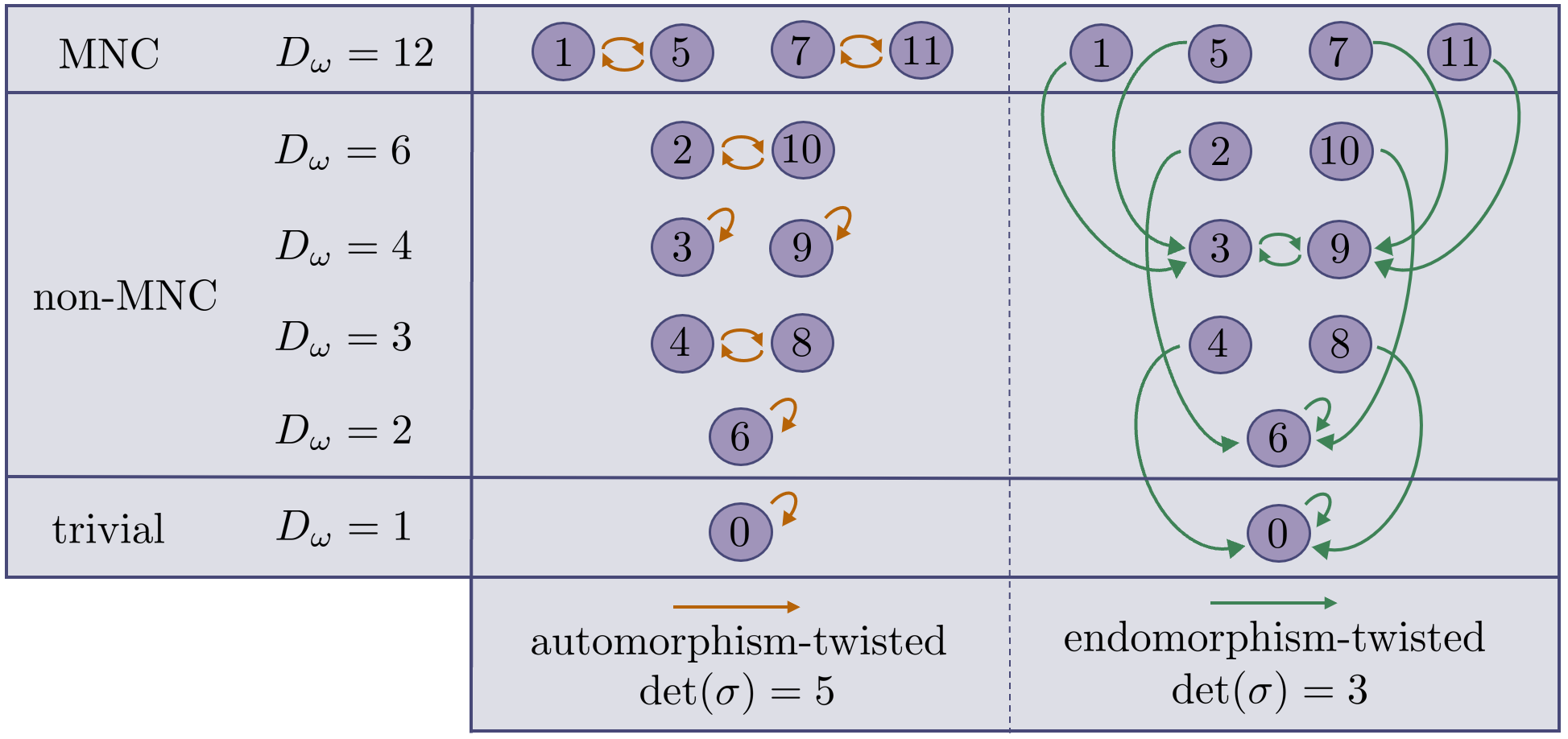}
    \caption{ \small The $12$ phases of the group $G=\ZZ_{12}\times\ZZ_{12}$ are depicted as nodes. Arrows represent the actions of two endomorphisms: on the left, an automorphism with $\det\sigma=5$; on the right, a noninvertible endomorphism with $\det\sigma=3$. Observe that the automorphism preserves complexity, as required by Claim \eqref{compclaim}, by preserving phases $3,6,9,0$ and exchanging the remaining phases with others of equal complexity. On the other hand, the noninvertible endomorphism reduces the SPT complexity of all MNC phases ($1,5,7,11$), as required by Claim \eqref{MNCclaim} but nevertheless preserves the SPT complexity of phases $3,6,9$, even fixing phase $6$.}
    \label{fig:Z12}
\end{figure}

An endomorphism $\sigma:G\ra G$ acts on the cocycle $\omega$ as a pullback. Concretely,
\begin{equation}\label{sigmaactsomega}
    \sigma:\omega\mapsto\sigma^*\omega~,\qquad (\sigma^*\omega)(g,h)=\omega(\sigma(g),\sigma(h))~.
\end{equation}
The action of endomorphisms has the following property:
\begin{equation}\label{compclaim}
\text{
\parbox{0.58\linewidth}{
\centering
    \emph{An automorphism preserves the complexity of phases.}
}
}
\end{equation}
This is because, if $\sigma$ is an automorphism, the transformed projective center
\begin{equation}
    K_{\sigma^*\omega}=\{\,g\,:\,\omega(\sigma(g),\sigma(h))=\omega(\sigma(h),\sigma(g))~\forall\,h\,\}~.
\end{equation}
equals $\sigma^*K_\omega$, since $\sigma(h)$ runs over the whole $G$, and this in turn is isomorphic to $K_\omega$ by $\sigma^*$. The converse to Claim \eqref{compclaim} is false because noninvertible endomorphisms may also preserve complexity. As a counterexample, take any $G,\sigma$ with $\omega=1$. Less trivially, take $G=H_1\times H_2$ and $\omega=P^*\omega_1$, where $P$ projects onto $H_1$ and $\omega_1$ is any cocycle on $H_1$. The endomorphism $\sigma=P$ is not an automorphism, yet it fixes $\omega$ and its complexity. Despite the lack of a full converse, one can make the following weaker claim:
\begin{equation}\label{MNCclaim}
\text{
\parbox{0.58\linewidth}{
\centering
    \emph{An endomorphism maps MNC phases, and only MNC phases, to MNC phases if and only if it is an automorphism.}
}
}
\end{equation}
In other words, an endomorphism preserves the distinction between MNC and non-MNC phases precisely when it is an automorphism. The `if' direction follows from Claim \eqref{compclaim}. To see the `only if' direction, note that the kernel of $\sigma$ is contained in $K_{\sigma^*\omega}$, so the MNC condition $K_{\sigma^*\omega}=\{1\}$ implies that $\ker\sigma=\{1\}$. The properties \eqref{compclaim} and \eqref{MNCclaim} appear in Figure \ref{fig:phases_compare} as constraints on the arrows between nodes. The special case of $G=\ZZ_{12}\times\ZZ_{12}$ is explored in complete detail in Figure \ref{fig:Z12}.

Let us examine in detail one of the most studied settings for investigations of one-dimensional SPT phases -- that of symmetry group $G=\ZZ_n\times\ZZ_n$, where phases are classified by $H^2(G;U(1))=\ZZ_n$.

Elements of $\ZZ_n\times\ZZ_n$ are ``vectors'' $(w,x)$ with $w,x\in\ZZ_n$. Endomorphisms of $\ZZ_n\times\ZZ_n$ are matrices with entries in $\ZZ_n$ that act on these vectors by matrix multiplication:
\begin{align}
\begin{split}
    &\,\End(\ZZ_n\times\ZZ_n)=M_2(\ZZ_n)=\left\{\left(\begin{array}{cc}a&b\\c&d\end{array}\right)\,:\,a,b,c,d\in\ZZ_n\right\}~,\\
    &g=\left(\begin{array}{c}w\\x\end{array}\right)~,\qquad\sigma(g)=\left(\begin{array}{cc}a&b\\c&d\end{array}\right)\left(\begin{array}{c}w\\x\end{array}\right)=\left(\begin{array}{c}aw+bx\\cw+dx\end{array}\right)~.
\end{split}
\end{align}
Automorphisms are those with invertible matrix, i.e. where the determinant $ad-bc$ is relatively prime to $n$.

Now let's discuss cocycles. The $n$ classes of $H^2(\ZZ_n\times\ZZ_n;U(1))=\ZZ_n$ are represented by cocycles
\begin{equation}\label{cocyclek}
    \omega_k[(w,x),(y,z)]=\exp(\tfrac{2\pi i}{n}k\,xy)~.
\end{equation}
Note that $\exp(\tfrac{2\pi i}{n}(-k)\,wz)$ is cohomologous to $\omega_k$ by the coboundary of $\phi[w,x]=\exp(\tfrac{2\pi i}{n}\,wx)$ and that $\exp(\tfrac{2\pi i}{n}\,wy)$ and $\exp(\tfrac{2\pi i}{n}\,xz)$ are trivialized by $\phi[w,x]=\exp(\tfrac{2\pi i}{n}\,w)$ and $\phi[w,x]=\exp(\tfrac{2\pi i}{n}\,x)$, respectively.

An endomorphism $\sigma$ on $\ZZ_n\times\ZZ_n$ induces an endomorphism $\sigma^*$ on $H^2=\ZZ_n$ as follows:
\begin{align}
\begin{split}
    (\sigma^*\omega_k)[(w,x),(y,z)]&=\omega_k[\sigma(w,x),\sigma(y,z)]\\
    &=\omega_k[(aw+bx,cw+dx),(ay+bz,cy+dz)]\\
    &=\exp(\tfrac{2\pi i}{n}k\,(cw+dx)(ay+bz))\\
    &=\exp(\tfrac{2\pi i}{n}k\,(ac\,wy+bc\,wz+ad\,xy+bd\,xz))\\
    &\sim\exp(\tfrac{2\pi i}{n}k\,(ad-bc)\,xy)\\
    &=\omega_{k(ad-bc)}[(w,x),(y,z)]~.
\end{split}
\end{align}
The penultimate line holds up to coboundaries. We conclude that the action of $\sigma^*$ on the group of SPT phases is multiplication of the SPT index $k$ by the determinant $(ad-bc)$ of $\sigma$.

Endomorphisms of $\ZZ_n$ are given by multiplication by an element of $\ZZ_n$, while automorphisms are those where the multiplication is by a generator of $\ZZ_n$, i.e. by a number relatively prime to $n$. This means that $\sigma^*$ is an automorphism of the group of SPT phases precisely when $\sigma$ is an automorphism of $G$.

For example, the automorphism $\sigma(w,x)=(x,w)$ that exchanges the two factors has the effect of inverting SPT phases since it has determinant $-1$. (For $n=2$, inversion is the identity, so the two phases -- trivial and Haldane -- are fixed by the exchange automorphism.) On the other hand, the endomorphism $\sigma(w,x)=(w,e)$ that collapses the second factor to the identity has determinant $0$, so it destroys all SPT phases.

Let us compute the projective center $K_{\omega_k}$, the set of elements $(w,x)$ such that
\begin{equation}
    \exp(\tfrac{2\pi i}{n}k\,xy)=\omega_k[(w,x),(y,z)]=\omega_k[(y,z),(w,x)]=\exp(\tfrac{2\pi i}{n}k\,wz)~,\quad\forall\,(y,z)~,
\end{equation}
i.e. such that $k\,xy\equiv k\,wz\mod n~,\,\forall\,y,z$. Taking $z=0$ while varying $y$ and vice versa, we find
\begin{equation}
    K_{\omega_k}=\{\,(w,x)\in\ZZ_n\times\ZZ_n\,:\,k\cdot(w,x)\equiv 0\}~.
\end{equation}
In particular, when $k$ is coprime to $n$, the projective center is trivial, so the cocycle $\omega_k$ is MNC. The invariant $(\det\sigma)\,k$ of the transformed phase is coprime to $n$ precisely when $k$ and $(\det\sigma)$ and both coprime to $n$; that is, when the original phase is MNC and $\sigma$ is an automorphism, in agreement with Claim \ref{MNCclaim}.

%%%%%%%%%%%%%%%%%%%%%%%%%

\subsection{Patterns of zeros under endomorphisms}\label{sec:patternsendos}

Let $G$ be abelian. The cocycle $\omega$ defines a ``pattern of zeros'' $\zeta_\omega:G\ra G^*$ given by
\begin{equation}
    \zeta_\omega:g\mapsto\chi_g^\omega(\cdot)=\frac{\omega(\cdot,g)}{\omega(g,\cdot)}~.
\end{equation}
The image is indeed linear characters (one-dimensional representations) since
\begin{equation}
    \chi_g^\omega(h)\chi_g^\omega(k)=\frac{\omega(h,g)}{\omega(g,h)}\frac{\omega(k,g)}{\omega(g,k)}=\frac{\omega(k,g)\omega(h,gk)}{\omega(g,h)\omega(hg,k)}\overset{\text{abelian}}{=}\frac{\omega(k,g)\omega(h,kg)}{\omega(g,h)\omega(gh,k)}=\frac{\omega(hk,g)}{\omega(g,hk)}=\chi_g^\omega(hk)~.
\end{equation}
The kernel of $\zeta_\omega$ is the projective center $K_\omega$ \eqref{projcenter}.

The pattern of zeros $\zeta_\omega$ determines the cohomology class $[\omega]$ of the cocycle $\omega$. To see this, note that the map $\omega\mapsto\zeta_\omega$ is a group homomorphism: $\zeta_{\omega_1}\zeta_{\omega_2}=\zeta_{\omega_1\omega_2}$; therefore, is suffices to check that its kernel consists of coboundaries. Suppose $\omega\mapsto 1$, i.e. $\omega(g,h)=\omega(h,g)$ for all $g,h$. Then any projective representation with class $\omega$ satisfies $V_gV_h=V_hV_g$. By Schur's lemma, $V_g$ must be proportional to the identity by a scalar $\lambda(g)$. Then $\omega$ is the coboundary of $\lambda$ since
\begin{equation}
    \lambda(g)\lambda(h)\mathds{1}=V_gV_h=\omega(g,h)V_{gh}=\lambda(gh)\omega(g,h)\mathds{1}~.
\end{equation}

It is convenient to represent a pattern of zeros $\zeta_\omega$ as a two-dimensional array with columns indexed by group elements $g$ and rows indexed by linear characters $\alpha$. The entry $(g,\alpha)$ in this array is $\star$ if $\zeta_\omega(g)=\chi_\alpha$ and zero otherwise. Since $\zeta_\omega(g)$ is a particular linear character, there is exactly one $\star$ per column. The row indexed by $\alpha$ has either $|K_\omega|$ or zero $\star$'s depending on whether $\chi_\alpha$ is in the image of $\zeta_\omega$. The rank of the array is $|G|/|K_\omega|$. For example, the two phases of symmetry $G=\ZZ_2\times\ZZ_2$ have patterns of zeros
\begin{equation}
    \zeta_\text{trivial} =            \left(\begin{array}{cccc}
         \star & \star & \star & \star \\
         0 & 0 & 0 & 0 \\
         0 & 0& 0 & 0 \\
         0 & 0 & 0 & 0
    \end{array}\right)~,
    \quad\zeta_\text{Haldane} = \left(\begin{array}{cccc}
         \star & 0 & 0 & 0 \\
         0 & \star & 0 & 0 \\
         0 & 0 & \star & 0 \\
         0 & 0 & 0 & \star
    \end{array}\right)~,
\end{equation}
with columns indexed by $g=(0,0),(0,1),(1,0),(1,1)$ and rows by $\alpha$ with $\tfrac{1}{\pi i}\log\chi_\alpha(w,x)=0,w,x,w+x$. 
Now consider the action of an endomorphism $\sigma$. It acts on $\omega$ as Eq.~\eqref{sigmaactsomega} and on $\zeta_\omega$ as
\begin{equation}
    \sigma\cdot\zeta_\omega=\zeta_{\sigma^*\omega}:g\mapsto\chi_g^{\sigma^*\omega}=\sigma^*\chi_{\sigma(g)}^\omega~.
\end{equation}
This rule tells us how the array for $\omega$ transforms into the array for $\sigma^*\omega$:
\begin{equation}\label{patternrule}
\parbox{0.7\linewidth}{
\centering
\emph{For each group element $g$, look up the unique row $\beta$ of the old pattern $\zeta_\omega$ such that the entry $(\sigma(g),\beta)$ is $\star$. Then compose $\beta$ with $\sigma$ to obtain the row $\alpha$ of the new pattern $\sigma\cdot\zeta_\omega$ such that the entry $(g,\alpha)$ is $\star$.}
}
\end{equation}

Using this rule for transforming patterns of zeros, one can check the examples of endomorphisms introduced above. The exchange automorphism swaps the middle two rows and swaps the middle two columns, fixing both the trivial and Haldane patterns. On the other hand, the endomorphism that collapses the second factor copies the first the third columns, which have $g$ in the image of $\sigma$, and moves their $\star$ entries up according to $\sigma$; the result is that both patterns are mapped to the trivial one.

The MNC property has a meaning in terms of patterns of zeros: the only column with a $\star$ in the $\alpha=1$ row is the $g=1$ column. Claim \ref{MNCclaim} can be shown in this language. Consider the `if' direction. We wish to find the entries $(g,1)$ of the new pattern that are $\star$. If $\sigma$ is an automorphism, these entries are the entries $(\sigma(g),(\sigma^{-1})^*1)=(\sigma(g),1)$ of the old pattern. Precisely when the old pattern is MNC, the only of these entries with $\star$ is the one with $\sigma(g)=1$ and so, since $\sigma$ is an automorphism, $g=1$ is the only solution and the new pattern is MNC. Consider the `only if' direction. The entry $(\sigma(g),1)$ of the old pattern is the entry $(g,\sigma^*1)=(g,1)$ of the new pattern. When the new pattern is MNC, it has a $\star$ in this row only for $h=1$. Precisely when the old pattern is MNC, it does only for $\sigma(g)=1$, which means the new pattern does for all $h\in\ker\sigma$; therefore, precisely in this case do we have $\ker\sigma=\{1\}$, which is to say that $\sigma$ is an automorphism.

%%%%%%%%%%%%%%%%%%%%%%%

\subsection{Twisted symmetric channels}

Having understood the action of group endomorphsisms on phases, we turn to studying the channels that implement it. Here, we introduce twisted symmetry conditions and discuss the structure of Kraus operators of twisted symmetric channels. Later we will argue that the $\sigma$-SS condition implements the action of $\sigma$.

The $\sigma$-twisted weak symmetry ($\sigma$-WS) condition is
\begin{equation}\label{tWSC}
    \boxed{
    \qquad\cU_g\circ\cE\circ\cU_{\sigma(g)}^\dagger=\cE~,\quad\forall\,g~.\qquad\text{($\sigma$-twisted weak symmetry condition)}\qquad
    }
\end{equation}
By setting the phases $\theta_i$ equal as before, we obtain the $\sigma$-twisted strong symmetry ($\sigma$-SS) condition:
\begin{equation}\label{tSSC}
    \boxed{
    \qquad U_gK_iU_{\sigma(g)}^\dagger=e^{i\theta(g)}K_i~,\quad\forall\,i,g~.\qquad\text{($\sigma$-twisted strong symmetry condition)}\qquad
    }
\end{equation}
Using the argument from before but with $X=U_g$, $Y=e^{i\theta(g)}U_{\sigma(g)}$, we obtain the alternative statement
\begin{equation}\label{tSSCalt}
    \boxed{
    \qquad\cE^\dagger(U_g)=e^{i\theta(g)}U_{\sigma(g)}~,\quad\forall\,g~.\qquad\text{($\sigma$-twisted strong symmetry condition)}\qquad
    }
\end{equation}

The untwisted SS condition means that the channel decomposes as a sum of channels on irrep blocks, acting only within multiplicity spaces. A similar statement holds for $\sigma$-SS channels: each Kraus operator has a block decomposition $K_i=\oplus_{\alpha\beta}K_i^{\alpha\beta}$ such that the component $K_i^{\alpha\beta}$, which is a map from the multiplicity space of $\beta$ to that of $\alpha$, vanishes unless $\alpha=\sigma^*\beta$. This is because the $\sigma$-SS condition says that $K_i$ maps to a space where $g$ acts as $\sigma(g)$ did before mapping. This means the completeness condition on $\cE$ implies
\begin{equation}\label{twistcomp}
    \sum_i(K_i^{\sigma^*\alpha,\alpha})^\dagger K_i^{\sigma^*\alpha,\beta}=\sum_i(K_i^\dagger K_i)^{\alpha\beta}=\mathds{1}\delta^{\alpha\beta}~,
\end{equation}
which in particular enforces a completeness condition on the channels
\begin{equation}
    \cE_\alpha(\rho)=\sum_i (K_i^{\sigma^*\alpha,\alpha})^\dagger\rho\,K_i^{\sigma^*\alpha,\alpha}~.
\end{equation}
When $\sigma$ is not an automorphism, each term $K_i^\dagger K_i$ may have off-diagonal components: $K_i$ maps $\alpha$ to $\sigma^*\alpha$, which $K_i^\dagger$ maps back to any $\beta$ in the preimage. Eq. \eqref{twistcomp} implies these must cancel in the sum.

Extending the untwisted class sdQC of causal channels, one can define $\sigma$-sdQC as the set of channels with a purification that is both causal and $\sigma$-twisted symmetric under an on-site symmetry. The folded MPS of the MPU representing the purification has a symmetry $U_g\otimes U_{\sigma(g)}^\dagger\otimes U_g^A\otimes (U_{\sigma(g)}^A)^\dagger$, which defines a projective representation $Q$. Local realization of the symmetry is again the condition that $Q$ is linear.

While strong symmetry twisted by an automorphism is possible in reversible channels, strong symmetry twisted by a noninvertible endomorphism is not.\footnote{A stronger statement also holds: noninvertible twists are impossible not just in reversible channels but in all channels that are invertible as linear maps. To see this, note that $\cE^\dagger$ annihilates $U_g-\mathds{1}$ for $g\in\ker\sigma$ and that $\cE$ is invertible iff $\cE^\dagger$ is.} To see this, suppose $g$ belongs to the kernel of $\sigma$. Then the single Kraus operator $K$ of the reversible channel satisfies $K^\dagger U_gK=\mathds{1}$, but this implies $U_g=\mathds{1}$, so $g=1$ by faithfulness. In light of \hyperref[theorem2]{Theorem 2} (below), this means that reduction of SPT complexity -- as opposed to change of phase at a fixed SPT complexity level -- is a phenomenon unique to irreversible channels. On the other hand, any automorphism can be realized by a reversible channel: let $U_g$ contain one copy of each irrep and let $K$ be the permutation matrix that implements the induced action $\sigma^*$ on irreps.

The impossibility of noninvertible twists for reversible channels is reflected in purifications. If $\sigma$ is an automorphism, the construction in \cref{sec:purif} yields purifications $W$ of $\sigma$-SS channels that satisfy
\begin{equation}
    (U_g\otimes\mathds{1}^A)W=W(U_{\sigma(g)}\otimes\mathds{1}^A)~.
\end{equation}
However, if $\sigma$ is not invertible, the irrep block structure of the Kraus operators means that some rows of $W$ constructed this way must be zero, meaning it is not unitary and so not a valid purification.

Endomorphisms compose contravariantly under the composition of channels. If $\cE$ is a $\sigma$-WS channel and $\cE'$ a $\sigma'$-WS channel, their composition $\cE\circ\cE'$ has WS twisted by $\sigma'\circ\sigma$, as can be seen by
\begin{equation}
    \cU_g\circ\cE\circ\cE'\circ\cU_{(\sigma'\circ\sigma)(g)}^\dagger=\cU_g\circ\cE\circ\cU_{\sigma(g)}^\dagger\circ\cE'=\cE\circ\cE'~.
\end{equation}
If both $\cE$ and $\cE'$ have $\sigma$- and $\sigma'$- twisted-SS, respectively, the composition $\cE\circ\cE'$ has $(\sigma'\circ\sigma)$-SS since
\begin{equation}
    (\cE\circ\cE')^\dagger(U_g)=(\cE'^\dagger\circ\cE^\dagger)(U_g)=\cE'^\dagger(U_{\sigma(g)})=U_{(\sigma'\circ\sigma)(g)}~.
\end{equation}
We also note that convex combinations and tensor products of $\sigma$-WS/SS channels are $\sigma$-WS/SS.

%%%%%%%%%%%%%%%%%%%%%%

\subsection{Transmutation of SPT phases}\label{necsuff}

This section is dedicated to showing that certain $\sigma$-twisted strongly symmetric channels have the effect of transmuting one coherent SPT phase into another, according to the action of $\sigma$ on the SPT invariants.

It is not the case that all $\sigma$-SS channels perform transmutation of SPT phases, but most of them do. We state our result for $\sigma$-SS channels satisfying a genericness condition
\begin{equation}\label{channel-generic}
    \Phi_\alpha\ne 0~,\quad\forall\,\alpha\in\im\,\sigma~,
\end{equation}
where $\Phi_\alpha$ is the $\alpha$-labeled component of $\cE$ in the irrep block decomposition of $\cE$ discussed in \cref{sec:weakstrong}. This condition excludes, for example, the fully dephasing channel of \cref{stringordersimulations}, which destroys SPT order despite having strong symmetry. It includes all channels generated by Lindbladians in finite time. In the following theorem, ``coherent SPT phase'' means a class of coherent SPT states with a given pattern of zeros (which, by \hyperref[theorem1]{Theorem 1} and the conjecture of \cref{loclindstring}, is a phase defined by strongly symmetric Lindbladians). Mapping an SPT phase to another refers to mapping every state in one phase to a state in the other.
\begin{center}\label{theorem2}
\fbox{
$\qquad$\parbox{0.55\linewidth}{
\centering
\textbf{Theorem 2:} \emph{Generic locally $\sigma$-twisted strongly symmetric channels map the coherent SPT phase with invariant $\omega$ to the phase with invariant $\omega'=\sigma^*\omega$. \\ \, \\For uncorrelated noise, the converse holds: \\ if a channel maps the coherent SPT phase with invariant $\omega$ to the phase with invariant $\omega'$, it is $\sigma$-twisted strongly symmetric for some $\sigma$ with $\sigma^*\omega=\omega'$.
}
}$\qquad$
}
\end{center}
As we will discuss later, the converse statement is false for causal channels.\footnote{We also remark about the converse statement that the genericness condition \eqref{channel-generic} is sufficient but not necessary: a weaker $\omega'$-dependent genericness condition on only the subset of the $\alpha$ that appear in the pattern of $\omega'$ is enough.}

A consequence of \hyperref[theorem2]{Theorem 2} is that, among channels of uncorrelated noise, twisted strongly symmetric channels are precisely those that map within the space of SPT states. Focusing on the MNC phases discussed in \cref{sec:patternsendos}, we can also conclude that, among channels of uncorrelated noise, \emph{automorphism-twisted} strongly symmetric channels are precisely those that map within the space of MNC SPT states.

\hyperref[theorem2]{Theorem 2} tells us when a channel preserves a given SPTO:
\begin{center}
\fbox{
$\qquad$\parbox{0.55\linewidth}{
\centering
\textbf{Corollary:} \emph{The SPTO with invariant $\omega$ is preserved by generic locally $\sigma$-twisted strongly symmetric channels with $\sigma$ that fix $\omega$. Among channels of uncorrelated noise, $\sigma$-twisted strongly symmetric channels with such $\sigma$ are the only channels that preserve this SPTO.}
}$\qquad$
}
\end{center}

To prove the theorem, we need the following lemma, which generalizes \hyperref[lemma1]{Lemma 1} (for uncorrelated noise, not necessarily symmetric) and \hyperref[lemma2]{Lemma 2} (for channels in $\sigma$-sdQC, which in particular are $\sigma$-WS) by adding a twist $\sigma$. Let $s(g,\alpha_l,\alpha_r)$ denote a string operator with end operators transforming in $\alpha_l$ and $\alpha_r^*$, respectively.\footnote{We previously considered only string operators with $\alpha_l=\alpha_r$ since these are the ones with nonvanishing patterns of zeros.}
\begin{center}\label{lemma3}
\fbox{
$\qquad$\parbox{0.55\linewidth}{
\centering
\textbf{Lemma 3:} \emph{Consider either a channel of uncorrelated noise or a translation-invariant causal channel in $\sigma$-sdQC. \\ \, \\ The channel satisfies the $\sigma$-twisted (local) strong symmetry condition if and only if it maps each string operator $s(g,\alpha,\alpha)$ to a sum of string operators $s(\sigma(g),\beta_l,\beta_r)$, where $\sigma^*\beta_{l,r}=\alpha$ (if no $\beta_{l,r}$ exists, the sum is empty).}
}$\qquad$
}
\end{center}

Let us now prove the lemma. The label on the string bulk is changed from $g$ to $\sigma(g)$ if and only if $\cE$ is $\sigma$-SS. In the case of uncorrelated noise, this is because each $\cE_s$ is $\sigma$-SS and so $\cE_s^\dagger(U_g)=e^{i\theta(g)}U_{\sigma(g)}$. For a channel in $\sigma$-sdQC, the argument is essentially that of \cref{stringlocal}. The label on the ends of the string are changed from $\alpha$ to a sum of irreps $\beta_{l,r}$ satisfying $\sigma^*\beta_{l,r}=\alpha$. This is because the superoperators $S_{l,r}^g$ \eqref{superop} (which are simply $\cE_{l,r}^\dagger$ for uncorrelated noise) are invariant under acting on the inner legs with $U_h$ and on the outer legs with $U_k$ for $\sigma(k)=h$, as can be seen by an argument like Eq. \eqref{superoptrans}. This means that they map the representation space $\alpha$ on the inner legs to its preimage under $\sigma^*$ on the outer legs. We have shown that the string operator evolves into an operator $s(\sigma(g),\cS_l^g(\cO^l_\alpha),\cS_r^g(\cO^r_\alpha))$ and that $\cS_{l,r}^g(\cO^{l,r}_\alpha)$ is a sum of end operators that transform with $\beta_{l,r}$ such that $\sigma^*\beta_{l,r}=\alpha$, proving the lemma. For uncorrelated noise, an alternative way of understanding the change in representation labeling the end operators is with the block decomposition of the Kraus operators. Since $\cE_{l,r}$ are $\sigma$-SS, their Kraus operators vanish outside of the irrep blocks $K_i^{\lambda\tau}$ with $\tau$ in the preimage of $\lambda$ under $\sigma^*$. Meanwhile $\cO_\alpha^{l,r}$ have nonvanishing blocks for irreps $\lambda',\lambda$ such that $\lambda'\otimes\lambda^*=\alpha$. Putting these together, the nonvanishing blocks of each term $K_i^\dagger\cO_\alpha^{l,r}K_i$ in the evolved end operator occur at irreps $\tau',\tau$ in the preimage of $\lambda',\lambda$ with $\lambda'\otimes\lambda^*=\alpha$. Each of these blocks has $\tau'\otimes\tau^*$ in the preimage of $\alpha$, so we conclude that $\cE_{l,r}^\dagger(\cO_\alpha^{l,r})$ lives in the sum of irreps $\beta$ with $\sigma^*\beta=\alpha$.

With \hyperref[lemma3]{Lemma 3} in hand, let us turn toward proving \hyperref[theorem2]{Theorem 2} by first reformulating it in terms of patterns of zeros. The pattern $\zeta_\omega$ of a state is understood as the collection of pairs $(g,\alpha)$ such that $\langle s(g,\alpha,\alpha)\rangle$ is generically nonvanishing on the state. By the rule \eqref{patternrule}, the pattern of zeros $\sigma\cdot\zeta_\omega$ consists of pairs $(g,\alpha)$ such that $\alpha=\sigma^*\beta$ for the (unique) $\beta$ for which $(\sigma(g),\beta)$ appears in the pattern $\zeta_\omega$. The theorem demonstrates how this new pattern can be understood as expectation values of evolved operators $\cE^\dagger(s(g,\alpha,\alpha))$ evaluated on the original state. To be precise, the first half of the theorem states that, on an SPT state,
\begin{equation}\label{fact2}
\parbox{0.8\linewidth}{
\centering
\emph{If $\cE$ is a generic locally $\sigma$-SS channel, then generically $\langle\cE^\dagger(s(g,\alpha,\alpha))\rangle\ne 0$ precisely for the pairs $(g,\alpha)$ such that $\alpha=\sigma^*\beta$ for the (unique) $\beta$ with $\langle s(\sigma(g),\beta,\beta)\rangle\ne0$.}
}
\end{equation}
To see why this is true, apply \hyperref[lemma3]{Lemma 3} to write $\cE^\dagger(s(g,\alpha,\alpha))$ as a sum of terms $s(\sigma(g),\beta_l,\beta_r)$ with $\sigma^*\beta_{l,r}=\alpha$. Due to the condition \eqref{channel-generic}, these terms do not vanish (though the sum may be empty if no $\beta_{l,r}$ exist). A pattern of zeros of an SPT state has a unique entry $\beta$ per column, so the expectation values of the terms in the sum vanish unless $\beta_l=\beta_r=\beta$; either zero or one terms do not vanish. We have $\langle\cE^\dagger(s(g,\alpha,\alpha))\rangle\ne 0$ when the nonvanishing term $\langle s(\sigma(g),\beta,\beta)\rangle$ appears in the sum. This can only happen when $\alpha=\sigma^*\beta$, and in this case it generically happens, since generically the $\beta$-components of the end operators are nonzero.

It remains to prove the second half of \hyperref[theorem2]{Theorem 2}. The argument follows that of \cref{necsuff1}, except that $\sigma$ is no longer constrained to be connected to the identity endomorphism. Now the condition that the transfer matrix \eqref{transfer-single-eq} has $\lambda_\text{max}=1$ implies that $\cE_s^\dagger(U_g)=U_h$, which is to say that $\cE_s$ is $\sigma$-SS for some $\sigma$ with $\sigma(g)=h$. Then apply the first half of \hyperref[theorem2]{Theorem 2} to see that the channel maps the phase $\omega$ to the phase $\sigma^*\omega$, which by assumption is $\omega'$; therefore, $\sigma$ satisfies $\sigma^*\omega=\omega'$, as claimed.

As we mentioned earlier, the second half of \hyperref[theorem2]{Theorem 2} does not generalize from uncorrelated noise to all causal channels. This is because there are causal channels that are not locally $\sigma$-SS yet nevertheless transmute SPT phases. For example, the phase $\omega$ is mapped to $\omega'$ by convex combinations of locally $\sigma_i$-SS channels where each $\sigma_i^*\omega$ equals $\omega'$. Additionally, one can add to the convex combination a channel that is not $\sigma$-SS for any $\sigma$. This extra factor annihilates string operator expectation values and so does not alter the effect of the channel on string order. Finally, there are channels that are $\sigma$-SS but not \emph{locally} $\sigma$-SS. These change the bulk labels of strings from $g$ to $\sigma(g)$ and the end labels from $\chi_\alpha$ to $\chi^\nu_g\chi_\alpha$, where $\nu$ is the projectivity cocycle of $Q$. In doing so, they transform the pattern of zeros in a way that endomorphism actions cannot; for example, if $\sigma=1$, the cohomology invariant of the channel is simply added to that of the state: $\omega\mapsto\omega+\nu$ \cite{PhysRevLett.124.100402}.\footnote{This is not surprising. In closed systems, 1D SPT phases can be prepared by symmetric causal unitaries with index $\nu=\omega$.}

%%%%%%%%%%%%%%%%%%%

\subsubsection{Edge modes perspective}

The transformation of the SPT invariant $\omega$ under a $\sigma$-SS channel can also be seen in terms of edge modes:
\begin{equation}\begin{split}
    \includegraphics[width=0.65\textwidth]{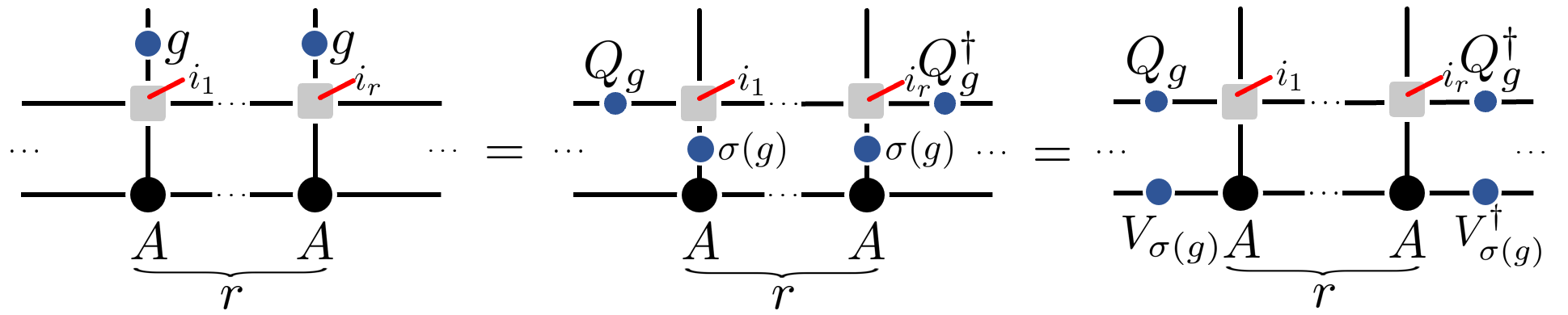}
\end{split}\end{equation}
The $\sigma$-SS condition means that $U_g$ is hit by $\sigma$ upon pulling through $K_i$, and local realization means that $Q$ is linear. Then the edge modes of the evolved MPS state transforms like $\sigma^*V\otimes Q$, which has cocycle $\sigma^*\omega$.

%%%%%%%%%%%%%%

\subsubsection{Irrep probabilities perspective}\label{twisted-probs}

Irrep probabilities were introduced in \cref{sec:irreps}, where it was shown that SS Lindbladian evolution preserves them while non-SS Lindbladian evolution maps them to the fully degenerate value $1/|G|$. In this section, we consider the effects of (not necessarily Lindbladian) causal channels on irrep probabilities.

We find that strongly symmetric channels twisted by automorphisms preserve the degeneracies of irrep probabilities (though permute the irrep probabilities themselves) for all SPT phases, while those twisted by noninvertible endomorphisms reduce the degeneracy if the initial state is in an MNC phase and either reduce or preserve it (depending on the phase) for non-MNC phases. In light of the claim of Ref. \cite{de_Groot_2020} that the degeneracy of irrep probabilities measures SPT complexity, this result reflects the behaviour of complexity that we observed in \cref{endaction}; in particular, in Figure \ref{fig:phases_compare}. Meanwhile, channels that are not $\sigma$-SS for any endormorphism $\sigma$ send the degeneracy to the maximal value $|G|$, regardless of the initial state. This result means that irrep probabilities and their degeneracy detect whether a channel is twisted strongly symmetric.

To see that automorphism-twisted SS channels permute irrep probabilities, use what we learned in \cref{necsuff} about the transformation of the string operators $\langle s(U_g,\mathds{1},\mathds{1})\rangle$ to compute
\begin{equation}
    p_\alpha\longmapsto\frac{1}{|G|}\sum_g\chi_\alpha(g)\langle s(U_{\sigma(g)},\mathds{1},\mathds{1})\rangle=\frac{1}{|G|}\sum_g\chi_\alpha(\sigma^{-1}(g))\langle s(U_g,\mathds{1},\mathds{1})\rangle=p_{(\sigma^{-1})^*\alpha}~.
\end{equation}
A similar computation can be performed for endomorphisms. Fix a set of elements $h\in G$ that represent the cosets of the quotient $G/\ker\sigma$. Then
\begin{align}\begin{split} \label{irrepprobsevolved}
    p_\alpha
    \longmapsto\,&\frac{1}{|G|}\sum_g\chi_\alpha(g)\langle s(U_{\sigma(g)},\mathds{1},\mathds{1})\rangle\\
    =&\left(\frac{|\ker\sigma|}{|G|}\sum_{h}\chi_\alpha(h)\langle s(U_{\sigma(h)},\mathds{1},\mathds{1})\rangle\right)\left(\frac{1}{|\ker\sigma|}\sum_{k\in\ker\sigma}\chi_\alpha(k)\right)~.
\end{split}\end{align}
Orthogonality of characters means that the sum over $k\in\ker\sigma$ enforces the constraint that $\alpha$ restricted to $\ker\sigma$ is trivial. For the characters $\alpha$ without this property (of which there is at least one if $\sigma$ is noninvertible), the corresponding irrep probability $p_\alpha$ must vanish. When at least one of the $p_\alpha$'s vanishes, they cannot be fully degenerate, so there is less degeneracy than for a MNC state. We conclude that SS channels twisted by noninvertible endomorphisms reduce the degeneracy of MNC phases. Finally, channels that are not $\sigma$-SS for any $\sigma$ annihilate the string order parameters without $g=1$, so we get again the result \eqref{maxdegen} that the irrep probabilities become maximally degenerate, regardless of the initial state.

\begin{figure}[t]
\centering
\begin{minipage}{.49\textwidth}
   \centering
    \includegraphics[width=\linewidth, height=.75\linewidth]{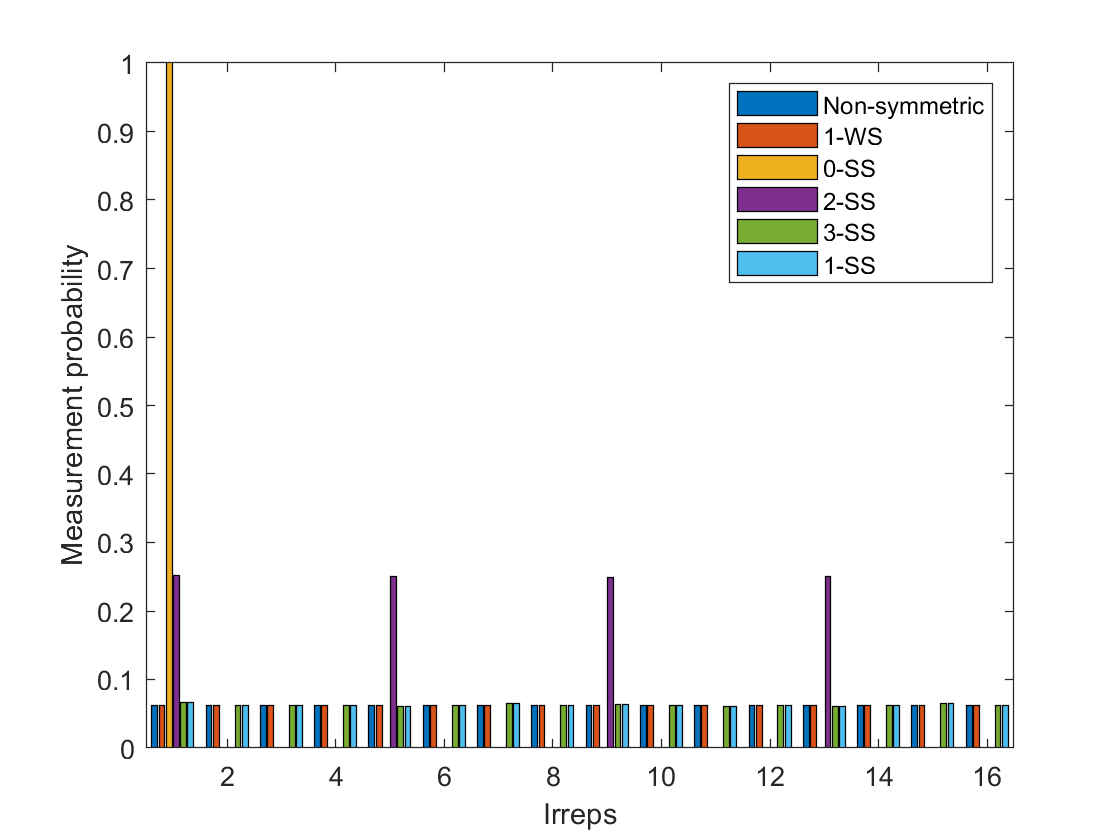}
   \caption{A random state in the trivial phase of $\ZZ_4\times\ZZ_4$, which has nondegenerate irrep probabilities. Under non-$\sigma$-SS channels the irrep probabilities become exactly fully degenerate. For $\sigma$-SS channels, the nondegeneracy of the trivial phase is preserved. The irrep numbered by $i=n+4m$ is given by $(a,b)\mapsto\exp(\pi i (na+mb))$. The channels are defined in the text below.}
    \label{fig:trivial}
\end{minipage}\hfill
\begin{minipage}{.49\textwidth}
  \centering
  \includegraphics[width=\linewidth, height=.75\linewidth]{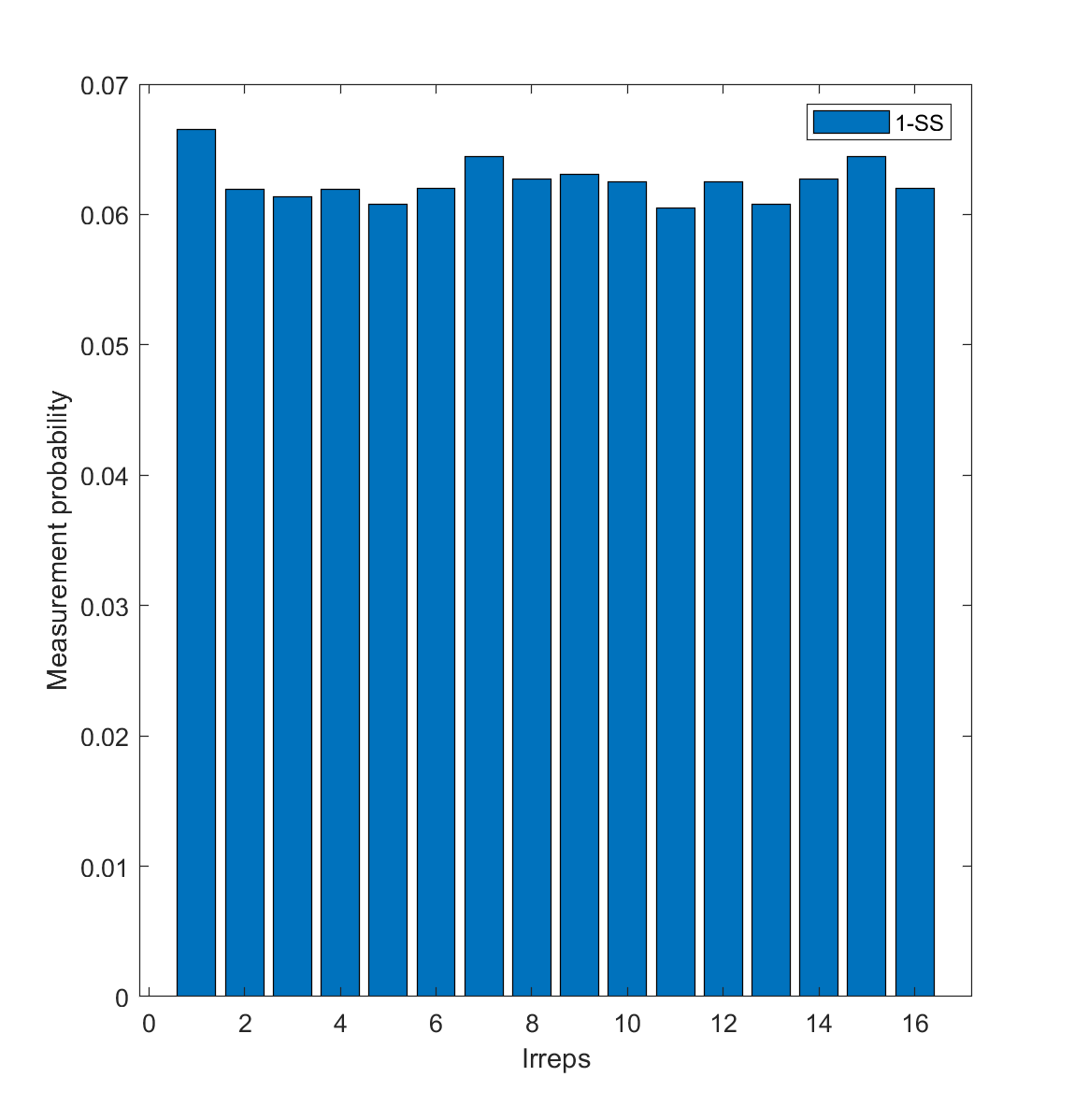}
  \caption{A close-up of the irrep probabilities of a random state in the trivial phase evolved by a $\sigma$-SS channel (in this case, the $1$-SS dephasing channel) shows that they have no enforced degeneracy.\\\,\\\,\\\,\\}
  \label{fig:trivialclose}
\end{minipage}
\end{figure}

To illustrate these results, we perform numerical checks on random example SPT states with symmetry $G=\ZZ_4\times\ZZ_4$. We generate random, injective, symmetric MPS of bond dimension $D = 16$ in a particular SPT phase, with support in all irrep sectors. Then we act on them with channels satisfying various symmetry conditions. The resulting irrep probabilities are depicted in Figures 4, 5 (trivial phase) and Figures 6, 7 (nontrivial phases). For $Z_4\times\ZZ_4$, the automorphisms are those $\sigma$ with $\det\sigma=1,3$, while the noninvertible endomorphisms are those with $\det\sigma=0,2$.

We now give the explicit constructions for the example channels used for this numerical investigation. We denote these twisted strongly symmetric channels by $k$-SS for $k=\det\sigma\in\{0,1,2,3\}$. Each Kraus operator of our example channels is a $16 \times 16$ matrix. We consider one WS but non-SS channel, which is given by the depolarising channel (denoted 1-WS) \eqref{depolarising}, and the SS channel we consider is the dephasing channel (denoted 1-SS) \eqref{DEPHASING}. The channel which is denoted 0-SS, that is SS twisted by the constant endormorphism $\det\sigma=0$, is given by Kraus operators $(K_i)_{ab}=\delta_{ai}\delta_{b1}$. The channel which is denoted 3-SS, that is SS twisted by an automorphism with $\det\sigma=3$, is given by two Kraus operators, each expressed in terms of $4\times 4$ blocks as $K_i=(\tilde K_i)^{\oplus 4}$, where
\begin{equation}
     \quad \tilde K_0 =  \left(\begin{array}{cccc}
         1 & 0 & 0 & 0 \\
         0 & 0 & 0 & 0 \\
         0 & 0& 0 & 0 \\
         0 & 0 & 0 & 0
    \end{array}\right)~,
     \quad \tilde K_1 = \left(\begin{array}{cccc}
         0 & 0 & 0 & 0 \\
         0 & 0 & 1 & 0 \\
         0 & 1 & 0 & 0 \\
         1 & 0 & 0 & 0
    \end{array}\right)~,
\end{equation}
Finally, the channel which is denoted 2-SS, that enacts a $\det(\sigma)=2$ endomorphism twist, is given by four Kraus operators with blocks
\begin{equation}
      \tilde K_i = \Bigg\{~\left(\begin{array}{cccc}
         1 & 0 & 0 & 0 \\
         0 & 0 & 0 & 0 \\
         0 & 0 & 0 & 0 \\
         0 & 0 & 0 & 0
    \end{array}\right),\quad
    \left(\begin{array}{cccc}
         0 & 0 & 0 & 0 \\
         0 & 0 & 0 & 0 \\
         0 & 1 & 0 & 0 \\
         0 & 0 & 0 & 0
    \end{array}\right),\quad
         \left(\begin{array}{cccc}
         0 & 0 & 1 & 0 \\
         0 & 0 & 0 & 0 \\
         0 & 0 & 0 & 0 \\
         0 & 0 & 0 & 0
    \end{array}\right),\quad
     \left(\begin{array}{cccc}
         0 & 0 & 0 & 0 \\
         0 & 0 & 0 & 0 \\
         0 & 0 & 0 & 1 \\
         0 & 0 & 0 & 0
    \end{array}\right)~  \Bigg\}.
\end{equation}

\begin{figure}
\centering
\begin{minipage}{.49\textwidth}
   \centering
    \includegraphics[width=\textwidth]{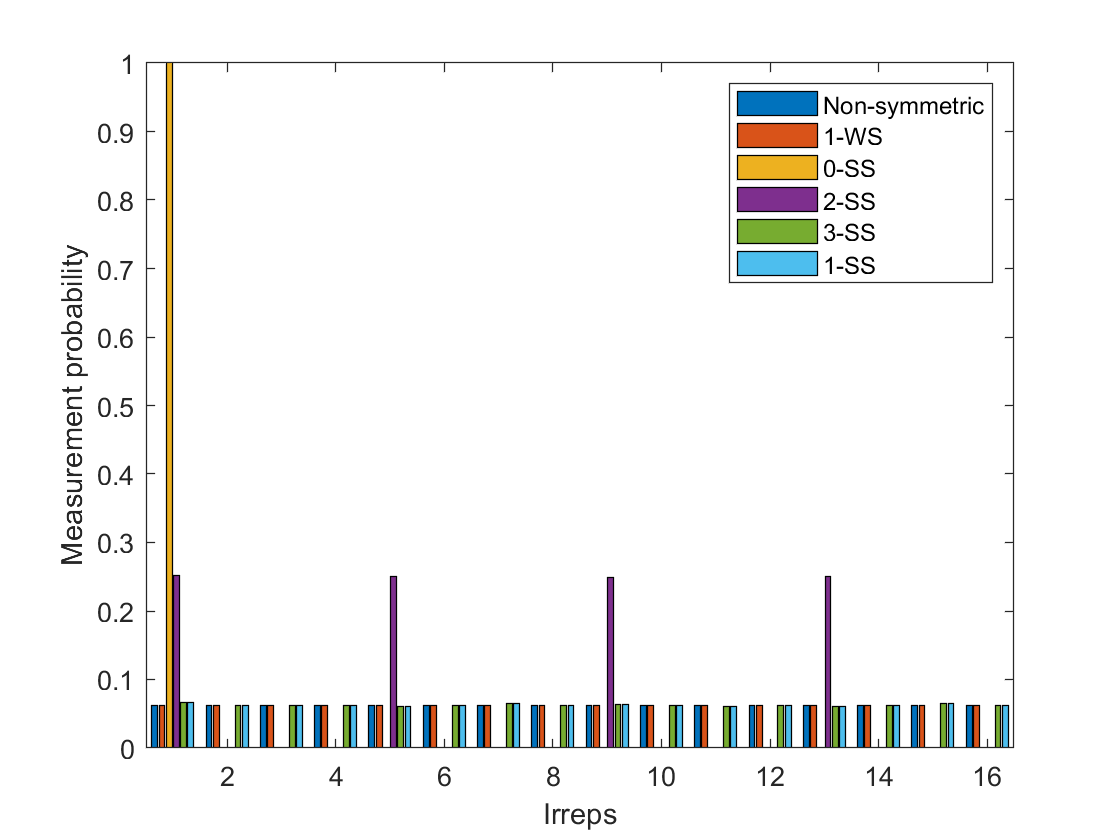}
    \captionof{figure}{Irrep probabilities for a random state in a $\ZZ_4\times\ZZ_4$ MNC SPT phase after evolution by channels satisfying various symmetry conditions. The initial state has maximum degeneracy.}
    \label{fig:irrepsMNC}
\end{minipage}\hfill
\begin{minipage}{.49\textwidth}
  \centering
  \includegraphics[width=\textwidth]{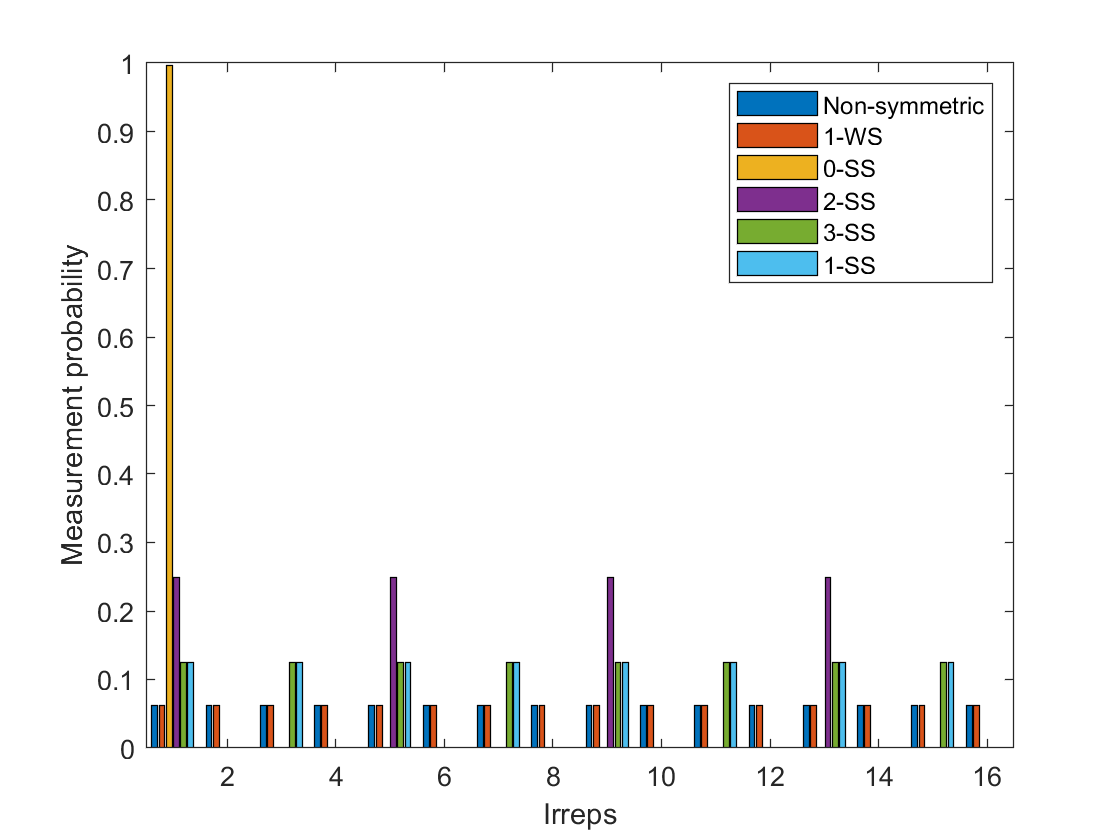}
  \captionof{figure}{Irrep probabilities for a random state in a $\ZZ_4\times\ZZ_4$ non-MNC SPT phase after evolution by channels satisfying various symmetry conditions. The initial state has partial degeneracy.}
  \label{fig:irrepsnonmnc}
\end{minipage}
\end{figure}

Irrep probabilities let one compute the entanglement of a state that is not accessible to local operations and classical communication (LOCC) that respect the symmetry $G$ \cite{de_Groot_2020}. The inaccessible entanglement is given by the entropy of the irrep probabilities 
\begin{equation}\label{inacc}
    E_\text{inacc}=-\sum_\alpha p_\alpha\log_2 p_\alpha.
\end{equation}
The lower bound on this quantity over pure states in an SPT phase is determined by the SPT complexity of the phase and is $\log_2(D_\omega^2)=\log_2(|G|/|K_{\omega}|)$, while the upper bound is given by $\log_2(|G|)$.

While the upper bound on $E_\text{inacc}$ is unchanged by evolution under any (weakly symmetric) channel, the lower bound may decrease or remain the same according to the symmetry condition satisfied by the channel. If the channel is $\sigma$-SS, it changes the SPTO from $[\omega]$ to $[\sigma^*\omega]$ and the lower bound to $\log_2(D_{\sigma^*\omega}^2)$. Then, since the lower bound decreases as SPT complexity decreases, SS and automorphism-twisted SS channels leave the lower bound unchanged, while the lower bound is decreased under SS channels twisted by noninvertible endomorphisms. In particular, channels twisted by the constant endomorphism $\sigma:g\mapsto e$ send the lower bound to zero (since $D_{\sigma^*\omega}=1$). In fact, by Eq. \eqref{irrepprobsevolved}, states evolved by $e$-SS channels saturate the lower bound by concentrating their support in the trivial irrep.

%%%%%%%%%%%%%%%%%%%%%%%%%

\subsubsection{Revisiting the example of the SPTO-destroying Lindbladian}\label{coser-revisited}

Our investigation into the strong symmetry condition was motivated in \cref{coserexample} by Coser and P\'erez-Garc\'ia's example \cite{coser2019classification} of a weakly symmetric Lindbladian that destroys SPTO. Let us now revisit this example and discuss how it fits into the theory of twisted symmetric channels that we have developed in this section.

Recall that this Lindbladian \eqref{coser} is not strongly symmetric, which means by \hyperref[theorem1]{Theorem 1} that it does not preserve SPTO. Moreover, by \hyperref[theorem2]{Theorem 2}, the only channels that map within the space of SPT states are $\sigma$-SS channels, so the channels this Lindbladian generates at finite times must destroy SPTO altogether. To see how this assertion is consistent with the claim of Ref. \cite{coser2019classification} that their Lindbladian maps one SPT phase to another in finite time, notice that they are interested in matching states only approximately (albeit exponentially well), whereas we require exact matching in order to preserve string order.

At infinite time, however, the channel generated by this Lindbladian maps arbitrary states \emph{exactly} to a product state, which has a well-defined SPTO -- the trivial order. This means that the infinite time evolution satisfies the strong symmetry condition twisted by the constant endomorphism $\sigma:g\mapsto e$. This can be seen explicitly by writing out the channel on each site $s$:
\begin{equation}
    \cE_{s,t}=e^{t\cL_s}=e^{t(\cT_s-\mathds{1})}=\cT_s-e^{-t}\cL_s\xrightarrow[]{t\ra\infty}\cT_s~,\qquad\text{for }\cT_s(\rho)=\Tr[\rho]|\phi\rangle\langle\phi|~,
\end{equation}
where we used $\cT_s^2=\cT_s$. This channel has dual $\cT_s^\dagger(X)=\langle\phi|X|\phi\rangle\mathds{1}$, which, since the state $|\phi\rangle$ is chosen to be symmetric, satisfies $\cT_s^\dagger(U_g)=\mathds{1}$, the twisted strong symmetry condition for $\sigma:g\mapsto e$. We remark that the channel $\cT_s$ is generic, in the sense of Eq. \eqref{channel-generic}, despite arising as an infinite time evolution.

The transformation of one SPT phase into another by this Lindbladian is considered in Ref. \cite{coser2019classification}. Let $|\omega\rangle$ denote a state in the phase labeled by $\omega$. Any state $|\omega_1\rangle$ may be transformed into any $|\omega_2\rangle$ by appending a $|\omega_2^{-1}\rangle\otimes|\omega_2\rangle$, then acting with $\cE_\infty\otimes\mathds{1}$, and finally discarding the product state that is reached after infinite time:
\begin{equation}\label{w1tow2}
 |\omega_1\rangle\sim|\omega_1\rangle\otimes|\omega_2^{-1}\rangle\otimes|\omega_2\rangle\stackrel{\cE_\infty}{\longmapsto}|\phi\rangle^{\otimes L}\otimes|\omega_2\rangle\sim|\omega_2\rangle~.
\end{equation}
At first glance, this procedure may seem to suggest that the channel is capable of transforming between arbitrary $G$-SPTOs $\omega_1$ and $\omega_2$, in violation of \hyperref[theorem2]{Theorem 2} and the rule that SPT complexity cannot be increased. This apparent paradox is dissolved by realizing that the full symmetry group of these states is $G\times G\times G$, rather than just the diagonal subgroup $G$. The fact that $\cE_\infty\otimes\mathds{1}$ changes the $G\times G\times G$-SPTO according to the action of the endomorphism $\sigma:(g_1,g_2,g_3)\mapsto (e,e,g_3)$ is consistent with it being strongly symmetric with this twist. It is not possible to reduce the symmetry group to the diagonal factor: either $|\phi\rangle$ is taken to be symmetric (resulting in a copy of $G$ on each factor) or it is taken to be nonsymmetric, in which case the full symmetry group is a copy of $G$ on the third factor. If the states are chosen so that only the third factor of $G$ remains a symmetry, the channel (which acts trivially on the third factor) may be regarded as having untwisted strong symmetry; indeed, the $G_3$-SPTO is $\omega_2$ on either side of the transformation.

%%%%%%%%%%%%%%%
%%%%%%%%%%%%%%%

\section{Conclusion and outlook}\label{sec:conc}

\noindent Our main result is to show that SPTO is preserved by fast evolution of a local Lindbladian precisely if the Lindbladian satisfies a strong symmetry condition; in other words, SPTO is robust to coupling to an environment if and only if the coupling is strongly symmetric. There are at least two ways to interpret this finding. First, it may be taken simply as a rule for determining how order parameters such as string order will transform under a coupling of interest: rather than calculate the full dynamics of a system, one need only look at the symmetry of its generator. Second, the result may be taken as motivation for strong symmetry as the appropriate symmetry condition for classifying symmetry protected phases of open systems. Just as the Lindbladian phase equivalence of Ref.~\cite{coser2019classification} was designed so that local observables are analytic within phases, the strong symmetry condition is chosen so that SPT order parameters are constant within phases.

We have focused on a special class of mixed states: coherent SPT mixtures. These are mixed states that, according to the phase diagram defined by strongly symmetric evolution, lie in the same phase as some pure SPT state and share its SPT invariant $[\omega]\in H^2(G,U(1))$. A question left for the future is what the rest of the phase diagram looks like in one dimension. It would also be interesting to study strong symmetry in higher dimensions, where symmetry-enriched topological orders are present alongside SPT phases.

As a separate result, we determined how causal channels, including those not generated by Lindbladians, interact with SPTO. We found that those satisfying twisted symmetry conditions map between coherent SPT phases, sometimes decreasing but never increasing their complexity. Since the SPT complexity of an SPT phase determines its computational power in measurement-based quantum computing \cite{Stephen_2017}, it would be interesting to study twisted strongly symmetric channels as equivalence relations for a resource theory.

This research raises several other questions for future investigation. Firstly, since string orders are experimentally tractable \cite{Smith_2019, azses2020identification}, one can ask how to detect mixed state SPTO in experiment. Secondly, our work considers SPTO at finite time, as at infinite times fingerprints of SPTO such as string order get washed out. This raises the question of whether coherent SPT mixtures arise as steady states of Lindbladians. Also, what are the implications of our findings for SPTO at finite temperature? This would clarify further the nature of SPT mixed states. And finally, it would be interesting to explore further how our findings relate to other properties of SPT phases, their boundaries and transitions, as studied in previous work \cite{coser2019classification,PhysRevLett.125.240405,Bardyn_2013}. 

%%%%%%%%%%%%%%%%%

\section{Acknowledgements}

We are grateful for discussions with \'A.~Capel Cuevas, S.~Lieu, D.~Malz, D.~P\'erez-Garc\'ia, and D.~T.~Stephen. 
C.d.G.\ and N.S.\ have been supported 
by the Deutsche Forschungsgemeinschaft (DFG) under Germany's Excellence Strategy through the Munich Center for Quantum Science and Technology MCQST (EXC-2111 -- 390814868). 
A.T.\ acknowledges support from the Max Planck Harvard Research Center for Quantum Optics (MPHQ). 
N.S.\ further acknowledges support by the European Research Council (ERC) under the European Union's Horizon 2020 research and innovation programme through the ERC-CoG SEQUAM (Grant No.~863476).

\bibliographystyle{quantum}
\bibliography{bibliography.bib}

\begin{thebibliography}{10}

\bibitem{HALDANE1983464}
F.D.M. Haldane.
\newblock ``Continuum dynamics of the 1-d heisenberg antiferromagnet:
  Identification with the $o(3)$ nonlinear sigma model''.
\newblock
  \href{https://dx.doi.org/https://doi.org/10.1016/0375-9601(83)90631-X}{Physics
  Letters A {\bf 93}, 464--468}~(1983).

\bibitem{PhysRevLett.50.1153}
F.~D.~M. Haldane.
\newblock ``Nonlinear field theory of large-spin heisenberg antiferromagnets:
  Semiclassically quantized solitons of the one-dimensional easy-axis n\'eel
  state''.
\newblock \href{https://dx.doi.org/10.1103/PhysRevLett.50.1153}{Phys. Rev.
  Lett. {\bf 50}, 1153--1156}~(1983).

\bibitem{PhysRevLett.59.799}
Ian Affleck, Tom Kennedy, Elliott~H. Lieb, and Hal Tasaki.
\newblock ``Rigorous results on valence-bond ground states in
  antiferromagnets''.
\newblock \href{https://dx.doi.org/10.1103/PhysRevLett.59.799}{Phys. Rev. Lett.
  {\bf 59}, 799--802}~(1987).

\bibitem{PhysRevB.40.4709}
Marcel den Nijs and Koos Rommelse.
\newblock ``Preroughening transitions in crystal surfaces and valence-bond
  phases in quantum spin chains''.
\newblock \href{https://dx.doi.org/10.1103/PhysRevB.40.4709}{Phys. Rev. B {\bf
  40}, 4709--4734}~(1989).

\bibitem{PhysRevB.45.304}
Tom Kennedy and Hal Tasaki.
\newblock ``Hidden $\mathbb{Z}_2\times\mathbb{Z}_2$ symmetry breaking in
  haldane-gap antiferromagnets''.
\newblock \href{https://dx.doi.org/10.1103/PhysRevB.45.304}{Phys. Rev. B {\bf
  45}, 304--307}~(1992).

\bibitem{pollmann2012detection}
Frank Pollmann and Ari~M. Turner.
\newblock ``Detection of symmetry-protected topological phases in one
  dimension''.
\newblock \href{https://dx.doi.org/10.1103/PhysRevB.86.125441}{Phys. Rev. B
  {\bf 86}, 125441}~(2012).

\bibitem{pollmann:1d-sym-protection-prb}
F.~{Pollmann}, A.~M. {Turner}, E.~{Berg}, and M.~{Oshikawa}.
\newblock ``{Entanglement spectrum of a topological phase in one dimension}''.
\newblock \href{https://dx.doi.org/10.1103/PhysRevB.81.064439}{Phys. Rev. B
  {\bf 81}, 064439}~(2010).
\newblock  \href{http://arxiv.org/abs/0910.1811}{arXiv:0910.1811}.

\bibitem{schollwock}
Ulrich Schollwöck.
\newblock ``The density-matrix renormalization group in the age of matrix
  product states''.
\newblock
  \href{https://dx.doi.org/https://doi.org/10.1016/j.aop.2010.09.012}{Annals of
  Physics {\bf 326}, 96--192}~(2011).

\bibitem{cirac2021matrix}
Ignacio Cirac, David Perez-Garcia, Norbert Schuch, and Frank Verstraete.
\newblock ``Matrix product states and projected entangled pair states:
  Concepts, symmetries, and theorems''.
\newblock Rev. Mod. Phys. {\bf 93}, 045003~(2021).
\newblock  \href{http://arxiv.org/abs/2011.1212}{arXiv:2011.12127}.

\bibitem{hastings2007area}
M~B Hastings.
\newblock ``An area law for one-dimensional quantum systems''.
\newblock \href{https://dx.doi.org/10.1088/1742-5468/2007/08/p08024}{Journal of
  Statistical Mechanics: Theory and Experiment {\bf 2007},
  P08024–P08024}~(2007).

\bibitem{verstraete:faithfully}
F.~Verstraete and J.~I. Cirac.
\newblock ``Matrix product states represent ground states faithfully''.
\newblock Phys. Rev. B {\bf 73}, 094423~(2006).
\newblock
  \href{http://arxiv.org/abs/cond-mat/0505140}{arXiv:cond-mat/0505140}.

\bibitem{schuch:mps-entropies}
Norbert Schuch, Michael~M. Wolf, Frank Verstraete, and J.~Ignacio Cirac.
\newblock ``Entropy scaling and simulability by matrix product states''.
\newblock Phys.\ Rev.\ Lett. {\bf 100}, 30504~(2008).
\newblock  \href{http://arxiv.org/abs/0705.0292}{arXiv:0705.0292}.

\bibitem{molnar:normal-peps-fundamentalthm}
Andras {Molnar}, Jos{\'e} {Garre-Rubio}, David {P{\'e}rez-Garc{\'\i}a}, Norbert
  {Schuch}, and J.~Ignacio {Cirac}.
\newblock ``{Normal projected entangled pair states generating the same
  state}''.
\newblock \href{https://dx.doi.org/10.1088/1367-2630/aae9fa}{New J. Phys. {\bf
  20}, 113017}~(2018).
\newblock  \href{http://arxiv.org/abs/1804.0496}{arXiv:1804.04964}.

\bibitem{PhysRevB.85.075125}
Frank Pollmann, Erez Berg, Ari~M. Turner, and Masaki Oshikawa.
\newblock ``Symmetry protection of topological phases in one-dimensional
  quantum spin systems''.
\newblock \href{https://dx.doi.org/10.1103/PhysRevB.85.075125}{Phys. Rev. B
  {\bf 85}, 075125}~(2012).

\bibitem{chen2011classification}
Xie Chen, Zheng-Cheng Gu, and Xiao-Gang Wen.
\newblock ``Classification of gapped symmetric phases in one-dimensional spin
  systems''.
\newblock \href{https://dx.doi.org/10.1103/PhysRevB.83.035107}{Phys. Rev. B
  {\bf 83}, 035107}~(2011).

\bibitem{schuch2011classifying}
Norbert Schuch, David P\'erez-Garc\'{\i}a, and Ignacio Cirac.
\newblock ``Classifying quantum phases using matrix product states and
  projected entangled pair states''.
\newblock \href{https://dx.doi.org/10.1103/PhysRevB.84.165139}{Phys. Rev. B
  {\bf 84}, 165139}~(2011).

\bibitem{chen:spt-bosons}
Xie Chen, Zheng-Cheng Gu, Zheng-Xin Liu, and Xiao-Gang Wen.
\newblock ``Symmetry protected topological orders in interacting bosonic
  systems''.
\newblock Science {\bf 338}, 1604~(2012).
\newblock  \href{http://arxiv.org/abs/1301.0861}{arXiv:1301.0861}.

\bibitem{PhysRevA.71.062313}
Robert Raussendorf, Sergey Bravyi, and Jim Harrington.
\newblock ``Long-range quantum entanglement in noisy cluster states''.
\newblock \href{https://dx.doi.org/10.1103/PhysRevA.71.062313}{Phys. Rev. A
  {\bf 71}, 062313}~(2005).

\bibitem{Hastings_2011}
Matthew~B. Hastings.
\newblock ``Topological order at nonzero temperature''.
\newblock \href{https://dx.doi.org/10.1103/physrevlett.107.210501}{Physical
  Review Letters{\bf 107}}~(2011).

\bibitem{Roberts_2017}
Sam Roberts, Beni Yoshida, Aleksander Kubica, and Stephen~D. Bartlett.
\newblock ``Symmetry-protected topological order at nonzero temperature''.
\newblock \href{https://dx.doi.org/10.1103/physreva.96.022306}{Physical Review
  A{\bf 96}}~(2017).

\bibitem{Diehl_2011}
Sebastian Diehl, Enrique Rico, Mikhail~A. Baranov, and Peter Zoller.
\newblock ``Topology by dissipation in atomic quantum wires''.
\newblock \href{https://dx.doi.org/10.1038/nphys2106}{Nature Physics {\bf 7},
  971–977}~(2011).

\bibitem{Bardyn_2013}
C-E Bardyn, M~A Baranov, C~V Kraus, E~Rico, A~İmamoğlu, P~Zoller, and
  S~Diehl.
\newblock ``Topology by dissipation''.
\newblock \href{https://dx.doi.org/10.1088/1367-2630/15/8/085001}{New Journal
  of Physics {\bf 15}, 085001}~(2013).

\bibitem{Kraus_2008}
B.~Kraus, H.~P. Büchler, S.~Diehl, A.~Kantian, A.~Micheli, and P.~Zoller.
\newblock ``Preparation of entangled states by quantum markov processes''.
\newblock \href{https://dx.doi.org/10.1103/physreva.78.042307}{Physical Review
  A{\bf 78}}~(2008).

\bibitem{zhou2017symmetryprotected}
Leo Zhou, Soonwon Choi, and Mikhail~D. Lukin.
\newblock ``Symmetry-protected dissipative preparation of matrix product
  states''~(2017).
\newblock  \href{http://arxiv.org/abs/1706.01995}{arXiv:1706.01995}.

\bibitem{PhysRevLett.125.240405}
Simon Lieu, Ron Belyansky, Jeremy~T. Young, Rex Lundgren, Victor~V. Albert, and
  Alexey~V. Gorshkov.
\newblock ``Symmetry breaking and error correction in open quantum systems''.
\newblock \href{https://dx.doi.org/10.1103/PhysRevLett.125.240405}{Phys. Rev.
  Lett. {\bf 125}, 240405}~(2020).

\bibitem{albert2018lindbladians}
Victor~V. Albert.
\newblock ``Lindbladians with multiple steady states: theory and
  applications''~(2018).
\newblock  \href{http://arxiv.org/abs/1802.00010}{arXiv:1802.00010}.

\bibitem{Bu_a_2012}
Berislav Bu{\v{c}}a and Toma{\v{z}} Prosen.
\newblock ``A note on symmetry reductions of the lindblad equation: transport
  in constrained open spin chains''.
\newblock \href{https://dx.doi.org/10.1088/1367-2630/14/7/073007}{New Journal
  of Physics {\bf 14}, 073007}~(2012).

\bibitem{PhysRevA.89.022118}
Victor~V. Albert and Liang Jiang.
\newblock ``Symmetries and conserved quantities in lindblad master equations''.
\newblock \href{https://dx.doi.org/10.1103/PhysRevA.89.022118}{Phys. Rev. A
  {\bf 89}, 022118}~(2014).

\bibitem{lieu2020albertQECsym}
Simon Lieu, Ron Belyansky, Jeremy~T. Young, Rex Lundgren, Victor~V. Albert, and
  Alexey~V. Gorshkov.
\newblock ``Symmetry breaking and error correction in open quantum systems''.
\newblock \href{https://dx.doi.org/10.1103/physrevlett.125.240405}{Physical
  Review Letters{\bf 125}}~(2020).

\bibitem{coser2019classification}
Andrea Coser and David P{\'e}rez-Garc{\'\i}a.
\newblock ``Classification of phases for mixed states via fast dissipative
  evolution''.
\newblock
  \href{https://dx.doi.org/https://doi.org/10.22331/q-2019-08-12-174}{Quantum
  {\bf 3}, 174}~(2019).

\bibitem{verstraete2006matrix}
F.~Verstraete and J.~I. Cirac.
\newblock ``Matrix product states represent ground states faithfully''.
\newblock \href{https://dx.doi.org/10.1103/PhysRevB.73.094423}{Phys. Rev. B
  {\bf 73}, 094423}~(2006).

\bibitem{biamonte2017tensor}
Jacob Biamonte and Ville Bergholm.
\newblock ``Tensor networks in a nutshell''~(2017).
\newblock  \href{http://arxiv.org/abs/1708.00006}{arXiv:1708.00006}.

\bibitem{ORUS2014117}
Román Orús.
\newblock ``A practical introduction to tensor networks: Matrix product states
  and projected entangled pair states''.
\newblock
  \href{https://dx.doi.org/https://doi.org/10.1016/j.aop.2014.06.013}{Annals of
  Physics {\bf 349}, 117--158}~(2014).

\bibitem{handwaving}
Jacob~C. Bridgeman and Christopher~T. Chubb.
\newblock ``Hand-waving and interpretive dance: An introductory course on
  tensor networks''.
\newblock \href{https://dx.doi.org/10.1088/1751-8121/aa6dc3}{J. Phys. A: Math.
  Theor.{\bf 50}}~(2017).
\newblock  \href{http://arxiv.org/abs/1603.0303}{arXiv:1603.03039}.

\bibitem{perez2006matrix}
D.~Perez-Garcia, F.~Verstraete, M.~M. Wolf, and J.~I. Cirac.
\newblock ``Matrix product state representations''.
\newblock
  \href{https://dx.doi.org/https://doi.org/10.48550/arXiv.quant-ph/0608197}{Quantum
  Info. Comput. {\bf 7}, 401–430}~(2007).

\bibitem{nielsen2002quantum}
Michael~A. Nielsen and Isaac~L. Chuang.
\newblock ``Quantum computation and quantum information: 10th anniversary
  edition''.
\newblock Cambridge University Press. ~(2010).

\bibitem{wolf}
Michael~M. Wolf.
\newblock ``Quantum channels and operations: Guided tour''~(2012).

\bibitem{benenti-casati-strini}
Giuliano Benenti, Giulio Casati, and Giuliano Strini.
\newblock ``Principles of quantum computation and information''.
\newblock \href{https://dx.doi.org/10.1142/5528}{World Scientific}. ~(2004).
\newblock
  \href{http://arxiv.org/abs/https://www.worldscientific.com/doi/pdf/10.1142/5528}{arXiv:https://www.worldscientific.com/doi/pdf/10.1142/5528}.

\bibitem{fultonharris}
W.~Fulton and J.~Harris.
\newblock ``Representation theory: A first course''.
\newblock Springer New York. ~(2013).
\newblock
  url:~\href{https://books.google.de/books?id=6TwmBQAAQBAJ}{books.google.de/books?id=6TwmBQAAQBAJ}.

\bibitem{breuer2002theory}
Heinz-Peter Breuer and Francesco Petruccione.
\newblock ``{The Theory of Open Quantum Systems}''.
\newblock
  \href{https://dx.doi.org/10.1093/acprof:oso/9780199213900.001.0001}{Oxford
  University Press}. ~(2007).

\bibitem{Haegeman_2012}
Jutho Haegeman, David Pérez-García, Ignacio Cirac, and Norbert Schuch.
\newblock ``Order parameter for symmetry-protected phases in one dimension''.
\newblock \href{https://dx.doi.org/10.1103/physrevlett.109.050402}{Physical
  Review Letters{\bf 109}}~(2012).

\bibitem{shiozakiryu}
Ken Shiozaki and Shinsei Ryu.
\newblock ``Matrix product states and equivariant topological field theories
  for bosonic symmetry-protected topological phases in (1+1) dimensions''.
\newblock \href{https://dx.doi.org/10.1007/JHEP04(2017)100}{J. High Energ.
  Phys.{\bf 100}}~(2017).

\bibitem{PhysRevB.96.075125}
Anton Kapustin, Alex Turzillo, and Minyoung You.
\newblock ``Topological field theory and matrix product states''.
\newblock \href{https://dx.doi.org/10.1103/PhysRevB.96.075125}{Phys. Rev. B
  {\bf 96}, 075125}~(2017).

\bibitem{else2012symmetry}
Dominic~V Else, Stephen~D Bartlett, and Andrew~C Doherty.
\newblock ``Symmetry protection of measurement-based quantum computation in
  ground states''.
\newblock
  \href{https://dx.doi.org/https://doi.org/10.1088/1367-2630/14/11/113016}{New
  Journal of Physics {\bf 14}, 113016}~(2012).

\bibitem{de_Groot_2020}
Caroline de~Groot, David~T Stephen, Andras Molnar, and Norbert Schuch.
\newblock ``Inaccessible entanglement in symmetry protected topological
  phases''.
\newblock \href{https://dx.doi.org/10.1088/1751-8121/ab98c7}{Journal of Physics
  A: Mathematical and Theoretical {\bf 53}, 335302}~(2020).

\bibitem{berkovich1998}
I.A.G. Berkovich, L.S. Kazarin, and E.M. Zhmud.
\newblock ``Characters of finite groups''.
\newblock De Gruyter. ~(2018).

\bibitem{PhysRevLett.125.190402}
Lorenzo Piroli and J.~Ignacio Cirac.
\newblock ``Quantum cellular automata, tensor networks, and area laws''.
\newblock \href{https://dx.doi.org/10.1103/PhysRevLett.125.190402}{Phys. Rev.
  Lett. {\bf 125}, 190402}~(2020).

\bibitem{Ignacio_Cirac_2017}
J~Ignacio Cirac, David Perez-Garcia, Norbert Schuch, and Frank Verstraete.
\newblock ``Matrix product unitaries: structure, symmetries, and topological
  invariants''.
\newblock \href{https://dx.doi.org/10.1088/1742-5468/aa7e55}{Journal of
  Statistical Mechanics: Theory and Experiment {\bf 2017}, 083105}~(2017).

\bibitem{PhysRevB.98.245122}
M.~Burak \ifmmode \mbox{\c{S}}\else \c{S}\fi{}ahino\ifmmode~\breve{g}\else
  \u{g}\fi{}lu, Sujeet~K. Shukla, Feng Bi, and Xie Chen.
\newblock ``Matrix product representation of locality preserving unitaries''.
\newblock \href{https://dx.doi.org/10.1103/PhysRevB.98.245122}{Phys. Rev. B
  {\bf 98}, 245122}~(2018).

\bibitem{gnvw}
D.~Gross, V.~Nesme, and H.~Vogts.
\newblock ``Index theory of one dimensional quantum walks and cellular
  automata''.
\newblock \href{https://dx.doi.org/10.1007/s00220-012-1423-1}{Commun. Math.
  Phys. {\bf 310}, 419–454}~(2012).

\bibitem{PhysRevLett.124.100402}
Zongping Gong, Christoph S\"underhauf, Norbert Schuch, and J.~Ignacio Cirac.
\newblock ``Classification of matrix-product unitaries with symmetries''.
\newblock \href{https://dx.doi.org/10.1103/PhysRevLett.124.100402}{Phys. Rev.
  Lett. {\bf 124}, 100402}~(2020).

\bibitem{Stephen_2017}
David~T. Stephen, Dong-Sheng Wang, Abhishodh Prakash, Tzu-Chieh Wei, and Robert
  Raussendorf.
\newblock ``Computational power of symmetry-protected topological phases''.
\newblock \href{https://dx.doi.org/10.1103/physrevlett.119.010504}{Physical
  Review Letters{\bf 119}}~(2017).

\bibitem{Smith_2019}
Adam Smith, M.~S. Kim, Frank Pollmann, and Johannes Knolle.
\newblock ``Simulating quantum many-body dynamics on a current digital quantum
  computer''.
\newblock \href{https://dx.doi.org/10.1038/s41534-019-0217-0}{npj Quantum
  Information{\bf 5}}~(2019).

\bibitem{azses2020identification}
Daniel Azses, Rafael Haenel, Yehuda Naveh, Robert Raussendorf, Eran Sela, and
  Emanuele~G. Dalla~Torre.
\newblock ``Identification of symmetry-protected topological states on noisy
  quantum computers''.
\newblock \href{https://dx.doi.org/10.1103/PhysRevLett.125.120502}{Phys. Rev.
  Lett. {\bf 125}, 120502}~(2020).

\end{thebibliography}

\appendix

\end{document}